%% file: main.tex
\documentclass[
aps,
prx,
showpacs,
preprintnumbers,
twocolumn,
superscriptaddress,
10pt,
longbibliography,
nofootinbib,
]{revtex4-2}

\usepackage[T1]{fontenc}
\usepackage{newtxtext}
\usepackage{newtxmath}

\usepackage{amsmath}
\usepackage{graphicx}
\usepackage[normalem]{ulem}

\usepackage{algorithm}
\usepackage{algorithmic}

\usepackage{enumitem}

\usepackage{makecell}

\usepackage{physics}
\usepackage{amsmath}
\usepackage{mathtools}
\usepackage{dsfont}

\usepackage{nicematrix} 

\usepackage{appendix}

\usepackage{placeins}

\usepackage{color}
\usepackage[dvipsnames]{xcolor}
\usepackage{colortbl}
\usepackage{hyperref}
\hypersetup{colorlinks,citecolor=blue,linkcolor=blue,urlcolor=blue}

\definecolor{darkgreen}{rgb}{0.55, 0.71, 0.00}
\definecolor{Gray}{gray}{0.9}

\newcommand{\melem}[3]{\langle #1 \vert #2 \vert #3 \rangle}

\usepackage{graphicx}
\usepackage[all]{hypcap}
\graphicspath{{visual_elements/supp_figs/}{visual_elements/main_figs/}}

\usepackage{tikz}
\usetikzlibrary{shapes.geometric, decorations.pathmorphing, decorations.pathreplacing, arrows.meta}

\tikzset{
  every path/.style={line width=1.4pt},
  photon/.style={red, line width=0.7pt, decorate,
    decoration={snake, amplitude=1.5pt, segment length=5pt}},
  vleg/.style={green!60!black, line width=1.4pt},
  bleg/.style={blue, line width=1.4pt}
}

\newcommand{\nocontentsline}[3]{}
\newcommand{\tocless}[2]{\bgroup\let\addcontentsline=\nocontentsline#1{#2}\egroup}

\usepackage{comment}
\usepackage{todonotes}

\definecolor{nqdcolor}{rgb}{0.5586, 0.0586, 0.4219}

\input{preamble/macros}

\newcommand{\PKS}{Max Planck Institute for the Physics of Complex Systems, MPI-PKS, N\"othnitzerstr.~38, 01187 Dresden, Germany}

\begin{document}

\title{Floquet ground state order}

\author{Suman Mondal}
\email{mondal@pks.mpg.de}
\author{Paul M.~Schindler}
\author{Marin Bukov}
\email{mgbukov@pks.mpg.de}
\affiliation{\PKS}

\begin{abstract}
Clean, interacting, periodically driven systems are believed to thermalize to featureless, infinite-temperature states that preclude the formation of stable ordered states.
Here, we show stable quantum order, distinct from prethermal order, and its phase transitions in the ground state of the average-energy operator \AE. 
Decomposing the time evolution into geometric and average-energy dynamics, we apply Geometric Floquet theory to identify the Floquet ground state. We derive a general variational principle to compute the Floquet ground state in large interacting periodically driven systems, and use it to develop a matrix product state algorithm.    
In driven nonintegrable chains of more than 100 interacting spins, our numerical simulations exhibit a gapped, area-law entangled Floquet ground state in extended parameter regimes; the ground-state manifold can support both dressed equilibrium order (e.g., Floquet antiferromagnets) and inherently nonequilibrium order (e.g., ground-state time crystals). This quantum order is protected by a finite average-energy gap, which closes at phase transition points upon changing model parameters. 
We decompose Floquet phases of matter into their equilibrium and nonequilibrium elements: whereas the Floquet ground state at any fixed time is a representative of an equilibrium phase of matter, the nonequilibrium content is entirely produced by the drive holonomy and is geometric.  
Finally, we identify quasilocal counterdiabatic driving as a mechanism for the stability against heating of gapped Floquet ground states in nonintegrable quantum systems, and argue why Floquet ground states evade the Eigenstate Thermalization Hypothesis.  
Our work enables the study of interacting nonequilibrium matter deep into the many-body regime, and lays the foundations for the engineering of strongly correlated ordered quantum states.
\end{abstract}

\maketitle

\section{\label{sec:intro}Introduction}
A core principle of modern physics is understanding how macroscopic order self-organizes from individual particle interactions.
Ordered states explain the emergence of complex large-scale structures from simple microscopic rules.
Exploring this collective behavior reveals entirely new physical laws that do not exist at the individual particle level~\cite{anderson1972more}. 
For instance, the modern classification of phases of matter, such as liquids, solids, or superconductors, relies on the specific type of internal order they possess~\cite{fisher1998renormalization}.
Statistical physics distinguishes between thermodynamic order arising from a competition between energy and entropy at finite temperature, and quantum order occurring in ground states at zero temperature as a result of competing interactions obeying Heisenberg's uncertainty principle~\cite{Sachdev2011Quantum}. 

\begin{figure}[t!]
    \centering
    \includegraphics[width=\linewidth]{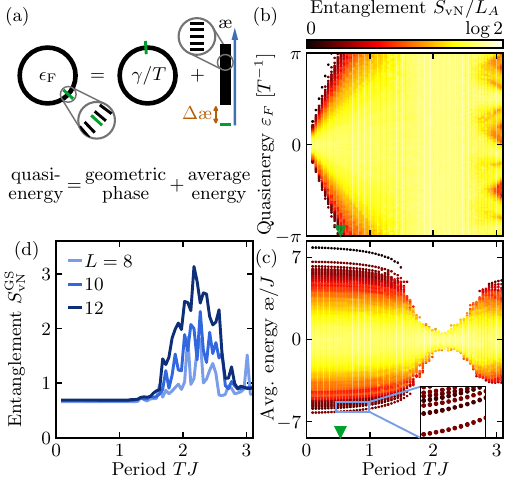}
    \caption{
    \textbf{Geometric Floquet theory reveals a gapped low-entangled Floquet ground state.}
    (a) For large quantum many-body systems, quasienergies $\varepsilon_F$ occupy the unit circle densely, and no distinguished Floquet state can be identified. Decomposing them into geometric phase $\gamma$ and average energy $\mae$ can single out a distinct lowest-lying Floquet state (green), protected by a finite $\mae$-gap from a continuum of excitations -- the Floquet ground state candidate. 
    Quasienergy (b) and average energy (c) sorting of the exact Floquet eigenstates against driving period in the open nonintegrable Floquet Ising chain~\eqref{eq:Ising_chain} of $L{=}12$ spins, obtained using exact diagonalization; colorbar shows half-chain entanglement per spin. 
    Average energy is agnostic to the quasienergy folding threshold (green triangle); it exhibits a gapped, weakly-entangled ground state in both the high-frequency regime (small $JT$) and the nonperturbative DTC regime ($JT{\approx}\pi$), surrounding a gapless ergodic regime ($JT{\approx}2$). The ground state [dark red] is not visible in the quasienergy spectrum, which is exponentially dominated by infinite-temperature states [bright yellow].   
    (d) Half-chain von Neumann entanglement entropy against the period of the Floquet ground state. Parameters: $h_x/J=0.6$, $h_z/J=0.3$. 
    }
    \label{fig:GFT_ED}
\end{figure} 

A major frontier in contemporary research is to extend the notion of ordered quantum states away from equilibrium~\cite{moessner2017equilibration}. However, driven systems do not conserve energy, which poses a formidable challenge to the design of self-contained frameworks for nonequilibrium-ordered states: in experiments, external drives cause the system to heat and render its state unstable at long times~\cite{dalessio2016quantum}. This is rooted in a fundamental obstruction: the absence of a well-defined energy obviates the notion of a stable lowest-energy ground state. 
To avoid this, a new form of eigenstate quantum order was proposed~\cite{huse2013localization}, which occurs at arbitrary energy density and is stabilized by many-body localization~\cite{abanin2019mbl}. In the absence of a stabilizing mechanism, heating can be at best delayed, and the system exhibits prethermal order~\cite{ho2023quantum} -- a long-lived metastable variant of thermodynamic order which is ultimately destined to melt.  
By contrast, here we show that clean (disorder-free) periodically driven (Floquet) systems can host a new kind of stable nonequilibrium quantum order, which we name Floquet ground state order.  

To do this, we overcome a major conceptual challenge: generic Floquet systems do not possess a ground state since quasienergy is defined up to arbitrary integer shifts of the drive frequency~\cite{sambe1973steady}. In the thermodynamic limit, an exponentially large number of states in the many-body Hilbert space causes the quasienergy spectrum to populate the unit circle densely, and precludes the appearance of spectral gaps, cf.~Fig.~\ref{fig:GFT_ED}(a).   
Na\"ively, it appears as if one cannot define stable quantum order in Floquet eigenstates, which requires gap protection. 
By developing a different gauge-equivalent description, we demonstrate the existence of stable parameter regimes where these issues can be resolved. 

We use Geometric Floquet theory -- a new perspective on the theory of periodically driven systems based on adiabaticity~\cite{schindler2025geometric,perez2026floquet}. Instead of working with folded quasienergies, decomposing the dynamics of Floquet eigenstates into a geometric and dynamical phase allows for their unambiguous sorting using a concept called average energy. Exact simulations on small spin chains indicate the formation of parameter regimes where the lowest average-energy state (henceforth called the Floquet ground state) is gapped and weakly entangled, see Fig.~\ref{fig:GFT_ED}(b,c), which can explain previous observations of its anomalous stability~\cite{heyl2019quantum,ikeda2024robust}.   
A natural question is whether this behavior persists for larger system sizes. If yes, can the Floquet ground state be ordered, and does it possess new forms of inherently nonequilibrium order? Can it support stable quantum phases of matter and exhibit proper phase transitions?

Here we answer these questions in the affirmative by showing that the average-energy ground state of nonintegrable periodically driven spin chains can be gapped and area-law entangled in the many-body regime of more than one hundred spins. We achieve this by formulating a general variational principle for the average energy operator and its ground state, and applying it to one-dimensional Floquet spin chains using a matrix product state (MPS) ansatz. An extrapolation in the system size is consistent with a finite average-energy gap above the area-law ground state in the thermodynamic limit. 

These features allow us to introduce the concept of Floquet ground-state order -- a generalization of zero-temperature quantum order to out-of-equilibrium periodically driven systems. 
To illustrate it, we exhibit examples of symmetry breaking in the Floquet ground-state, including long-range ordered equilibrium-like dressed antiferromagnets and a new kind of inherently nonequilibrium ground-state time crystal without a static counterpart. 
We further demonstrate the stability of the average energy gap protecting the ordered state to changes in the couplings of the driven Hamiltonian. This leads us to define the notion of Floquet ground-state phases of matter and their phase transitions in clean (disorder-free) systems. We exemplify it by providing a paradigmatic exactly solvable driven XY spin chain and analyzing its Floquet ground-state phase diagram; however, the concept is generic and applies to nonintegrable Floquet systems, as we demonstrate numerically. 

We apply Geometric Floquet theory to deconstruct each Floquet phase of matter into its equilibrium-like and nonequilibrium elements.  
The average-energy ground state at any fixed time is indistinguishable from the ground state of a static system, and can therefore be connected to representatives of an equilibrium phase of matter using quasiadiabatic continuation without closing the average energy gap. 
The inherently nonequilibrium nature of certain Floquet phases is then entirely borne by the holonomy (i.e., the geometric phases); the latter provides a dynamical obstruction to connecting a nonequilibrium-ordered phase to an equilibrium-ordered phase in real time without closing this gap.  

Finally, we argue that certain gapped area-law Floquet ground states are infinitely long-lived, and hence the corresponding phases of matter are stable against energy absorption from the drive that creates them. We exhibit quasilocal counterdiabatic driving as the mechanism that prevents energy absorption by exactly eliminating all transitions out of the ground-state manifold. In contrast to quantum many-body scars, embedded in the bulk of the spectrum with no gap or transport structure, a Floquet ground state occurs at the edge of the average-energy spectrum; the gap above it renders the counterdiabatic gauge potential that parallel-transports the state quasilocal.
In turn, this stability, whenever present in the thermodynamic limit, implies the existence of a well-defined ergodic phase with finite boundaries, where Floquet's theorem breaks down, and no physically meaningful Floquet ground state exists. 
In the broad context of quantum thermalization, our work demonstrates that not all Floquet eigenstates in nonintegrable periodically driven many-body systems, are locally equivalent to featureless infinite-temperature states: Floquet ground states evade the Eigenstate Thermalization Hypothesis (ETH) protected by a finite average-energy gap. 

The variational principle we develop opens new avenues for investigating Floquet ground-state physics in systems of hundreds of driven interacting particles. 
We anticipate this to remove an existing bottleneck with Floquet-engineering strongly-correlated quantum matter. 
Our work, therefore, paves the way for the engineering of ordered nonequilibrium states in quantum simulators and solid-state platforms deep into the many-body regime.

\section{\label{sec:GFT}Many Body Floquet Ground State}

\subsection{\label{subsec:GFT}Geometric Floquet Theory Revisited}

Geometric Floquet theory (GFT) derives Floquet theory from first principles using the adiabatic theorem applied to time-periodic protocols~\cite{schindler2025geometric}. Based on counter-diabatic driving w.r.t.~the initial time $t_0$ (or the phase) of the drive, this derivation holds exactly, without approximations; in particular, it is valid for any drive frequency $\omega=2\pi/T$ and amplitude. 
The evolution operator for the lab-frame Hamiltonian $H(t)=H(t+T)$ takes the form
\begin{eqnarray}
    \label{eq:GFT}
    \!\!\!\!\!\!\!\!\!  
    U(t,t_0) \!\!\!&=&\!\!\! \mathcal{T}\exp\left({-}i\int_{t_0}^t   \mathrm ds H(s) \right ) = W(t,0)\mathrm e^{-i(t-t_0)\mAE[t_0]} ,
\end{eqnarray}
where $\mAE[t_0] = \sum_n \mae_n \Pi_n[t_0]$ is the average energy operator with Floquet eigenstate projectors $\Pi_n[t_0]$ and eigenvalues $\mae_n = T^{-1}\int_0^T\mathrm dt\, \mathrm{tr}(H(t)\Pi_n[t])$ which produce a dynamical phase $\mae_n T$. The square bracket $[t_0]$ denotes a parametric dependence on the starting time of the periodic drive. The corresponding adiabatic gauge potential (AGP) is given by $A_\mAE(t) {=} H(t) {-} \mAE[t]$, and generates the Wilson line operator $W(t,t_0)=\mathcal{T}\exp\left({-}i\int_{t_0}^t \mathrm ds A_\mAE(s) \right )$, whose eigenvalues at time $t{=}t_0{+}T$ give the (negative) geometric phases $\gamma_n$ associated with the Floquet states~\cite{peralta2025streda}.
Reduced to eigenvalues, the geometric Floquet decomposition takes the form $\epsilon_{\F,n}{=} (\mae_n {-} T^{-1}\gamma_n)\!\!\mod \omega$~\cite{fainshtein1978some,grifoni1998contemporary,le2022missing}, see Fig.~\ref{fig:GFT_ED}(a). The Floquet unitary is $U_F[t_0]{=}U(T+t_0,t_0)$. We refer the reader to Appendix~\ref{app:sec:background} for a detailed discussion and comparison to conventional Floquet theory. 

The adiabaticity perspective behind Geometric Floquet theory reveals that quasienergy spectral folding, $\epsilon_\F{\rightarrow}\epsilon_\F{+}m\omega$ ($m\in\mathbb{Z}$), appears as an artifact of partially fixing the gauge freedom associated with the adiabatic gauge potential~\cite{schindler2025geometric}. 
Instead, using the gauge of Eq.~\eqref{eq:GFT} results in an unambiguously sorted average-energy spectrum $\mae_n= T^{-1} \int_0^T\mathrm dt\, \text{tr}(H(t)\Pi_n[t])$, since the lab-frame Hamiltonian $H(t)$ is extensive. The lowest average-energy state $\Pi_0[t]$ defines a candidate for a physically meaningful Floquet ground state.

To illustrate these concepts, consider the open driven Ising chain
\begin{eqnarray}
\label{eq:Ising_chain}
    H(t) &=& (1+\cos\omega t)H_{zz+z} + (1-\cos\omega t)H_{x},
\end{eqnarray}
where $H_{zz+z} = \sum_{n} 2J\, S^z_n S^z_{n+1} + h_z\,  S^z_n$ describes nearest neighbor Ising interactions of strength $J$ in a longitudinal field $h_z$, and $H_{x} = \sum_{n} h_x\,  S^x_n$ is a transverse field of magnitude $h_x$; here, $[S^\alpha_n,S^\beta_m]=i\delta_{mn}\epsilon^{\alpha\beta\gamma}S^\gamma_n$ denote the spin-$1/2$ matrices.  

We use exact diagonalization to compute the Floquet states $\Pi_n[0]$ and the corresponding quasienergies $\epsilon_{\F,n}$ and average energies $\mae_n$, cf.~Fig.~\ref{fig:GFT_ED}(b,c). 
Whereas the quasienergy $\epsilon_\F$ exhibits a clear folding threshold at some system-size-dependent driving period [green triangle], the average energy $\mae$ is, by construction, agnostic to the folding ambiguity~\cite{schindler2025geometric}. We emphasize that the average energy eigenstates are the exact Floquet states $\Pi_n$, so Figs.~\ref{fig:GFT_ED}(b,c) contain the same data just sorted differently. 

At a fixed driving period $T$, increasing the system size $L$ amounts to an exponential increase in the number of eigenstates that need to fit within the Floquet zone $[-\pi,\pi)$. Therefore, for generic nonintegrable models, the quasienergy spectrum is gapless, and the corresponding density of states is constant, Fig.~\ref{fig:GFT_ED}(b). 
Moreover, at any finite system size $L$, maximally entangled (locally infinite-temperature) states exponentially dominate the quasienergy for periods larger than the folding period.   

By contrast, the average energy spectrum reveals that the system exhibits three distinct regimes as a function of the drive period $T$. 
In the generic ergodic regime ($JT{\approx}2$), marked by a proliferation of many-body resonances~\cite{bukov2016heating}, all Floquet eigenstates are maximally entangled and locally indistinguishable from infinite-temperature states. 
In turn, the other two regimes ($JT{\lesssim}1$, $JT{\approx}3$) are structured. The average energy spectrum exhibits striking similarities to static systems: it features (i) a clearly defined bulk consisting of infinite-temperature states which dominates the density of states, Fig.~\ref{fig:GFT_ED}(c). This bulk is surrounded by two low-entangled spectral edges that support, (ii), clearly discernible spectral gaps [inset], that cannot be resolved inside the dense quasienergy spectrum. 
Whereas the high-frequency behavior is generic, the `low'-frequency regime is model-dependent. 

The system size scaling of the ground-state entanglement shows a stable behavior in the high-frequency regime with little-to-no system-size dependence, and the expected clear trend towards the maximally entangled value of $\log(2)$ in the heating regime, see Fig.~\ref{fig:GFT_ED}(d). Remarkably, in the `low'-frequency regime, the trend reverts: entanglement decreases with increasing system size $L$. The behavior suggests an area-law scaling in both the `low'- and high-frequency regimes, and a volume-law scaling in the ergodic regime. 
Hence, the fate of the ergodic regime, surrounded by the two structured regimes on either sides, remains unclear in the thermodynamic limit.

Although this ED data is exact and contains no approximations, for nonintegrable models it is inevitably limited to small system sizes.
At the same time, the existence of a spectral gap hiding a low-entangled state beneath, suggests the possibility of designing approximate variational descriptions for the Floquet ground state $\Pi_0$; this would allow us to channel limited computational resources towards simulating larger system sizes, as we now explicitly demonstrate.  

\begin{figure*}[ht!]
    \centering
    \includegraphics[width=\textwidth]{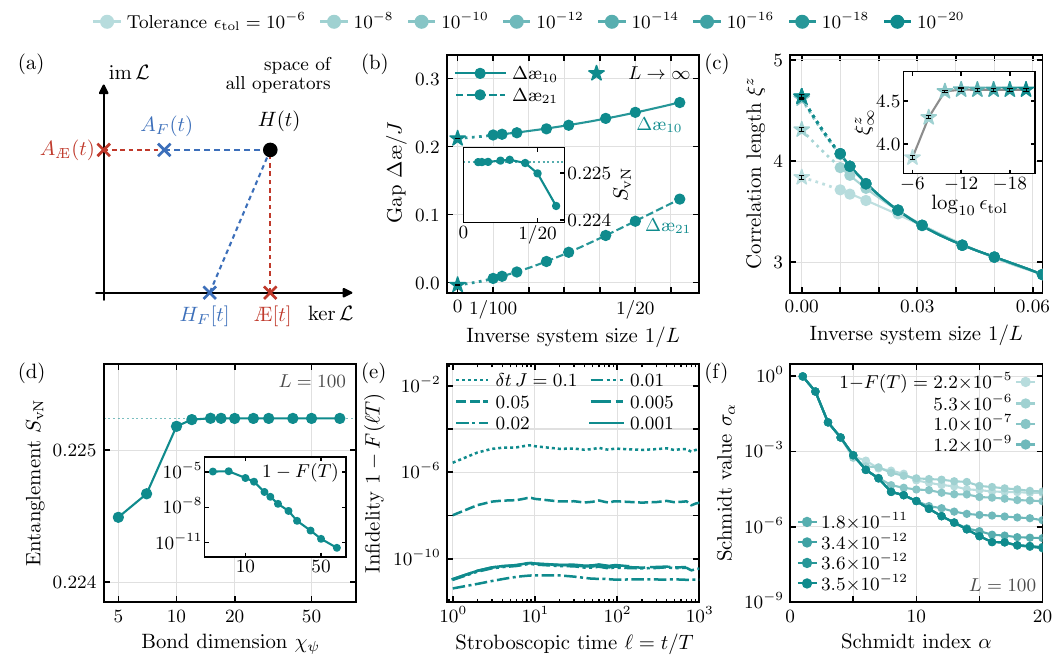}
    \caption{
\textbf{Variational principle captures Floquet ground state} for the nonintegrable Ising chain~\eqref{eq:Ising_chain}.
(a)~Key enabling principle: the average energy $\mAE[t]$ is the orthogonal projection of the lab-frame Hamiltonian $H(t)$ onto the kernel of the Liouvillian, $\ker\mathcal{L}$, which defines the space of dynamical invariants, see also Fig.~\ref{fig:local_vs_global-problem}. 
(b)~Two lowest average energy gaps, $\Delta\mae_{10}{=}\mae_1{-}\mae_0$ (solid) and $\Delta\mae_{21}{=}\mae_2{-}\mae_1$ (dashed), against inverse system size at tolerance $\epsilon_\mathrm{tol}{=}10^{-20}$.  
Curves are polynomial fits in $1/L$ with stars marking the $1/L=0$ extrapolation intercepts. 
Inset: entanglement entropy, $S_\text{vN}$ versus $1/L$ showing area-law scaling.
(c)~Correlation length in the Floquet ground state against inverse system size $1/L$, extracted from the connected correlator $C^z_{c}(r){=}\langle S^z_{L/4}S^z_{L/4+r}\rangle_c{\propto}\,e^{-r/\xi^z}$. Lines are polynomial fits in $1/L$. 
Inset: $L\to \infty$-extrapolated correlation length $\xi^\infty_{z}$ against the Krylov-Arnoldi solver tolerance $\epsilon_\text{tol}$ saturates with decreasing tolerance.
(d)~Half-chain entanglement $S_{\mathrm{vN}}$ converges against the state bond dimension $\chi_\psi$. 
Inset: one-cycle infidelity $1{-}F(T)$, with $F(\ell T){=}\text{tr}(U_F^{\,\ell}\tilde\Pi_0[0] )$, improves exponentially.
(e)~Stroboscopic infidelity $1-F(\ell T)$ improves with decreasing Trotterization step size $\delta t$.
(f)~Half-chain Schmidt spectrum $\sigma_\alpha$ at $L=100$, normalised to $\sum_\alpha\sigma_\alpha^2=1$. The spectral tail floor decreases with decreasing tolerance, giving monotonic convergence.
Infidelities listed in the legend correspond to the tolerance $\epsilon_\mathrm{tol}$ in the topmost legend. 
The sweep variable in (c) and (f) is the Krylov tolerance, $\epsilon_\mathrm{tol}=10^{-6}$ to $10^{-20}$ [topmost legend].
Parameters are $TJ=0.5$, $h_x/J=1.2$, $h_z/J=0.3$. 
Unless otherwise stated, we use $L=100$ spins, tolerance $\epsilon_\mathrm{tol}{=}10^{-20}$, maximum Krylov subspace dimension $m{=}30$, and MPS bond dimension $\chi_\psi=70$; we use Krylov-converged states. 
}    
    \label{fig:GFT_DMRG}
\end{figure*} 

\subsection{\label{subsec:varl}Variational Principle}

The immediate problem with scaling up the simulations to large system sizes lies in the definition of the average energy $\mAE$: it depends explicitly on the Floquet states $\Pi_n$, which appear inaccessible without real-time evolution (time-ordering) and exact diagonalization. We now derive a basis-independent definition, based on which we develop a variational principle controlled by the average energy. 

Observe first that the eigendecomposition of the average energy consists of time-independent eigenvalues $\mae_n$ and time-dependent Floquet eigenprojectors $\Pi_n[t]$. Because Floquet states are stationary states of the dynamics generated by the Hamiltonian $H(t)$, the average energy $\mAE$ obeys the equation: 
\begin{eqnarray}
\label{eq:Liouvillian}
    [H(t), \mAE[t]] - i\partial_t \mAE[t] \equiv \mathcal{L}(\mAE[t]) = 0\, , 
\end{eqnarray}
where $\mathcal{L}(\cdot) = [H(t), \cdot] - i\partial_t(\cdot)$ is the Floquet Liouvillian. Operators $O[t]$\footnote{We use the square bracket $[t]$ to indicate that only the eigenvectors of an operator depend on time $t$, whereas the eigenvalues are $t$-independent. This coincides with the definition of the Floquet gauge in Floquet theory.}, satisfying $\mathcal{L}(O[t]){=}0$ are called dynamical invariants: they represent the generalization of symmetries to dynamical systems. Mathematically, dynamical invariants constitute the kernel of the Liouvillian $\mathcal{L}$.

Since the geometric decomposition~\eqref{eq:GFT} is an instance of counterdiabatic driving~\cite{schindler2025geometric}, the latter imposes a stringent relation~\cite{Kolodrubetz2017Geometry,Sels2017Minimizing} between the average energy $\mAE$ and its adiabatic gauge potential $A_\mAE$: $[\mAE[t], A_\mAE(t)] {+} i\partial_t \mAE[t]{=}0$, where the generalized force on the right-hand side vanishes since the average energies $\mae_n$ are time-independent. Substituting the infinitesimal decomposition $H(t) {=} \mAE[t]{+}A_\mAE(t)$, we eliminate the average energy and arrive at the defining equation for the AGP: $\mathcal{L}(A_\mAE) {=} {-}i\partial_t H$. To write down the solution, note that the Liouvillian $\mathcal{L}$ is not invertible (its kernel contains all dynamical invariants); hence, using the Moore-Penrose pseudoinverse\footnote{One can think of $\mathcal{L}^+$ as a generalized antiderivative along the flow generated by $H(t)$ which fixes the integration constant so the period-average vanishes.} $\mathcal{L}^+$, we find $A_\mAE(t) = \mathcal{L}^+(-i\partial_t H(t)) = \mathcal{L}^+\mathcal{L} (H(t))$. Applying the infinitesimal GFT decomposition one last time, we derive
\begin{eqnarray}
\label{eq:AE_ker_proj}
    \mAE[t] = (1 - \mathcal{L}^+\mathcal{L})H(t). 
\end{eqnarray}
Identifying $1 - \mathcal{L}^+\mathcal{L}$ as the orthogonal projector onto the kernel of the Liouvillian $\mathcal{L}$, it follows that the average energy $\mAE$ is the orthogonal projection of the lab-frame Hamiltonian $H(t)$ onto the space of dynamical invariants, see Fig.~\ref{fig:GFT_DMRG}(a). 

Since $\mathcal{L}$ depends only on the lab-frame Hamiltonian $H(t)$, cf.~Eq.~\eqref{eq:Liouvillian}, all objects on the right-hand side of Eq.~\eqref{eq:AE_ker_proj} are known a priori. 
Therefore, Eq.~\eqref{eq:AE_ker_proj} provides the sought-after basis-independent definition of the average energy. Moreover, kernel projectors provide a well-suited starting point for variational principles. 

In Appendix~\ref{app:sec:varl} we design a variational principle to compute the average energy $\mAE$ from the corresponding gauge potential $A_\mAE$~\cite{kim2024variational,mckeever2024towards}. The key challenge lies in enforcing the pseudo-inverse gauge onto the variational family. To resolve this, we solve for an intermediate variable $X$ that obeys the linear normal equation
\begin{equation}
\label{eq:normal_eq}
    \mathcal{L}^2 X = \mathcal{L} H = -i\partial_t H.
\end{equation}
The sought-after approximation to the AGP is then given by $\mathcal{L} X = \mathcal L^+ \mathcal L H = \tilde A_\mAE$; by construction, it has no component along the space of dynamical invariants, thus enforcing the minimum-norm average-energy gauge. 

Although our variational principle can be applied with any suitable ansatz in any spatial dimension, in this work, we focus on one-dimensional chains using a matrix product operator (MPO) ansatz for $X$ in Sambe superspace~\cite{cheng2022matrix,keliri2026sambe}. 
Once the variational approximation to the average energy operator $\tilde\mAE=H-\tilde A_\mAE$ has been determined, we apply conventional DMRG to obtain the average-energy ground state as a matrix product state (MPS, bond dimension $\chi_\psi$), cf.~Appendix~\ref{app:sec:mps_algo}. Finally, if necessary for high precision, we fine-tune the Floquet ground state using a Krylov-Arnoldi algorithm~\cite{luitz2021polynomial}, see Appendix~\ref{app:sec:krylov}; all physical quantities we show in the figures are evaluated on Krylov-refined states. 

We now apply our variational procedure to construct an approximation to the Floquet ground state\footnote{We use the projector notation to capture any possible degenerate ground-state manifolds arising due to symmetry breaking; in practice we use pure states represented as an MPS, see Appendix~\ref{app:sec:mps_algo}.}, $\tilde\Pi_0[0]\approx \Pi_0[0]$, of the Ising chain~\eqref{eq:Ising_chain}. 
Figure~\ref{fig:GFT_DMRG}(b) shows the average energy gap $\Delta\mae_{mn} = \mae_m-\mae_n$ as a function of the inverse system size $1/L$. We find that the gap $\Delta\mae_{10}$ between the first excited state and the ground state remains finite for systems of up to $L=100$ spins. An extrapolation (star marker) to larger system sizes using a third-order polynomial fit is consistent with a finite gap in the thermodynamic limit, additionally corroborated by the convexity of the curve. By contrast, the gap between the next two states, $\Delta\mae_{21}$, closes and extrapolates to zero as $L\to\infty$. We thus find a unique average energy ground state, followed by a continuum of excitations above the stable $\mae$-gap. 

The existence of a finite gap in one-dimensional chains is inherently related to a finite correlation length, as we consistently find in Fig.~\ref{fig:GFT_DMRG}(c). Moreover, as anticipated by the ED data in Fig.~\ref{fig:GFT_ED}, the ground state is area-law entangled, as shown by the half-chain entanglement in Fig.~\ref{fig:GFT_DMRG}(b, inset), whose value is independent of system size. Notice that all these results have converged with the bond dimension of the MPS approximating the Floquet ground state [Fig.~\ref{fig:GFT_DMRG}(d)] and the numerical tolerance [Fig.~\ref{fig:GFT_DMRG}(c, inset)]. 

The variational approximation to the average energy $\mAE$ does not allow us to reliably measure its variance in the Floquet ground state [cf.~App.~\ref{app:sec:varl:diagnostics}]; we therefore measure the stroboscopic infidelity, $1-F(\ell T)$, of the evolved variational ground state with its initial value, $F(\ell T) = \text{tr}(\tilde\Pi_0[0]\tilde\Pi_0[\ell T]) = \text{tr}(U_F^{\,\ell}\tilde\Pi_0[0] )$. Time evolution is performed using a fourth-order Trotter-Suzuki decomposition, with a time step small enough to keep the overall error below $10^{-8}$. Figures~\ref{fig:GFT_DMRG}(d, inset) and (e) shows that the infidelity $1-F(\ell T)$ remains time-independent within the precision of the variational approximation. 

Finally, Fig.~\ref{fig:GFT_DMRG}(f) displays the normalized Schmidt values $\sigma_\alpha$ across the half-chain cut ordered by their magnitude against their Schmidt index $\alpha$. Using the Krylov-Arnoldi solver [App.~\ref{app:sec:krylov}] with a tunable tolerance $\epsilon_\text{tol}$ allows us to controllably improve the approximation of the Floquet ground state, as measured by the one-cycle infidelity [see legend], down to double precision. 
A bond dimension of about $\chi_\psi=70$ suffices to capture the state for the parameter regime of interest. We note that these numbers depend on the choice of model parameters and are also model-dependent, much like in MPS approximations of ground states in static spin chains.

\begin{figure*}[t!]
  \centering
  \includegraphics[width=1\linewidth]{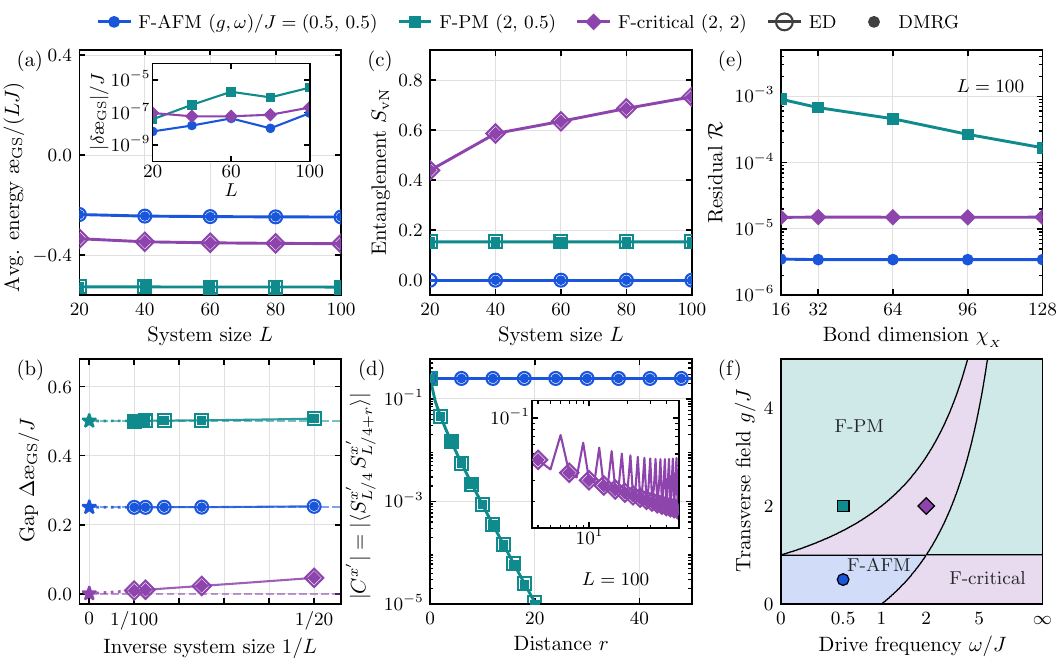}
  \caption{
  \textbf{Circularly driven XY spin chain~\eqref{eq:XY_chain}.}
  Comparison between Floquet-DMRG and exact results calculated using exact diangonalization (ED) methods. 
  In all panels, open symbols joined by solid lines are exact numerical data and filled symbols of the same shape and colour are the Floquet-DMRG data.  
  (a)~Ground-state average energy density $\mae_{\mathrm{GS}}/L$ as a function of chain length $L$. Inset shows the deviation from exact values $|\delta\mae_{\mathrm{GS}}|{=}|\mae^{\mathrm{ED}}_{\mathrm{GS}}-\mae^{\mathrm{DMRG}}_{\mathrm{GS}}|$ for different system sizes. 
  (b)~Ground state average energy gap $\Delta\mae_{\mathrm{GS}}$ versus invserse system size $1/L$. Solid lines are quadratic least-squares fits in $1/L$ to the DMRG data with stars marking their $1/L\to0$ extrapolations; dashed lines show the analytic $L\to\infty$ gaps. 
  The $L=60$ F-critical phase point, where the two lowest levels are accidentally degenerate, is excluded from the fit.
  (c) Half-chain von Neumann entanglement entropy $S_{\mathrm{vN}}$ remains constant with system size $L$ for the F-AFM and F-PM phase, and grows with $L$ in the F-critical phase. 
  (d) Correlation function $|C^{x'}(r)|=|\langle S^{x'}_{L/4} S^{x'}_{L/4+r}\rangle|$. The AFM point retains long-range order, the PM decays exponentially, and the critical point decays as a power law [inset]. 
  (e) Variational residual $\mathcal{R}$ versus the bond dimension $\chi_X$ of the MPS representation of the auxiliary operator $X$, cf.~Eq.~\eqref{eq:normal_eq}. In all the cases, the $\mathcal{R}$ is small for the largest $\chi_X=128$ considered that results in the $\delta \mae_{GS}< 10^{-5}$ for $L=100$ as shown in (a).
  (f) Floquet ground-state phase diagram spanned by drive frequency $\omega$ and transverse field $g$, distinguishing Floquet paramagnetic (F-PM, teal), Floquet critical (F-critical, purple), and Floquet antiferromagnetic (F-AFM, blue) phases. 
  Symbols mark the parameter sets used in (a)--(e). 
  The parameters are $A/J=0.5$ and $(g,\omega)/J=(0.5,\,0.5)$ (AFM), $(2.0,\,0.5)$ (PM), and $(2.0,\,2.0)$ (critical), marked in the phase diagram (f).
  Unless otherwise stated, we use $L=100$, and $\chi_X {=} 128$.
  }
  \label{fig:XY}
\end{figure*}

\section{\label{sec:FGS_order}Floquet Ground-State Order}

The existence of gapped, area-law entangled Floquet ground states raises the question of whether they can support quantum order.
We answer it in the affirmative by studying an exactly solvable driven XY chain, and the nonintegrable Ising chain~\eqref{eq:Ising_chain}. 

\subsection{\label{sec:XY}Exactly Solvable Periodically Driven XY Chain}

The circularly driven open XY chain is defined by
\begin{eqnarray}
\label{eq:XY_chain}
    H(t) {=} \frac{1}{2}\sum_{n} 
            \left[
            \left( \frac{J}{2} S^+_{n+1} S^-_{n} {+} A i e^{-i\omega t} S^+_{n+1} S^+_{n} {+} \mathrm{h.c.}\right) 
            {+} g S_n^z
            \right].
\end{eqnarray}
A Jordan-Wigner transformation, followed by a Fourier transform, maps the Hamiltonian to a collection of $L$ circularly-driven two-level systems, for which exact closed-form expressions for the average energy $\mAE[t]$, its AGP $A_\mAE(t)$, and the Floquet ground state $\Pi_0[t]$ can be obtained, cf.~Appendix~\ref{app:subsec:XY}.
This allows us to benchmark our variational algorithm. 

We fix three representative points in parameter space. 
Figure~\ref{fig:XY}(b) shows that the average energy gap can either stay finite (teal, blue) or vanish (purple) in the thermodynamic limit, suggesting the ground state $\Pi_0[0]$ can be both correlated and critical; this is corroborated by the behavior of the half-chain entanglement in Fig.~\ref{fig:XY}(c). 

From the analytical solution, we identify $x'=(x-y)/\sqrt{2}$ as the ordering axis; hence, we use the correlation function $C^{x'}(r) {=} \text{tr}(\Pi_0[0] S^{x'}_{L/4} S^{x'}_{L/4+r})$ to examine the ordered properties of the state, and find three distinct behaviors, see Fig.~\ref{fig:XY}(d): 
(i) transient decay to a constant value (blue), indicating the existence of long-range order in the $x$-direction: the Floquet ground state exhibits antiferromagnetic (AFM) correlations. This is a direct confirmation that long-range order can be protected by the average energy gap $\Delta\mae$, and dressed in the presence of the periodic drive;
(ii) exponential decay to zero (teal), indicating the absence of long-range order and hence the state $\Pi_0[0]$ is paramagnetic;
(iii) power-law decay to zero (purple, inset), characteristic of a critical Floquet ground state featuring logarithmic growth of entanglement with system size.   

We emphasize that our variational results (solid markers) are in excellent agreement with the exact solution data (empty markers). This is corroborated by the operator-norm residual $\mathcal{R} ={\norm{\mathcal{L}^2 X+i\partial_t H}}/{\norm{\partial_t H}}$, cf.~Fig.~\ref{fig:XY}(e), which measures the extent to which the approximate average energy $\tilde\mAE$ is a dynamical invariant of the Hamiltonian $H(t)$ [Sec.~\ref{app:sec:varl:diagnostics}]. Increasing the MPO bond dimension $\chi_X$ allows us to improve the quality of the approximation for the paramagnet (teal) whereas the other two phases saturate much faster to the underlying precision floor. Because the circularly driven XY model is integrable, small values of $\chi_X$ suffice, and there is no need to use the Krylov-Arnoldi solver.

\subsection{\label{sec:MFIM}Nonintegrable Driven Ising Chain}

The presence of long-range order in the Floquet ground state of the XY Hamiltonian~\eqref{eq:XY_chain} raises the question about how generic the phenomenon is. Because exactly solvable models are fine-tuned, we now turn back to the nonintegrable driven Ising chain~\eqref{eq:Ising_chain}. 

\begin{figure*}[ht!]
  \centering
  \includegraphics[width=1\linewidth]{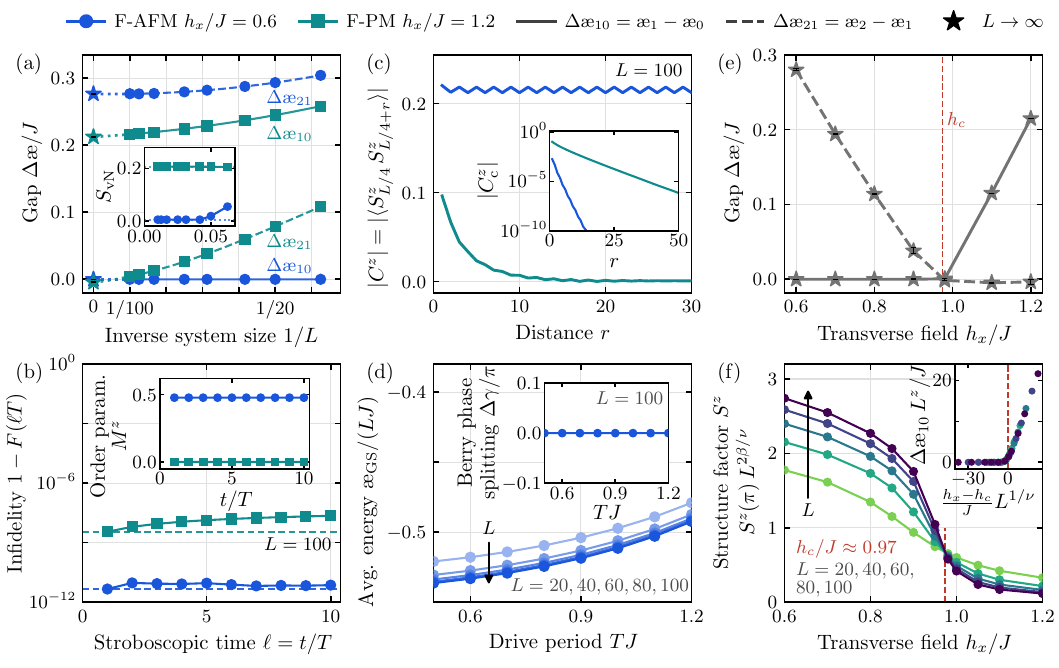}
  \caption{
  \textbf{Floquet ground state phases and phase transition in mixed-field Ising chain~\eqref{eq:Ising_chain} in the high-frequency regime.}
    Panels (a)--(c) show results for the Floquet antiferromagnetic (F-AFM) point (blue, circles, $h_x/J=0.6$) and the Floquet paramagnetic (F-PM) point (teal, squares, $h_x/J=1.2$).
    Stars mark polynomial extrapolation in $1/L$ in the TD limit [App.~\ref{app:sec:Ising}]. 
    {(a)} Average energy gap versus inverse system size survives the extrapolation in the TD limit for both the F-AFM and the F-PM state, for chains of $L=18$ to $L=100$ spins; 
    $\Delta\mae_{10}=\mae_1-\mae_0$ (solid lines) and $\Delta\mae_{21}=\mae_2-\mae_1$ (dashed lines) are the gaps between the lowest two and the next two average-energy eigenstates, respectively. 
    In the F-AFM $\Delta\mae_{10}\approx 0$ indicates a degenerate Floquet ground-state doublet. 
    Inset: the half-chain von Neumann entanglement entropy $S_{\mathrm{vN}}$ in the Floquet ground state versus $1/L$ saturates to a system-size-independent constant, exhibiting area-law scaling.
    {(b)} The stroboscopic infidelity $1-F(\ell T) =1-\text{tr}(U_F^{\,\ell}\tilde\Pi_0[0] )$ of the Floquet ground state over the first ten drive periods remains of order $10^{-11}$ ($10^{-8}$) for the F-AFM (F-PM) state.
    Time evolution is performed using TEBD ($4$-th order Trotter-Suzuki decomposition) under the exact Hamiltonian~\eqref{eq:Ising_chain}, with truncation error $10^{-10}$ and timestep $\delta t=0.001/J$.
    Inset: AFM bulk order parameter $M^{z}{=}\sum_{n=L/4}^{3L/4} (-1)^n\langle S^z_n\rangle$ is preserved.
    {(c)} Correlation function $|C^{z}(r)|=|\langle S^z_{n_0}S^z_{n_0 +r}\rangle|$ in the Floquet ground state versus distance $r$, with $n_0=L/4$. The F-AFM state exhibits long-range order (blue); F-PM correlations decay to zero (teal) showing absence of symmetry-breaking order. 
    Inset: connected correlators $|C_{\mathrm{c}}^{z}(r)|=|\langle S^z_{n_0}S^z_{n_0+r}\rangle-\langle S^z_{n_0}\rangle \langle S^z_{n_0+r}\rangle|$ decay exponentially in space, consistent with finite correlation length in a gapped ground state manifold.
    {(d)} Ground-state average-energy density $\mae_{\mathrm{GS}}/(LJ)$ for the F-AFM versus the drive period shows the stability of the ordered state; blue colors indicate different system sizes $L=20$--$100$. 
    Inset: absence of Berry phase splitting $\gamma_{\mathrm{GS}}/\pi$ in the AFM Floquet ground state doublet shows the Floquet phase is not geometric [cf.~Fig.~\ref{fig:Ising_GS-DTC}f].
    Panels (e) and (f) show the F-AFM--F-PM phase transition as a function of transverse field $h_x$.
    {(e)} Extrapolated average-energy gap closes at the critical point $h_c$.
    {(f)} Static $zz$-structure factor $S^{z}(\pi)L^{2\beta/\nu}$ rescaled with Ising-universality exponents $\beta=1/8$, $\nu=1$ and $z=1$, for $L=20$--$100$ (arrow indicates increasing $L$). Curves cross at $h_c/J\approx0.97$ (red dashed vertical line), which determines the critical field strength. 
    Inset: finite-size data collapse of the rescaled average-energy gap $\Delta\mae_0L^{z}/J$ against $(h_x-h_c)L^{1/\nu}/J$.
    The parameters are $TJ=0.7$, $h_z/J=0.3$, and $L=100$, unless otherwise stated.
  }
  \label{fig:Ising_FAFM}
\end{figure*}

\textbf{Floquet Antiferromagnets (F-AFMs).---}%
In the perturbative high-frequency regime, the period-averaged Hamiltonian exhibits a spectral gap, protecting an antiferromagnetic (AFM) state at $T=0$. We now show that this order survives for finite drive periods $JT\lesssim 1.2$. 

Figure~\ref{fig:Ising_FAFM}(a, blue data) shows the average energy gaps $\Delta\mae$ between the lowest three average-energy states. Extrapolation to the thermodynamic limit is consistent with a doubly-degenerate ground-state manifold $\Pi_0$, separated from the excitations by a finite average-energy gap. Consistent with this, the half-chain entanglement remains area-law (inset). 

We checked that the gap between the lowest two states closes exponentially, indicating symmetry breaking in the thermodynamic limit. Indeed, the bulk $zz$-correlation function $C^\text{z}(r) = (-1)^r\text{tr}(\Pi_0[0] S^z_{L/4} S^z_{L/4+r})$ decays to a constant value, revealing the presence of long-range order in the Floquet ground state at a finite drive period [Fig.~\ref{fig:Ising_FAFM}(c)]. The Floquet drive dresses the AFM axis and rotates it inside the $xz$-plane; the ground state $\Pi_0$ is therefore a Floquet AFM (F-AFM). The inset shows the exponential decay of the connected correlation function, whose length scale gives a finite correlation length, consistent with the finite average energy gap. 

\begin{figure*}
  \centering
  \includegraphics[width=1\linewidth]{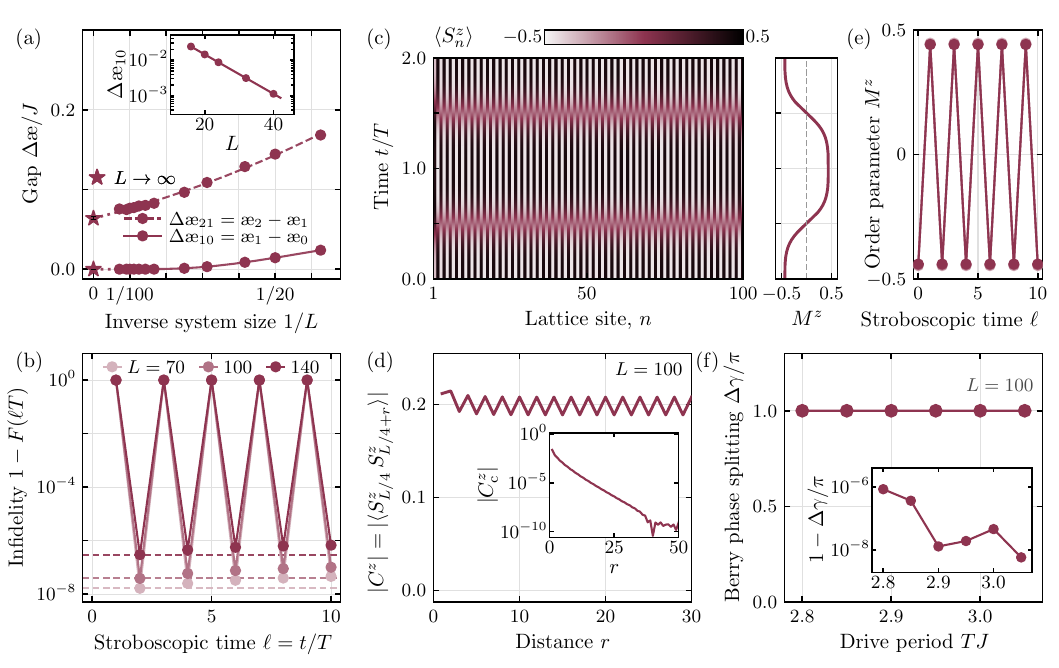}
  \caption{
    \textbf{Discrete ground-state time crystal in the periodically driven mixed-field Ising chain~\eqref{eq:Ising_chain}.} 
    $\Delta\mae_0=\mae_1-\mae_0$ (solid line interpolation) and $\Delta\mae_1=\mae_2-\mae_1$ (dashed line interpolation). 
    \textbf{(a)} Average energy gaps versus inverse system size for chains of $L=18$ to $L=140$ spins. 
    $\Delta\mae_{10}$ (solid line interpolation) is the splitting of the symmetry-broken Floquet doublet which closes exponentially with system size (inset); 
    $\Delta\mae_{21}$ (dashed line interpolation) is the excitation gap, which remains finite after extrapolation to $L=\infty$.
    \textbf{(b)} The stroboscopic infidelity $1-F(\ell T) =1-\text{tr}(U_F^{\,\ell}\tilde\Pi_0[0] )$ of the Floquet ground state over the first ten drive periods alternates between values close to unity at odd $\ell$ when the system transitions into the partner symmetry-broken state, and $2$--$8\times10^{-7}$ at even $\ell$.
    Time evolution is performed using TEBD ($4$-th order Trotter-Suzuki decomposition) under the exact Hamiltonian~\eqref{eq:Ising_chain}, with truncation error $10^{-10}$ and timestep $\delta t=0.02/J$.
    \textbf{(c)} Space-time heatmap of the local magnetisation $\langle S^z_i(t) \rangle$ [colorbar] within two drive periods. Alternating stripes show antiferromagnetic order; the spatial pattern flips every drive cycle $T$. 
    Right: Intra-period evolution of AFM bulk order parameter $M^{z}{=}\sum_{n=L/4}^{3L/4} (-1)^n\langle S^z_n\rangle$, showing period-doubling oscillations.
    \textbf{(d)} Absolute value of the ground-state correlation function $|C^z(r)|=|\langle S^z_{n_0}S^z_{n_0+r}\rangle|$ as a function of distance $r$ for $n_0=L/4$. Saturation to a finite value shows long-range order. 
    Inset: connected correlator $|C_{\mathrm{c}}^{z}(r)|=|\langle S^z_{n_0}S^z_{n_0+r}\rangle-\langle S^z_{n_0}\rangle \langle S^z_{n_0+r}\rangle|$ decays exponentially, implying the existence of a finite correlation length.
    \textbf{(e)} Stroboscopic evolution of the DTC order parameter $M^z(\ell)$ flipping sign at every period.
    \textbf{(f)} Berry-phase splitting $\Delta\gamma/\pi$ of the Floquet degenerate doublet versus driving period $T$ indicates the GS-DTC phase is stable over a finite region, and shows the geometric origin of GS-DTC order.
    Inset: deviation from perfect $\pi$-splitting within the precision set by the infidelity. 
    The parameters are $h_x/J=1.0$, $h_z/J=0.5$, $TJ=3.0$, and $L=100$ unless otherwise stated.
  }
  \label{fig:Ising_GS-DTC}
\end{figure*}

\textbf{Discrete Ground-State Time Crystals (GS-DTCs).---}%
Although the drive dresses the F-AFM order, AFM states also exist in equilibrium. We now turn our attention to inherently nonequilibrium ordered states -- discrete time crystals. 
To this end, we consider a nonperturbative `low'-frequency regime, where the Hamiltonian~\eqref{eq:Ising_chain} has no obvious small parameter\footnote{The existence of a finite $\mae$-gap suggests that a small parameter should exist; however, it is not obvious how to identify it.}. 

Similar to the F-AFM, Fig.~\ref{fig:Ising_GS-DTC}(a) shows the existence of a finite average energy gap above the ground-state manifold for system sizes of up to $L=140$ spins. Again, extrapolating to the thermodynamic limit is consistent with a doubly degenerate ground-state manifold, separated by a finite gap from a continuum of excitations. Similar to the F-AFM state, the correlation function supports the existence of long-range AFM order in the $z$-direction, cf.~Fig.~\ref{fig:Ising_GS-DTC}(d). Since we find no two-body correlations along the $x$- and $y$-directions and because the magnitude of the two-body $zz$-correlations is not maximal, the drive also induces higher-body correlations. 

However, the local $z$-magnetization $M^z_n(t) =\text{tr}(\tilde\Pi_0[t] S^z_n)$ (i.e., the order parameter) exhibits characteristic period-doubled oscillations [Fig.~\ref{fig:Ising_GS-DTC}(c,e)]. Indeed, after one drive cycle, the symmetry-broken Floquet ground state transitions completely into its partner state within the degenerate Floquet ground-state manifold, as can be seen from the evolution of the infidelity, Fig.~\ref{fig:Ising_GS-DTC}(b). 

According to Geometric Floquet theory, inherently nonequilibrium phenomena in periodically driven systems (i.e., phenomena without a static counterpart) are of geometric origin~\cite{schindler2025geometric}. 
In the case of DTCs, the characteristic $\pi$-splitting is completely absent from the average energy spectrum, which is perfectly degenerate. 
From the perspective of adiabatic theory, a degenerate energy manifold carries a nonabelian Wilczek-Zee connection. Evolving each of the Floquet ground states over one drive cycle and removing the dynamical phase $e^{-i\mae_0 T}$ with the average energy $\mae$ known from the variational principle, we numerically extract the corresponding many-body nonabelian Floquet ground-state holonomy as
\begin{equation*}
    W(T,0) {=} e^{i\chi}\begin{pmatrix} 0 & e^{i\theta} \\ e^{-i\theta} & 0 \end{pmatrix} + \delta W,
    \quad \chi {=} 0.3822,\quad \theta {=} 0.0056,
\end{equation*}
where the off-diagonal phases $e^{\pm i\theta}$ equal unity to within $10^{-6}$, and the error $\delta W$ is diagonal with entries of absolute value below $4\times10^{-6}$ (the numerical accuracy of the Floquet ground states). 
The holonomy is thus given by $\sigma^x$ up to overall phases: it exchanges the two symmetry-broken ground states every drive period; 
its eigenvalues $\pm e^{i\chi}$ are the exponentials of the two geometric phases; crucially, their ratio is fixed by the off-diagonal structure alone, independent of the values of $\chi$ and $\theta$.
We thus find that the many-body geometric phase difference between the two symmetry-broken degenerate ground states is precisely $\Delta\gamma = \pi$, see Fig.~\ref{fig:Ising_GS-DTC}(f), confirming the geometric origin of DTCs. 
By contrast, the F-AFM is geometrically trivial, cf.~Fig.~\ref{fig:Ising_FAFM}(d, inset). 

Our data are therefore consistent with the presence of a discrete time crystal in the Floquet ground state $\tilde\Pi_0[t]$. This demonstrates that the latter can exhibit nonequilibrium order, not present in any static system, enriching the paradigm of ground-state quantum order. We denote this state GS-DTC: unlike previously proposed MBL-DTCs that exhibit a $\pi$-pairing across all localized Floquet states~\cite{sacha2018time,khemani2019brief,else2020discrete,zaletel2023colloquium}, the GS-DTC is expected to show the geometric $\pi$-pairing only in its average-energy ground states\footnote{and potentially up to some finite critical energy density, in case Floquet order exists in the thermodynamic sense.}. Moreover, in Sec.~\ref{sec:no_heating} we argue that the stability mechanism for GS-DTCs is adiabaticity protected by the finite $\mae$-gap, in contrast to MBL-DTCs which are stabilized by many-body Anderson localization; therefore, GS-DTCs can exist in higher dimensions where MBL is unstable.
We also emphasize that the GS-DTC is not prethermal: as an exact Floquet eigenstate manifold, it is infinitely long-lived despite the system being clean (i.e., disorder-free)~\cite{huang2018clean,pizzi2020time}. 
It would be interesting to investigate whether strong $\pi$-modes~\cite{yates2019almost} can also be understood as ground states of the average energy operator.

\section{\label{sec:phases_of_matter}Floquet Ground-State Phases of Matter. Phase Transitions}

The above results demonstrate Floquet ground-state order at isolated points in parameter space. We now show its stability to quasistatic parameter deformations, which allows us to exhibit the existence of Floquet ground-state phases of nonequilibrium matter and investigate their phase transitions. 

Similar to the XY and the transverse-field Ising chains in static systems, the circularly driven XY model~\eqref{eq:XY_chain} provides a paradigmatic example to investigate Floquet ground-state phases of matter analytically because it admits a closed-form solution [App.~\ref{app:subsec:XY}].
Figure~\ref{fig:XY}(f) shows the Floquet ground-state phase diagram in the $(g,\omega)$-plane. As anticipated, it features three distinct phases: (i) a long-range ordered F-AFM phase at small drive frequencies $\omega/J$ and small transverse fields $g/J$, which is continuously connected to the ordered phase of the static limit of Eq.~\eqref{eq:XY_chain};\footnote{For $A/J=0.5$, the static model is the transverse-field Ising chain with the Ising axis in the $xy$-plane, cf.~Appendix~\ref{app:subsec:XY_phases}.} it is separated from (ii) an F-PM phase at large values of $g/J$ by (iii) a critical phase with algebraic correlations. 
The many-body phase diagram is obtained by filling the corresponding average-energy single-particle states. Whereas for drive frequencies above the single-particle bandwidth this coincides with filling the states according to the quasienergy ordering, we find that in the low-frequency regime the average-energy ground state need not coincide with the quasienergy `ground state' (defined w.r.t.~some fixed quasienergy-gap-preserving spectral unfolding). Hence, there are parameter regimes for which the ground state of $\mAE$ can be F-AFM whereas the `ground state' of the Floquet Hamiltonian -- critical (and vice versa), see Appendix~\ref{app:subsec:XY}. 

Away from fine-tuned exactly solvable models, we can numerically determine the Floquet ground-state phase diagram using the variational principle [Sec.~\ref{subsec:varl}]. For the driven Ising chain~\eqref{eq:Ising_chain}, Fig.~\ref{fig:Ising_FAFM}(e) shows the average energy gaps for different values of the transverse field $h_x$ in the high-frequency regime for $JT=0.7$, extrapolated to the thermodynamic limit. The gap $\Delta\mae$ closes and then reopens again, reminiscent of a phase transition between an AFM and a PM state. Guessing that the transition belongs to the Ising universality class (which it does in the infinite frequency limit, $JT{=}0$), we rescale the static structure factor $S^{z}(q) = {L}^{-2} \sum_{m,n} \mathrm e^{-i \, q\, (r_m-r_n)}\, \text{tr}(\tilde\Pi_0[0] S^z_m\, S^z_n )$ and read off the critical transverse field as $h_c/J{\approx} 0.97$, Fig.~\ref{fig:Ising_FAFM}(f). Using this value results in an almost perfect scaling collapse of the average-energy gap data [inset]. This analysis provides a strong indication for the existence of a quantum phase transition in the Floquet ground state at a finite drive period $JT{=}0.7$. 

In addition, we probe the stability of the F-AFM phase against changes in the drive period $T$ in Fig.~\ref{fig:Ising_FAFM}(d). We find that the ground-state average-energy density $\mae/L$ remains stable with increasing system size $L$ over the entire range $TJ\le1.2$ accessible to the variational algorithm. 
By quasistatically continuing the F-AFM state to larger periods $T$ using the Krylov-Arnoldi method only (see App.~\ref{app:sec:Ising}), we find that the attainable stroboscopic infidelity at fixed bond dimension deteriorates continuously beyond $TJ\approx1.5$ and, in contrast to the F-AFM regime, increases with $L$; we take this as the finite-size precursor of the ergodic regime~\cite{moessner2017equilibration} seen in Fig.~\ref{fig:GFT_ED}(c,d), where the lowest eigenstate of $\mAE$ is volume-law entangled for any finite system size. 
We could not resolve the F-AFM phase boundary sharply: a fixed bond dimension causes the simulations to degrade before the putative boundary is reached, whereas finite-size rounding lets them persist beyond it; within these uncertainties, the estimated boundary lies around $TJ\approx1.5$.

The exact diagonalization data in Fig.~\ref{fig:GFT_ED}(c) in the ergodic regime around $JT{\approx}2$, exhibit a so-called spectral implosion~\cite{schindler2025geometric} where $\mae_n{\to}0$ all vanish as the system size increases. A degenerate spectrum with all $\mae_n{=}0$ does not allow for the identification of individual eigenstates. As a result, we conjecture that, in the thermodynamic limit of the ergodic regime, the Floquet eigenspaces are no longer rank-one (invalidating the concept of individual Floquet eigenstates) and merge into a single giant superprojector $\Pi{=}\mathds{1}$ with $\mAE{\propto}\mathds{1}$, ultimately trivializing the content of the average energy decomposition $H(t){=}\mAE[t]{+}A_\mAE(t)$. If correct, this conjecture would reveal emergent ergodicity as the breakdown of Floquet's theorem in the ergodic phase.

Finally, we mention in passing that the GS-DTC state is also stable to small deviations in the driving period $T$, cf.~Fig.~\ref{fig:Ising_GS-DTC}(f); our data is therefore consistent with the existence of a stable GS-DTC phase in the thermodynamic limit. 

\begin{figure}[t!]
    \centering
    \includegraphics[width=\linewidth]{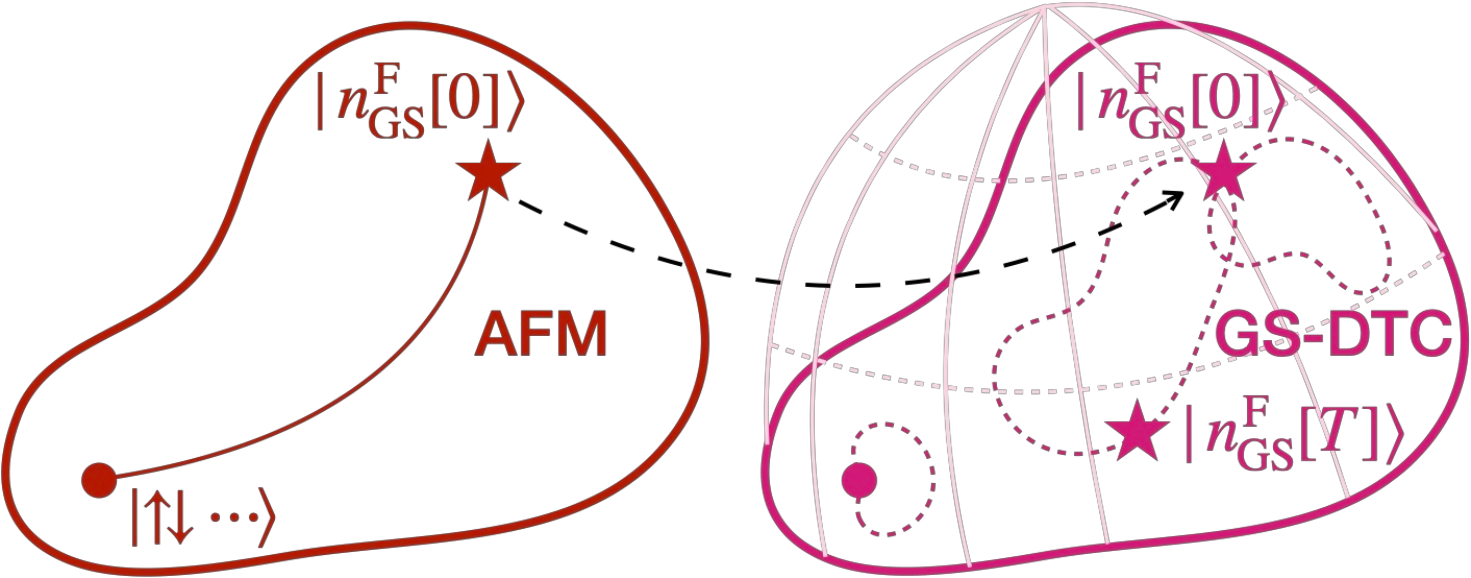}
    \caption{
    \textbf{Equilbrium vs.~Floquet ground state phases of matter.}
    Left: equilibrium phases of matter are defined starting from a reference state by quasiadiabatic continuation without closing the control-Hamiltonian gap. Each state is represented by a point.
    Right: we define Floquet ground-state phases of matter by quasiadiabatic continuation w.r.t.~the average energy gap $\Delta\mae$; each instance is a closed loop $\left\{\Pi_0[t]\!: t\in[0,T)\right\}$ of ground-state projectors defining an eigenspace of the Floquet unitary $U_F[t]$\footnote{Within a degenerate ground-state manifold, symmetry-broken states $|n^\mathrm{F}_\text{GS}[t]\rangle$ need not return to themselves after one drive cycle, but the eigenspace spanned by the Floquet eigenprojector $\Pi_0[t]$ does.}. 
    Geometric Floquet theory~\eqref{eq:GFT} decomposes Floquet phases of matter into an equilibrium-like ground-state element and an inherently nonequilibrium geometric element: at any fixed time $t$, the ground state $\Pi_0[t]$ of $\mAE[t]$ is an isolated point on the loop that represents an equilibrium phase of matter; these points are linked together by the Wilson line operator $W(t,0)$, forming the loop that defines the representative of the nonequilibrium phase. The loop is parallel-transported by $W(t,0)$ without leaving the gapped manifold.
    Inherently nonequilibrium Floquet ground-state phases of matter (e.g., the GS-DTC) can be strictly larger than their equilibrium counterparts; they are 
    stable to both symmetry-preserving and symmetry-breaking perturbations~\cite{vonkeyserlingk2016absolute}. 
    The dashed black line shows the quench protocol from Sec.~\ref{sec:prep}.
    }
    \label{fig:F_phases}
\end{figure} 

\section{\label{sec:prep}Floquet Ground State Preparation}

Studying Floquet ground state phases and their transitions requires experiments to prepare these states. We now briefly outline two protocols that leverage existing techniques for this. 

Recently, variational counterdiabatic (CD) driving was extended for the preparation of Floquet eigenstates~\cite{schindler2024counterdiabatic}. This generalization works in the same way as the corresponding variational CD principle for static systems, and can be applied to steer the population of eigenstates as the amplitude, frequency, or some other model parameter is changed in the presence of the Floquet drive. Imposing locality and experimental accessibility on the variational ansatz produces protocols that can prepare the Floquet ground state with fidelities an order of magnitude larger than what is obtainable in unassisted ramps. This works, provided the average energy gap $\Delta\mae$ stays open during the protocol. For instance, one can use Floquet-CD driving to prepare the F-AFM state starting from an equilibrium AFM state. However, equilibrium states of matter are not adiabatically connected to inherently nonequilibrium ordered states, since the average energy gap $\Delta\mae$ closes and the conditions for Floquet CD driving break down.

Alternatively, notice that the GFT factorization of the evolution operator, Eq.~\eqref{eq:GFT}, decomposes each Floquet ground-state phase of matter into a geometric and a dynamical building block.
Moreover, the Floquet state we want to prepare is the ground state of the static average energy operator $\mAE[t]$, for $t$ fixed. 
Putting the geometric part aside for a moment, the average-energy ground state $\Pi_0[t]$ at the fixed time $t$ is indistinguishable from (and therefore represents) an equilibrium phase of matter. For instance, for the GS-DTC state, $\Pi_0[t]$ is an ordinary dressed AFM for any fixed time $t$. Therefore, starting from some accessible reference state within the equilibrium phase of, say $\Pi_0[0]$, one can use the machinery of static CD driving~\cite{Kolodrubetz2017Geometry,claeys2019floquet,takahashi2024shortcuts,shrestha2026counterdiabatic} to prepare $\Pi_0[0]$, and then quench-start the periodic drive\footnote{This does not require that the average energy $\mAE[t]$ is a local operator; one only needs a (quasi-)local parent $\mAE^\text{q'loc}[t]$ for its gapped area-law entangled ground state $\Pi_0[t]$.}. The geometric component of the dynamics (the Wilson line $W$) will then lock in to the Floquet ground state, and produce the desired nonequilibrium phase of matter $\left\{\Pi_0[t]\!: t\in[0,T)\right\}$, cf.~Fig.~\ref{fig:F_phases}. The accuracy of this preparation protocol is determined entirely by the accuracy with which $\Pi_0[t]$ can be prepared in the absence of the periodic drive, and the freedom of choosing the fixed reference time $t$ can be exploited accordingly. 

This argument does not imply that the GS-DTC and the AFM belong to the same phase of matter. Indeed, there exists no protocol to prepare one state from the other in the presence of the Floquet drive without closing the average energy gap $\Delta\mae$, as expected for two distinct phases of matter. This is because, in the above protocol, the generator of dynamics connecting the accessible reference state to the Floquet ground state $\Pi_0[0]$ is not continuously connected to the generator $H(t)$ of the Floquet dynamics once the drive is quench-turned on\footnote{Adiabatic continuation within the equilibrium phase only concerns the dynamical part of the evolution, and completely misses the geometric part, cf.~Eq.~\eqref{eq:GFT}.}. Hence, the quench implements a discontinuous change of reference frames to produce the missing geometric component.  

Schematically, equilibrium phases of matter are defined by all states connected by quasiadiabatic continuation, i.e., CD evolution without closing the control-Hamiltonian gap. 
Each state in an equilibrium phase is represented by a point. It is stable w.r.t.~sufficiently small symmetry-preserving perturbations; perturbations that explicitly break the symmetry take the state outside the phase. 
By contrast, Floquet ground-state phases of matter are defined by quasiadiabatic continuation w.r.t.~the average energy gap. Each representative of a Floquet phase forms an entire closed loop $\left\{\Pi_0[t]\!: t\in[0,T)\right\}$, rather than isolated points $\Pi_0[t]$, see Fig.~\ref{fig:F_phases}. Moreover, inherently nonequilibrium Floquet phases of matter exhibit absolute stability~\cite{vonkeyserlingk2016absolute}: 
(i) perturbations that do not explicitly break the symmetry preserve the gap and the ground-state degeneracy, similar to equilibrium phases of matter. 
However, (ii) the component of any perturbation that transforms nontrivially under the symmetry is echoed out by the geometric part of the evolution (the Wilson line). 
Since each $\Pi_0[t]$ ($t$ fixed) lies in its corresponding equilibrium phase, a Floquet phase contains its static counterpart and can be strictly larger than it, cf.~Fig.~\ref{fig:F_phases}.
For instance, the GS-DTC phase has more stable states than the AFM, since any symmetry-breaking perturbation transforms odd under the $\mathbb{Z}_2$ lattice symmetry of the AFM, and is echoed out by the drive.

\section{\label{sec:no_heating}Stability against Energy Absorption and Heating}

Generic interacting Floquet systems are expected to absorb energy and heat up to a featureless infinite-temperature state~\cite{moessner2017equilibration, rodriguez2024quantifying,znidaric2026prethermalization}. One may therefore wonder why the Floquet ground state can be stable, and what the underlying mechanism to prevent heating is. 

Note first that the Floquet ground state is, by definition, a stationary state of the dynamics generated by the Hamiltonian $H(t)$. Consequently, its time evolution amounts to pure phase accumulation, preserving its average energy. What is perhaps surprising is that, for certain parameter regimes, the Floquet ground state is not a featureless infinite-temperature state, but only area-law entangled.

To see why, let us first recap the argument for heating~\cite{moessner2017equilibration}.
Intuitively, generic driven systems, which lack conservation laws, are bound to reach an infinite-temperature state by the maximum entropy principle of statistical mechanics. 
Specifically, since quasienergy is defined modulo the drive frequency, the quasienergy spectrum is folded into a finite Floquet zone. In the thermodynamic limit, the exponentially large number of eigenstates generically results in an effectively constant quasienergy density of states.
Hence, infinite-temperature Floquet eigenstates appear at all quasienergies; this, together with the absence of any quasienergy gaps in the thermodynamic limit, leaves the infinite-temperature ensemble as a natural candidate to describe the infinite-time dynamics, see Fig.~\ref{fig:GFT_ED}(b). 
This picture is corroborated by Floquet ETH~\cite{moessner2017equilibration,dalessio2016quantum}: apart from systems with additional constraints, such as integrability or many-body localization, generic Floquet eigenstates are believed to be locally indistinguishable from infinite-temperature states, although notable exceptions have been reported~\cite{haldar2021dynamical}. 

The heating mechanism can be understood perturbatively in the eigenbasis of the non-driven Hamiltonian. For a generic nonintegrable system, the periodic drive has nonzero matrix elements connecting eigenstates with different energies. Fermi's Golden Rule then gives a nonzero transition rate~\cite{rakcheev2022estimating} for absorbing energy from the drive. In the high-frequency regime, the heating rate is typically exponentially small in the drive frequency, giving rise to a long-lived prethermal plateau~\cite{ho2023quantum}. Nevertheless, in the absence of a mechanism that prevents heating, these transitions accumulate over sufficiently long times and drive the system toward a state whose local observables approach their infinite-temperature expectation values.

However, the above arguments do not rule out the existence of area-law Floquet ground states. 
First, note that the average energy spectrum can exhibit gaps at its edges, and that ETH is known to break down in their vicinity already for static systems: ground states do not obey ETH~\cite{dalessio2016quantum}. As Fig.~\ref{fig:GFT_ED}(b,c) shows, statistical probes that weigh all Floquet states equally, are naturally bound to miss a handful of structured eigenstates among a sea of infinite-temperature states.  
Second, recall that Floquet theory is an instance of exact counterdiabatic driving~\cite{schindler2025geometric} which is transitionless by construction: the gauge potential exactly cancels the off-diagonal terms responsible for transitions between the instantaneous ground-state manifold and excited states, providing a mechanism to evade Golden-rule heating; this is an exact nonperturbative effect which holds at any order of perturbation theory. By geometric Floquet theory~\cite{schindler2025geometric}, the lab-frame Hamiltonian $H(t)$ generates transitionless counterdiabatic driving for the Floquet states $\Pi_n[t]$, removing the transition channel from the ground state manifold before it can produce heating.
Thus, if the corresponding counterdiabatic gauge potential remains reasonably local in the thermodynamic limit, it provides a mechanism for protecting a Floquet eigenstate from heating.

To see this explicitly, let us briefly outline an exact procedure to embed area-law Floquet states in generic periodically driven many-body systems. Consider a local, nonintegrable, time-periodic Hamiltonian $H_\text{loc}(t)$ with an instantaneous area-law entangled ground-state manifold $\Pi_0[t]$ (degenerate in the case of spontaneously broken symmetry) and a finite spectral gap $\Delta{>}0$ throughout the period $T$.  
For gapped local Hamiltonians, Hastings' quasiadiabatic continuation provides a quasilocal time-periodic gauge potential, $A_\text{GS}(t) = \int\mathrm ds F_\Delta(s)e^{-i s H_\text{loc}(t)} \dot H_\text{loc}(t) e^{i s H_\text{loc}(t)}$, where the Osborne filter $F_\Delta$ is chosen to reproduce the inverse energy denominator for transitions across the gap~\cite{osborne2007simulating}. This construction implements exact parallel transport for the ground-state manifold~\cite{hastings2005Quasiadiabatic}. Moreover, the Osborne-Hastings AGP $A_\text{GS}$ is quasilocal in space with at worst stretched-exponentially decaying spatial tails~\cite{hastings2010locality}. It then follows that the CD Hamiltonian $H(t) = H_\text{loc}(t) + A_\text{GS}(t)$ is also a time-periodic quasilocal operator. By Geometric Floquet theory, the instantaneous ground state of $H_\text{loc}(t)$ is an exact Floquet eigenstate of $H(t)$. Because this state is also the ground state of the local gapped Hamiltonian $H_\text{loc}(t)$, which was assumed area-law, it gives an area-law entangled Floquet state. Its average energy is the period average of the instantaneous ground-state energy of $H_\text{loc}(t)$, $\mae_0=\bar E_0$; shifting the scalar $\bar E_0-E_0(t)$ from $H_\text{loc}$ to $A_\text{GS}$ then brings the decomposition into the average-energy gauge~\eqref{app:eq:AE_conditions} within the ground-state block\footnote{Whether or not $\Pi_0$ is separated from all other Floquet states by a finite $\mae$-gap is not implied by the construction.}, without affecting quasilocality or the parent. The constructed $H(t)$ thus admits a quasilocal decomposition of the form\footnote{The quasilocal parent Hamiltonian $\mAE^{\mathrm{q'loc}}(t)$ acts as a dynamical invariant only on the ground-state sector; in general, $\mathcal{L}(\mAE^{\mathrm{q'loc}})\neq 0$.} $H(t)=\mAE^{\mathrm{q'loc}}(t)+A^{\mathrm{q'loc}}_{\mAE}(t)$ discussed in Sec.~\ref{sec:outro}.
Alternative constructions of parent Hamiltonians~\cite{ljubotina2024tangent,fernandez2015frustration} can likely also be used to find $\mAE^{\mathrm{q'loc}}(t)$, although they may not exhibit the underlying gap protection mechanism based on quasilocal CD driving as clearly.

This argument applies to generic nonintegrable Floquet systems, 
constructing a specific class of Hamiltonians for which an exact transitionless channel protects a structured Floquet eigenstate in the thermodynamic limit.
Whereas it does not prove that the numerically obtained Floquet ground states of the specific driven Ising chain~\eqref{eq:Ising_chain} are stable in the thermodynamic limit, it demonstrates that 
(1) there exist nonintegrable quasilocal drives with an exact area-law Floquet eigenstate whose average energy is the period-averaged ground-state energy of a gapped local parent; and, 
(2) choosing $H_\text{loc}(t)$ inside an ordered phase gives exact Floquet eigenspaces that
are ordered at every fixed time and remain ordered under quasiadiabatic deformation of the parent; hence, ordered Floquet ground-state phases protected by this mechanism exist well beyond the class of integrable systems.

Let us point out that this stabilization mechanism does not invalidate the general energy absorption and heating arguments laid out above. 
Observe that generic initial states still heat up to infinite temperature, unless they have a finite overlap with an area-law Floquet ground state $\Pi_0[0]$, which may be difficult to sustain in the thermodynamic limit~\cite{Anderson1967Infrared,pizzi2020time}. 

Finally, we emphasize that Floquet ground-state order differs from prethermal order~\cite{ho2023quantum,khemani2019brief,else2020discrete} in that (i) it is zero-temperature (i.e., quantum) rather than thermodynamic, and (ii) it is infinitely (rather than exponentially) long-lived, and hence allows us to define stable phases of matter.
We also recall that the results we presented in this work apply to clean, disorder-free systems. In particular, unlike eigenstate order~\cite{huse2013localization,khemani2019brief}, Floquet ground-state phases do not require many-body Anderson localization as a stabilizing mechanism; hence, we expect them to exist in systems of any spatial dimension.

\section{\label{sec:outro}Discussion \& Outlook}

We derive a general variational principle for the average-energy ground state, applicable to any dimension, and allowing one to study Floquet many-body systems using arbitrary variational families~\cite{cheng2022matrix,vandewalle2024manybody}. 
Using a matrix product state ansatz, we demonstrate its utility for a driven nonintegrable Ising chain in both the perturbative high-frequency and a nonperturbative `low'-frequency regime. 
Numerical extrapolation in the system size is consistent with a finite average-energy gap in the thermodynamic limit, and the corresponding Floquet ground state retaining its area-law entanglement.

We show that the Floquet ground state can exhibit long-range order as a result of spontaneous symmetry breaking, introducing the concept of \textit{Floquet ground-state order}. In particular, the ground state of the periodically driven Ising chain hosts parameter regimes supporting equilibrium-like Floquet antiferromagnetic (F-AFM) and inherently nonequilibrium discrete-time-crystalline (GS-DTC) states. 
An interesting open problem is to investigate topologically ordered Floquet ground states~\cite{po2017radical,will2025probing,mambrini2026detuning}. 
Our DMRG data on up to 140 spins is converged in the bond dimension and is consistent with finite order in the thermodynamic limit; the one-cycle infidelity of the Floquet eigenstates improves systematically towards machine precision.

Exhibiting extended parameter regimes, in which the average energy gap remains open, allows us to define stable Floquet ground-state phases of matter in clean (disorder-free) systems. We illustrate the concept using both a paradigmatic exactly solvable circularly driven XY model and a nonintegrable Ising chain.
The average-energy ground state at any fixed time $\Pi_0[t]$ is indistinguishable from the ground state of a static system, and can be adiabatically connected to representatives of the corresponding equilibrium phase of matter. 
By contrast, the inherently nonequilibrium Floquet ground-state phases are induced by the geometric Wilson line operator; the latter provides a dynamical obstruction to connecting nonequilibrium-ordered states (the loops $\left\{\Pi_0[t]\!: t\in[0,T)\right\}$) to their equilibrium counterparts (the points $\Pi_0[t]$ with $t$ fixed).

Our results demonstrate explicitly that not all Floquet many-body states are locally equivalent to featureless infinite-temperature states. In particular, the Eigenstate Thermalization Hypothesis does not apply to all Floquet states. We have identified a gap protection mechanism enforced by quasilocal counterdiabatic driving, which eliminates transitions out of the Floquet ground state manifold, and provides the requisite stability for the underlying quantum phases of matter.
The notion of a ground-state gap together with the inner workings of this protection mechanism based on counterdiabatic evolution, set Floquet ground states apart from other ergodicity-resistant states, such as many-body scars~\cite{ljubotina2024tangent,iadecola2019quantum,maskara2021discrete,hudomal2022driving,ren2025scarfinder,bluvstein2021controlling}.
Because energy gaps do not exist in classical Hamiltonian systems, Floquet ground-state order, as a variant of quantum ground-state order, is inherently quantum mechanical. Whether a thermodynamic notion of Floquet order exists is currently an open question. 

We emphasize that the question about the infinite-time stability of the Floquet ground state of the driven Ising chain~\eqref{eq:Ising_chain} in the thermodynamic limit remains open. Although our numerical results strongly suggest a structured ground state, the MPS ansatz has a natural bias towards small-bond-dimension states; moreover, it may ignore any leftover exponentially small corrections in the drive frequency that cannot be resolved within machine precision. 
We also stress that the converse is equally open: since the available heating bounds are upper bounds in operator norm, there is at present no argument that the ground-state manifold of $\mAE$ must be destabilized in the thermodynamic limit.
These caveats notwithstanding, the stability mechanism we exhibited, based on counterdiabatic driving, shows that the ground-state problem is more structured and obeys different physics than arbitrary-state prethermalization. This also raises the question of whether the Floquet-Magnus expansion can converge for the gapped ground-state manifold only, as opposed to its asymptotic series behavior w.r.t.~the uniform operator norm dominated by infinite temperature states~\cite{abanin2015exponentially,mori2016rigorous}. 
We therefore anticipate gapped Floquet ground states to have provably enhanced and potentially infinite lifetimes. 

Irrespective of the fate of the Ising chain, the notions of Floquet ground state order and the corresponding quantum phases of matter remain generic: the forward construction of such phases based on ground-state geometric Floquet theory [Sec.~\ref{sec:no_heating}] demonstrates their existence in the thermodynamic limit of arbitrary instantaneously-gapped nonintegrable models. However, the prevalence of these systems (i.e., the solution of the inverse construction) still needs to be determined: more precisely, it remains unclear under which conditions local periodically driven Hamiltonians $H(t)$ decompose into a sum of a gapped quasilocal parent Hamiltonian $\mAE^\text{q'loc}(t)$ for the Floquet ground state and a quasilocal adiabatic gauge potential $A^\text{q'loc}_\mAE(t)$ which parallel-transports that same state: $H(t) {=} \mAE^\text{q'loc}(t) {+} A^\text{q'loc}_\mAE(t)$. Incidentally, this is the very same mathematical question of what becomes of Floquet's theorem in the thermodynamic limit within the space of quasilocal operators.

Experimentally, our work opens up the possibility for the realization of various kinds of nonequilibrium order~\cite{po2017radical,will2025probing,will2026chiral}, predicted to exist in Floquet systems in two and higher dimensions, since it relaxes the requirement for the system to be many-body localized to avoid heating. In this regard, we propose two generic protocols for Floquet ground state preparation: (i) explicit counterdiabatic driving of Floquet eigenstates~\cite{schindler2024counterdiabatic}, and (ii) a quench protocol that exploits the geometric decomposition of Floquet ground state phases of matter. 
Additionally, the variational techniques we have developed are expected to significantly boost the field of Floquet engineering for strongly correlated systems: Floquet ground states in the interacting many-body regime of large system sizes can now be efficiently accessed in classical numerical simulations. 
Whereas our numerical algorithms are sufficient to reveal the physics of the Floquet ground state, our implementation of the variational principle, cf.~Appendix~\ref{app:sec:mps_algo}, allows for plenty of room for future improvement and optimization. 

Despite this progress, a number of important questions remain unanswered. 
(i) Perhaps the most pressing one among them concerns the necessary and sufficient conditions on the lab-frame Hamiltonian $H(t)$ for a gapped area-law Floquet ground-state phase to occur.
We expect this to be generically the case in the vicinity of perturbative points (such as the infinite-frequency limit), if the effective Floquet-Magnus Hamiltonian truncated to the leading order exhibits a ground state gap. 
However, our results on the continuously driven Ising model~\eqref{eq:Ising_chain} in the nonperturbative `low'-frequency regime of the GS-DTC, where no obvious small parameter can be identified, imply that this criterion, albeit likely sufficient, may not be strictly necessary.  
(ii) Another question opening up an interesting future direction regards the definition of gapless Floquet ground state phases. Ongoing work on understanding Floquet systems using conformal field theories~\cite{fan2020emergent,han2020classification,lin2025local,lapierre2026nonequilibrium,mo2026observing} is likely to prove instrumental here.
And last but not least, (iii) the behavior of the Floquet ground state under coupling to a dissipative environment~\cite{ritter2025autonomous,raman2026dissipative} would be essential to understand before any practical applications of Floquet ground-state order in solid state systems are developed.

\textit{Acknowledgments.---}%
We thank Wojciech De Roeck, Roderich Moessner, Shivaji Sondhi, Miles Stoudenmire, Ruben Verresen, and in particular Frank Pollmann for valuable discussions.
We used the ITensor library~\cite{itensor, itensor-r0.3} for the tensor network simulations.
The authors used Claude Opus 5.5 and Fable 5.1 during the preparation of this manuscript to improve readability of appendices and assist with code optimization. 
The authors assume full responsibility for the scientific content, accuracy, and integrity of this paper.
MB was supported in part by grant NSF PHY-2309135 to the Kavli Institute for Theoretical Physics (KITP).
Funded by the European Union (ERC, QuSimCtrl, 101113633). Views and opinions expressed are, however, those of the authors only and do not necessarily reflect those of the European Union or the European Research Council Executive Agency. Neither the European Union nor the granting authority can be held responsible for them.

\textit{Data availability statement.---}%
The data and code associated with this manuscript are deposited on Zenodo~\cite{zenodo}, DOI: \href{https://doi.org/10.5281/zenodo.23497831}{10.5281/zenodo.23497831}.
Automatic access is restricted; access credentials will be provided to human researchers upon reasonable request via Zenodo for academic and verification purposes.

\clearpage
\appendix
\tableofcontents

\section{Floquet theory}
\label{app:sec:background}
 
In this appendix, we collect the background on Floquet theory underlying the variational principle of Appendix~\ref{app:sec:varl} and its matrix-product implementation in Appendix~\ref{app:sec:mps_algo}. 
We first recall conventional Floquet theory and fix our notation for Floquet states (Sec.~\ref{app:sec:Floquet}). 
We then introduce the three spaces in which we formulate the problem -- the physical Hilbert space, Sambe space, and Sambe superspace -- together with the matrix representation of the Floquet Liouvillian used in our numerical simulations (Sec.~\ref{app:sec:notation}). 
Section~\ref{app:sec:gft} reviews Geometric Floquet theory~\cite{schindler2025geometric} and the average-energy gauge, which we compare to the Floquet and Kato gauges in Sec.~\ref{app:sec:comparison}. 
Finally, in Sec.~\ref{app:sec:AE_operator} we derive basis-independent expressions for the average-energy operator $\mAE$ and its gauge potential $\agp_\mAE$, and prove the properties of the Floquet Liouvillian $\mathcal{L}$ used in the main text and in Appendix~\ref{app:sec:varl}.

\subsection{Conventional Floquet theory}
\label{app:sec:Floquet}
 
Consider a Hamiltonian $H(t)=H(t+T)$ with period $T=2\pi/\omega$, acting on a Hilbert space $\hilbert{H}$ of finite dimension $D=\dim\hilbert{H}$. 
Floquet's theorem guarantees that the evolution operator from the starting time $t_0$ to time $t$ factorizes as
\begin{equation}
  \label{app:eq:floquet_unitary}
  U(t,t_0)
  = \mathcal{T}\exp\left(-i\!\int_{t_0}^{t}\!\dd s\, H(s)\right)
  = P(t,t_0)\,e^{-i(t-t_0)\HF[t_0]},
\end{equation}
i.e., into a periodic micromotion operator $P(t,t_0)$ and a time-independent Floquet Hamiltonian $\HF[t_0]$. 
The micromotion operator is given by $P(t,t_0)=P(t)P^\dagger(t_0)$ with $P(t)\equiv P(t,0)=P(t+T)$ and $P(0)=\identity$; in particular, $P(t_0+T,t_0)=\identity$. 
The Floquet Hamiltonian depends on the starting time only parametrically, $\HF[t_1]=P(t_1,t_0)\HF[t_0]P^\dagger(t_1,t_0)$, and generates the stroboscopic dynamics, 
\begin{equation}
  \label{app:eq:floquet_stroboscopic}
  \UF[t_0] = U(t_0+T,t_0) = e^{-iT\HF[t_0]}.
\end{equation}
As in the main text, the square bracket $[t]$ indicates that only the eigenvectors of an operator depend on time, whereas its eigenvalues are $t$-independent; round brackets $(t)$ denote a generic time dependence.
 
The eigenvalues of $\HF[t_0]$ -- the quasienergies $\epsilon_{\F,n}$ -- do not depend on $t_0$. 
However, since Eq.~\eqref{app:eq:floquet_stroboscopic} fixes $\HF[t_0]$ only through its exponential, the quasienergies are defined modulo $\omega$: the replacement $\HF[t_0]\mapsto\HF[t_0]+\omega\sum_n\ell_n\Pi_n[t_0]$ with $\ell_n\in\mathbb{Z}$ leaves $\UF[t_0]$ invariant. 
This is the folding ambiguity in Floquet systems, whose origin can be traced back to an incomplete gauge fixing~\cite{schindler2025geometric}, see Sec.~\ref{app:sec:gft}.
 
We denote the eigenstates of $\HF[t]$ by $\ket{n_\F[t]}$, $\HF[t]\ket{n_\F[t]}=\epsilon_{\F,n}\ket{n_\F[t]}$, and fix their phases via $\ket{n_\F[t]}=P(t,t_0)\ket{n_\F[t_0]}$. 
These Floquet states are $T$-periodic, $\ket{n_\F[t+T]}=\ket{n_\F[t]}$, and evolve according to
\begin{equation}
  \label{app:eq:floquet_states}
  U(t,t_0)\ket{n_\F[t_0]} = e^{-i\epsilon_{\F,n}(t-t_0)}\ket{n_\F[t]}.
\end{equation}
Substituting Eq.~\eqref{app:eq:floquet_states} into the Schr\"odinger equation, we obtain the eigenvalue equation
\begin{equation}
  \label{app:eq:floquet_modes}
  \bigl(H(t)-i\partial_t\bigr)\ket{n_\F[t]} = \epsilon_{\F,n}\ket{n_\F[t]},
\end{equation}
which forms the starting point of the Sambe-space construction in Sec.~\ref{app:sec:varl:sambe}. 
The eigenprojectors $\Pi_n[t]=\ketbra{n_\F[t]}$ are independent of the phase convention, $T$-periodic, and satisfy $\Pi_n[t]=U(t,t_0)\Pi_n[t_0]U^\dagger(t,t_0)$; for a degenerate quasienergy, $\Pi_n[t]$ projects onto the entire eigenspace.
 
Finally, differentiating Eq.~\eqref{app:eq:floquet_unitary} w.r.t.~$t$ yields the \emph{Floquet decomposition}
\begin{equation}
  \label{app:eq:floquet_decomposition}
  H(t) = \HF[t] + \AF(t),
  \qquad
  \AF(t) = i\dot{P}(t)P^\dagger(t),
\end{equation}
where we refer to $\AF(t)$ as the Floquet adiabatic gauge potential; the reason for this name will become clear in Sec.~\ref{app:sec:gft}.

\subsection{Hilbert spaces and notation}
\label{app:sec:notation}
 
The variational principle requires us to introduce three different spaces: (i) the physical Hilbert space $\hilbert{H}$, (ii) the Sambe space $\hilbert{S}=\hilbert{H}\otimes\floquethilbert$ of time-periodic states, and (iii) the Sambe superspace $\hilbert{S}_S=\hilbert{H}^2\otimes\floquethilbert$ of time-periodic operators. 
Here, $\floquethilbert$ denotes the space of $T$-periodic functions, and $\hilbert{H}^2\equiv\hilbert{H}\otimes\hilbert{H}^*$ the space of operators on $\hilbert{H}$. 
We use plain symbols ($H$, $A$, \dots) for operators on $\hilbert{H}$, hats ($\sambe{O}$) for operators on Sambe space, and calligraphic symbols ($\mathcal{L}$, $\mathcal{C}_H$, \dots) for linear maps on Sambe superspace, i.e., superoperators. 
States in $\hilbert{H}$, $\hilbert{S}$, and $\hilbert{S}_S$ are denoted by $\ket{\psi}$, $\timeket{\psi}$, and $\floquetket{O}$, respectively.

\subsubsection{Physical Hilbert space}
\label{app:sec:varl:physical}
 
The physical Hilbert space $\hilbert{H}$ carries the states $\ket{\psi}$ and the operators of the driven system, including the lab-frame Hamiltonian $H(t)$. 
We expand $T$-periodic operators in harmonics of the drive frequency,
\begin{equation}
  \label{app:eq:fourier}
  O(t) = \sum_{n\in\mathbb{Z}} e^{in\omega t}\, O^{(n)},
  \qquad
  O^{(n)} = \frac{1}{T}\int_0^T\!\dd t\, e^{-in\omega t}\,O(t),
\end{equation}
and similarly for $T$-periodic states; hermiticity of $O(t)$ implies $O^{(-n)}=O^{(n)\dagger}$. 
In Fourier space, the product of two periodic functions becomes a convolution of their harmonics, $(AB)^{(n)}=\sum_m A^{(n-m)}B^{(m)}$, while the time derivative turns into multiplication by the harmonic index, $(-i\partial_t O)^{(n)}=n\omega\,O^{(n)}$. 
Both Sambe space and Sambe superspace use these two rules.

\subsubsection{Sambe space}
\label{app:sec:varl:sambe}
 
The Fourier-mode space $\floquethilbert=\mathrm{span}\{\ket{m}_\floquethilbert\}_{n\in\mathbb{Z}}$ represents the function $e^{in\omega t}$ by the basis vector $\ket{m}_\floquethilbert$,\footnote{Note the difference between the Floquet state $|n_\F\rangle\in\mathbb{H}$, and the Fourier mode $\ket{m}_\floquethilbert\in\mathbb{F}$.} with inner product $\braket{m}{n}_\floquethilbert=T^{-1}\int_0^T\dd t\,e^{-im\omega t}e^{in\omega t}=\delta_{mn}$~\cite{sambe1973steady}. 
A $T$-periodic state $\ket{\psi(t)}=\sum_m e^{im\omega t}\ket{\psi^{(m)}}$ thus maps to the Sambe-space vector
\begin{equation}
  \label{app:eq:sambe_state}
  \begin{aligned}
  \timeket{\psi} &= \sum_{m\in\mathbb{Z}} \ket{\psi^{(m)}}\otimes\ket{m}_\floquethilbert \in \hilbert{S},
  \\
  (\psi|\phi) &= \frac{1}{T}\int_0^T\!\dd t\,\braket{\psi(t)}{\phi(t)} ,
  \end{aligned}
\end{equation}
consistent with the photon-sector expansion of Appendix~\ref{app:sec:mps_algo}. 

Multiplication by a periodic operator $O(t)$ becomes a block-Toeplitz matrix $\sambe{O}$ with blocks $(\sambe{O})_{nm}=O^{(n-m)}$, such that $(\sambe{O}\timeket{\psi}^{(n)}=\sum_m O^{(n-m)}\ket{\psi^{(m)}}$, cf.~Eq.~\eqref{app:eq:conv1}; the derivative $-i\partial_t$, in turn, maps to $\omega\sambe{N}$, where the photon-number operator $\sambe{N}=\identity_\hilbert{H}\otimes\sum_m m\ketbra{m}_\floquethilbert$ is diagonal. 
Ordering rows and columns as $m=\dots,1,0,-1,\dots$, we have
\begin{widetext}
\begin{equation}
  \label{app:eq:sambe_mats}
  \sambe{H} =
  \begin{pmatrix}
    \ddots & \ddots & \ddots & \ddots & \\
    \ddots & H^{(0)}  & H^{(1)}  & H^{(2)} & \ddots \\
    \ddots & H^{(-1)} & H^{(0)}  & H^{(1)} & \ddots \\
    \ddots & H^{(-2)} & H^{(-1)} & H^{(0)} & \ddots \\
     & \ddots & \ddots & \ddots & \ddots
  \end{pmatrix},
  \qquad
  \sambe{N} =
  \begin{pmatrix}
    \ddots & \ddots & \ddots & \ddots & \\
    \ddots & \identity_\hilbert{H} & 0 & 0 & \ddots \\
    \ddots & 0 & 0 & 0 & \ddots \\
    \ddots & 0 & 0 & -\identity_\hilbert{H} & \ddots \\
     & \ddots & \ddots & \ddots & \ddots
  \end{pmatrix}.
\end{equation}
\end{widetext}
 
In Sambe space, Eq.~\eqref{app:eq:floquet_modes} turns into a time-independent eigenvalue problem 
\begin{equation}
    \sambe{Q}\timeket{n_\F} = \epsilon_{\F,n}\timeket{n_\F}
\end{equation}
for the quasienergy operator
\begin{equation}
  \label{app:eq:sambe_Q}
  \sambe{Q} = \sambe{H} + \omega\sambe{N},
\end{equation}
where $\timeket{n_\F}$ represents the Floquet state $\ket{n_\F[t]}$. 
This eigenvalue problem features an infinite degeneracy. 
To see this, let $\sambe{E}_m=\identity_\hilbert{H}\otimes\sum_k\ket{k+m}\bra{k}_\floquethilbert$ denote multiplication by $e^{im\omega t}$. 
Since $\sambe{E}_m$ commutes with the block-Toeplitz operator $\sambe{H}$ and $\sambe{N}\sambe{E}_m=\sambe{E}_m(\sambe{N}+m)$, the state $\sambe{E}_m\timeket{n_\F}$ is an eigenstate of $\sambe{Q}$ with eigenvalue $\epsilon_{\F,n}+m\omega$, for any $m\in\mathbb{Z}$. 
Hence, the spectrum $\{\epsilon_{\F,n}+m\omega\}_{n,m}$ of $\sambe{Q}$ is unbounded both from below and from above, and does not single out a lowest state; these replicas define different Floquet zones and reproduce the folding ambiguity of Sec.~\ref{app:sec:Floquet}.

\subsubsection{Sambe superspace}
\label{app:sec:varl:superspace}
 
Since the variational principle acts on operators rather than on states, we treat $T$-periodic operators as vectors,
\begin{equation}
  \label{app:eq:superspace_vector}
  \floquetket{O} = \sum_{n\in\mathbb{Z}} \floquetket{O^{(n)}}\otimes\ket{n}_\floquethilbert \in \hilbert{S}_S = \hilbert{H}^2\otimes\floquethilbert,
\end{equation}
where $\floquetket{O^{(n)}}\in\hilbert{H}^2$ is the column-stacked vectorization of the matrix $O^{(n)}$, which obeys $\floquetket{AXB}=(B^T\otimes A)\floquetket{X}$. 
We equip Sambe superspace with the period-averaged Hilbert-Schmidt inner product
\begin{equation}
  \label{eq:HS_inner_product}
  \floquetbraket{A}{B}
  = \frac{1}{T}\int_0^T\!\dd t\;\mathrm{tr}\bigl[A^\dagger(t)\,B(t)\bigr]
  = \sum_{n\in\mathbb{Z}}\mathrm{tr}\bigl[A^{(n)\dagger}B^{(n)}\bigr],
\end{equation}
where the second equality follows from Parseval's identity. 
Inside superoperator expressions, such as $\mathcal{L}(H)$, an operator $H$ and its vector $\floquetket{H}$ refer to the same element of $\hilbert{S}_S$.
 
The Floquet Liouvillian of the main text, Eq.~\eqref{eq:Liouvillian},
\begin{equation}
  \label{app:eq:liouvillian}
  \mathcal{L}(O) = \comm{H(t)}{O(t)} - i\partial_t O(t),
\end{equation}
defines a superoperator on $\hilbert{S}_S$. 
We denote by $\mathcal{C}_H$ the superoperator implementing the commutator with $H(t)$, and by $\omega\mathcal{N}$ the superoperator implementing $-i\partial_t$. 
The convolution rule and $(-i\partial_t O)^{(n)}=n\omega O^{(n)}$ then give the vectorized form
\begin{equation}
  \label{app:eq:liouvillian_vec}
  \mathcal{L} = \mathcal{C} + \omega\mathcal{N},
  \qquad
  \mathcal{C}^{(n)}=\mathcal{C}_H^{(n)} = \identity\otimes H^{(n)} - H^{(n)T}\otimes\identity,
\end{equation}
where $\mathcal{C}_H=\mathcal{C}$ is block-Toeplitz with blocks $(\mathcal{C}_H)_{nm}=\mathcal{C}_H^{(n-m)}$, and $\mathcal{N}=\identity_{\hilbert{H}^2}\otimes\sum_m m\ketbra{m}_\floquethilbert$ counts the harmonic of the operator; We sometimes drop the subscript $_H$ from the definition of $\mathcal{C}_H$ and its Fourier harmonics to avoid overcrowding notation.
Explicitly,
\begin{widetext}
\begin{equation}
  \label{app:eq:mats}
  \mathcal{C}_H =
  \begin{pmatrix}
    \ddots & \ddots & \ddots & \ddots & \\
    \ddots & \mathcal{C}^{(0)}  & \mathcal{C}^{(1)}  & \mathcal{C}^{(2)} & \ddots \\
    \ddots & \mathcal{C}^{(-1)} & \mathcal{C}^{(0)}  & \mathcal{C}^{(1)} & \ddots \\
    \ddots & \mathcal{C}^{(-2)} & \mathcal{C}^{(-1)} & \mathcal{C}^{(0)} & \ddots \\
     & \ddots & \ddots & \ddots & \ddots
  \end{pmatrix},
  \qquad
  \mathcal{N} =
  \begin{pmatrix}
    \ddots & \ddots & \ddots & \ddots & \\
    \ddots & \identity_{\hilbert{H}^2} & 0 & 0 & \ddots \\
    \ddots & 0 & 0 & 0 & \ddots \\
    \ddots & 0 & 0 & -\identity_{\hilbert{H}^2} & \ddots \\
     & \ddots & \ddots & \ddots & \ddots
  \end{pmatrix}.
\end{equation}
\end{widetext}
This is the structure we exploit in Appendix~\ref{app:sec:mps_algo}: only the blocks $\mathcal{C}_H^{(n)}$ need to be stored; $\mathcal{C}_H\floquetket{Y}$ is evaluated as a convolution rather than a full matrix-vector product. 
The zero-harmonic block $\mathcal{C}_H^{(0)}$ implements the commutator with the period-averaged Hamiltonian $H^{(0)}$, whereas the blocks with $n\neq0$ couple harmonics that differ by $n$.
 
The Floquet Liouvillian is self-adjoint w.r.t.~the inner product~\eqref{eq:HS_inner_product}. 
Indeed, cyclicity of the trace gives $\floquetbra{A}\mathcal{C}_H\floquetket{B}=\floquetbra{B}\mathcal{C}_H\floquetket{A}^*$, and integration by parts gives $\floquetbra{A}\mathcal{N}\floquetket{B}=\floquetbra{B}\mathcal{N}\floquetket{A}^*$, where the boundary term vanishes since $A$ and $B$ are $T$-periodic. 
In the vectorized form~\eqref{app:eq:mats} both properties are manifest: $\mathcal{N}$ is real and diagonal, and $\mathcal{C}_H^{(n)\dagger}=\mathcal{C}_H^{(-n)}$ follows from $H^{(n)\dagger}=H^{(-n)}$. 
Hence $\mathcal{L}=\mathcal{L}^\dagger$ is hermitian, and therefore $\ker\mathcal{L}\perp\im\mathcal{L}$.\footnote{The Liouville operator of Ref.~\cite{schindler2026thesis}, $i\comm{H}{\cdot}+\partial_t$, equals $i\mathcal{L}$ and is therefore anti-Hermitian; the two conventions lead to the same kernel, image, and projectors.} 
Because of the term $\omega\mathcal{N}$, $\mathcal{L}$ is an unbounded operator; its range is nonetheless closed, since $\mathcal{N}$ has compact resolvent and $\mathcal{C}_H$ is a bounded perturbation which preserves this property~\cite{schindler2026thesis,kato1980perturbation,reed1980methods}. 
The spectrum of $\mathcal{L}$ is therefore discrete, zero is an isolated eigenvalue, and the Moore--Penrose pseudo-inverse $\mathcal{L}^+$ exists as a bounded operator~\cite{groetsch1977generalized}.
 
Finally, note that the Floquet Liouvillian coincides with the adjoint action of the quasienergy operator,
\begin{equation}
  \label{app:eq:liouvillian_Q}
  \mathcal{L}(O) \;\hat{=}\; \comm{\sambe{Q}}{\sambe{O}},
\end{equation}
since $\comm{\sambe{H}}{\sambe{O}}$ represents $\comm{H(t)}{O(t)}$, and $\omega\comm{\sambe{N}}{\sambe{O}}$ has blocks $\omega(n-m)O^{(n-m)}$, which represent $-i\partial_t O$. 
Equation~\eqref{app:eq:liouvillian_Q} connects the two different constructions of Sambe superspace that we now introduce and compare.

\subsubsection{Order of the constructions}
\label{app:sec:varl:routes}
 
Sambe superspace can be defined in two ways: 
(i) by first going to Sambe space and then considering operators on it, or 
(ii) by first going to the space $\hilbert{H}^2$ of operators and then expanding in harmonics, as in Eq.~\eqref{app:eq:superspace_vector}. 
Throughout this work, including the implementation in Appendix~\ref{app:sec:mps_algo}, we use route (ii).
 
Route (i) results in the space of all operators on $\hilbert{S}$ whose elements $\sambe{O}$ carry two Fourier indices $(n,m)$, i.e., $\hilbert{S}\otimes\hilbert{S}^*\cong\hilbert{H}^2\otimes\floquethilbert\otimes\floquethilbert^*$. 
In this space, the Liouvillian is the adjoint action of $\sambe{Q}$, cf.~Eq.~\eqref{app:eq:liouvillian_Q}, with vectorized form
\begin{equation}
  \label{app:eq:route_i}
  \identity\otimes\sambe{Q}-\sambe{Q}^T\otimes\identity
  = \bigl(\identity\otimes\sambe{H}-\sambe{H}^T\otimes\identity\bigr)
  + \omega\bigl(\identity\otimes\sambe{N}-\sambe{N}\otimes\identity\bigr),
\end{equation}
where we used $\sambe{N}^T=\sambe{N}$; note that the photon number now enters via the difference $\identity\otimes\sambe{N}-\sambe{N}\otimes\identity$ rather than via a single number operator.
 
Nevertheless, the two routes agree on time-periodic operators. 
A time-periodic operator lifts to a block-Toeplitz element, $(\sambe{O})_{nm}=O^{(n-m)}$, which depends on $(n,m)$ only through the difference $n-m$. 
Products and commutators of block-Toeplitz operators are again block-Toeplitz, and so is $\comm{\sambe{N}}{\sambe{O}}$, with blocks $(n-m)O^{(n-m)}$; hence, the block-Toeplitz elements form an invariant subspace of the operator~\eqref{app:eq:route_i}. 
The map $\sambe{O}\mapsto\sum_n\floquetket{O^{(n)}}\otimes\ket{n}_\floquethilbert$ identifies this subspace with $\hilbert{S}_S$ of route (ii); under this identification, the photon-number difference acts as $\mathcal{N}$, and the commutator with $\sambe{H}$ acts as $\mathcal{C}_H$. 
Restricted to time-periodic operators, the operator~\eqref{app:eq:route_i} thus coincides with $\mathcal{L}=\mathcal{C}_H+\omega\mathcal{N}$, Eq.~\eqref{app:eq:liouvillian_vec}. 
The remaining (non-Toeplitz) directions of route (i) correspond to two-time kernels $O(t,t')$, which are not needed to represent $H(t)$, $\mAE[t]$, or $\agp_\mAE(t)$.
 
Route (i) is, however, not useful computationally, for three reasons. 
First, it enlarges the space: with a Fourier cutoff $|n|\leq n_c$ on the Fourier index, route (i) operates on $(2n_c+1)^2$ pairs of indices $(n,m)$ instead of the $2n_c+1$ Fourier harmonics of route (ii).
Second, the Hilbert-Schmidt inner product of operators on $\hilbert{S}$ diverges for block-Toeplitz elements, since the trace counts the same blocks once for every value of the harmonic index $n$; a finite inner product requires dividing by the number of Fourier modes, which reproduces the period average in Eq.~\eqref{eq:HS_inner_product}. 
Third, after truncation, the commutator of two truncated block-Toeplitz matrices is no longer block-Toeplitz close to the cutoff; thus, the invariant subspace identified above is preserved only approximately and has to be enforced by hand. 
By contrast, route (ii) truncates a single Fourier index and keeps the Toeplitz structure implicit in the convolutions, Eqs.~\eqref{app:eq:conv1} and~\eqref{app:eq:conv2}. We therefore stick to route (ii).

\subsection{Geometric Floquet theory}
\label{app:sec:gft}
 
We now review the basics of Geometric Floquet theory (GFT)~\cite{schindler2025geometric,schindler2026thesis} required for the variational construction. 
GFT identifies Floquet theory as an \emph{inverse counterdiabatic driving} problem: the lab-frame Hamiltonian $H(t)$ generates transitionless driving for the Floquet states~\cite{schindler2025geometric}. 
This places the machinery of adiabatic gauge potentials (AGPs) at our disposal. Analyzing the gauge freedom associated with rephasing the Floquet eigenstates reveals that quasienergy folding results from an incomplete fixing of the gauge freedom of the AGP, $U(1)\mapsto\mathbb{Z}$; fixing the gauge completely via parallel transport leads to the average-energy operator $\mAE$, whose spectrum is sorted (i.e., not folded). 
 
\subsubsection{Dynamical invariants and stationary states.}
 
The Floquet Hamiltonian $\HF[t]=P(t)\HF[0]P^\dagger(t)$ is a \emph{dynamical invariant}, i.e., an operator $I[t]$ satisfying
\begin{equation}
  \label{app:eq:IOM}
  \frac{\dd}{\dd t}I[t] + i\comm{H(t)}{I[t]} = 0
  \quad\Longleftrightarrow\quad
  \mathcal{L}(I[t]) = 0 .
\end{equation}
Equation~\eqref{app:eq:IOM} generalizes the conditions $\comm{H}{S}=0$, $\dot{S}=0$ defining a symmetry $S$ of a static Hamiltonian; unlike static symmetries, however, dynamical invariants have time-dependent eigenvectors, whereas their eigenvalues remain time-independent, $I[t]=\sum_n \iota_n\Pi_n[t]$. 
Just as a static symmetry defines invariant sectors, the evolution generated by $H(t)$ maps the instantaneous eigenspaces of a dynamical invariant onto each other, $U(t,t_0)\Pi_n[t_0]U^\dagger(t,t_0)=\Pi_n[t]$. 
For a periodic dynamical invariant, the eigenprojectors are periodic, $\Pi_n[t+T]=\Pi_n[t]$, and the corresponding states are stationary states of the drive: they return to themselves after each period, up to a phase.
 
\subsubsection{Inverse counterdiabatic driving and the Floquet problem.}
 
The same structure appears in adiabatic evolution. 
Consider a control Hamiltonian $\Hctrl(\lambda)$ with instantaneous eigenstates $\ket{n[\lambda]}$ and eigenvalues $E_n(\lambda)$. 
If the parameter $\lambda=\lambda(t)$ is varied sufficiently slowly compared to the inverse energy gap of $\Hctrl$, $\dot\lambda\to0$, a system prepared in the state $\ket{n[\lambda(0)]}$ follows the instantaneous eigenstate,
\begin{equation}
  \label{app:eq:adiabatic}
  \ket{n(t)} \to e^{i\gamma_n(t)}\, e^{i\phi_n(t)}\, \ket{n[\lambda(t)]},
\end{equation}
and accumulates the dynamical and the geometric (Berry) phase
\begin{subequations}
\label{app:eq:phases}
\begin{align}
  \label{app:eq:dynamical_phase}
  \phi_n(t) &= -\int_0^t\!\dd s\, E_n(\lambda(s)),
  \\
  \label{app:eq:geometric_phase}
  \gamma_n(t) &= \int_{\lambda(0)}^{\lambda(t)}\!\dd\lambda\, \braket{n[\lambda]}{i\partial_\lambda n[\lambda]};
\end{align}
\end{subequations}
the latter depends only on the path traced by $\ket{n[\lambda]}$ in Hilbert space, but not on the schedule $\lambda(t)$.
 
Counterdiabatic (CD) driving produces the evolution~\eqref{app:eq:adiabatic} at arbitrary speed by adding the AGP $\agp_\lambda$ to the control Hamiltonian, $\HCD=\Hctrl+\dot\lambda\agp_\lambda$~\cite{Kolodrubetz2017Geometry}, where the AGP contains the transition matrix elements between instantaneous eigenstates, $\bra{m}\agp_\lambda\ket{n}=i\braket{m}{\partial_\lambda n}$. 
Setting $\lambda=t$, the Floquet decomposition~\eqref{app:eq:floquet_decomposition} takes the form of a CD Hamiltonian~\cite{schindler2025geometric},
\begin{equation}
  \label{app:eq:floquet_CD}
  \underbrace{H(t)}_{\hat{=}\,\HCD} = \underbrace{\HF[t]}_{\hat{=}\,\Hctrl} + \underbrace{\AF(t)}_{\hat{=}\,\agp_\lambda},
\end{equation}
with $\HF[t]$ in the role of the control Hamiltonian and $\AF(t)=i\dot{P}P^\dagger$ in the role of the AGP. 
Finding the Floquet Hamiltonian is therefore an inverse CD problem: the CD Hamiltonian $H(t)$ is given, and one seeks the corresponding control Hamiltonian $H_\F[t]$.
 
\subsubsection{Gauge freedom and quasienergy folding.}
 
The AGP is defined only up to the $U(1)$ gauge transformation
\begin{equation}
  \label{app:eq:U1_gauge}
  \begin{aligned}
    \agp_\lambda &\mapsto \agp_\lambda - \sum_n \partial_\lambda\chi_n(\lambda)\,\Pi_n[\lambda],
    \\
    \ket{n_\F[\lambda]} &\mapsto e^{i\chi_n(\lambda)}\ket{n_\F[\lambda]},
  \end{aligned}
\end{equation}
which leaves $\HCD$ unchanged since the additional diagonal term is absorbed into $\Hctrl$. 
In the Floquet setting, the states $\ket{n_\F[t]}$ are required to be $T$-periodic, which restricts the gauge functions to $e^{i\chi_n(t+T)}=e^{i\chi_n(t)}$, i.e., $\chi_n(t)=\ell_n\omega t+\sum_{m\neq0}\chi_n^{(m)}e^{im\omega t}$ with $\ell_n\in\mathbb{Z}$~\cite{schindler2025geometric}. 
Requiring, in addition, that the control Hamiltonian remains a dynamical invariant, i.e., that its eigenvalues are time-independent, removes all harmonics, $\chi_n^{(m)}=0$, and leaves the discrete residual freedom $\chi_n(t)=\ell_n\omega t$:
\begin{equation}
  \label{app:eq:floquet_gauge}
  \HF[t]\;\mapsto\; \HF[t]+\omega\sum_n \ell_n\,\Pi_n[t],
  \qquad \ell_n\in\mathbb{Z}.
\end{equation}
Quasienergy folding is thus the result of an incomplete gauge fixing, which reduces the gauge group from $U(1)$ to $\mathbb{Z}$ rather than fixing the gauge completely~\cite{schindler2025geometric}.
 
\paragraph*{Kato gauge.}
 
The gauge freedom can be completely bypassed by working with the gauge-invariant Floquet eigenprojectors $\Pi_n[t]$. 
The parallel-transport (Kato) gauge~\cite{kato1950adiabatic,schindler2025geometric},
\begin{equation}
  \label{app:eq:kato_agp}
  \AK(t) = -\frac{1}{2}\sum_n\comm{\Pi_n[t]}{i\partial_t\Pi_n[t]},
\end{equation}
is fixed by the condition $\bra{n_\F[t]}\AK(t)\ket{n_\F[t]}=0$ at every time $t$, and yields the \emph{Kato decomposition}
\begin{equation}
  \label{app:eq:Kato_decomp}
  H(t) = \HK(t) + \AK(t),
  \qquad
  \HK(t) = \sum_n E_{\K,n}(t)\,\Pi_n[t],
\end{equation}
with Kato energies $E_{\K,n}(t)=\bra{n_\F[t]}H(t)\ket{n_\F[t]}$. 

The Kato Hamiltonian $\HK(t)$ shares its eigenvectors with $\HF[t]$ and is free of the folding ambiguity. Unlike $\HF[t]$, however, it is not a dynamical invariant, since the Kato energies $E_{\K,n}(t)$ are time-dependent in general. 
Because $\AK(t)$ has no diagonal elements at any time, the evolution it generates parallel-transports each Floquet eigenstate along its path, and returns it after one period up to its geometric phase, $\mathcal{T}\exp\bigl(-i\int_{t_0}^{t_0+T}\dd t\,\AK(t)\bigr)=e^{i\Gamma[t_0]}$, with $\Gamma[t_0]$ defined in Eq.~\eqref{app:eq:HK_stroboscopic} below. 
The Kato decomposition also reproduces the adiabatic limit, $\HK\to H$ and $\AK\to0$ for $\omega\to0$~\cite{schindler2025geometric}.
 
\paragraph*{Average-energy gauge.}
 
The average-energy operator combines the two desired properties: it is a dynamical invariant, and its spectrum is not folded~\cite{schindler2025geometric}. 
Its eigenvalues are the period-averaged Kato energies of the Floquet states, and we refer to them as average energies
\begin{equation}
  \label{app:eq:average-energies}
  \mae_n = \frac{1}{T}\int_0^T\!\dd t\, \bra{n_\F[t]} H(t)\ket{n_\F[t]} .
\end{equation}
The eigenvectors of the average energy operator are the Floquet states, $\mAE[t]=\sum_n\mae_n\Pi_n[t]$; hence, $\mAE[t]$ commutes with both $\HF[t]$ and $\HK(t)$. We call the lowest average-energy state the Floquet ground state.  

The corresponding \emph{average-energy decomposition},
\begin{equation}
  \label{app:eq:AE_decomposition}
  H(t) = \mAE[t] + \agp_\mAE(t),
\end{equation}
is fixed uniquely by the two conditions
\begin{equation}
  \label{app:eq:AE_conditions}
  \begin{aligned}
    \mathcal{L}\bigl(\mAE[t]\bigr) &= 0,
    \\
    \frac{1}{T}\int_0^T\!\dd t\,\bra{n_\F[t]}\agp_\mAE(t)\ket{n_\F[t]} &= 0
    \qquad \forall\, n,
  \end{aligned}
\end{equation}
i.e., $\mAE[t]$ is a dynamical invariant and $\agp_\mAE(t)$ satisfies the parallel-transport condition on average over one period; we give the proof in Sec.~\ref{app:sec:AE_projector}. 
In the infinite-frequency limit $\omega\to\infty$, the average-energy operator reduces to the period-averaged Hamiltonian, $\mAE[t_0]\to H^{(0)}=T^{-1}\int_0^T\dd t\,H(t)$; the decomposition~\eqref{app:eq:AE_decomposition} itself is exact at all frequencies.
 
\subsubsection{Wilson line and geometric phases.}
 
The average-energy decomposition factorizes the evolution operator as~\cite{schindler2025geometric}
\begin{equation}
  \label{app:eq:AE_evolution}
  \begin{aligned}
  U(t,t_0) &= W(t,t_0)\, e^{-i(t-t_0)\mAE[t_0]},
  \\
  W(t,t_0) &= \mathcal{T}\exp\left(-i\!\int_{t_0}^{t}\!\dd s\,\agp_\mAE(s)\right),
  \end{aligned}
\end{equation}
cf.~Eq.~\eqref{eq:GFT} of the main text for $t_0=0$. 
To verify Eq.~\eqref{app:eq:AE_evolution}, define $V(t)=U(t,t_0)e^{i(t-t_0)\mAE[t_0]}$; since $\mAE$ is a dynamical invariant, we have $U(t,t_0)\mAE[t_0]U^\dagger(t,t_0)=\mAE[t]$. Therefore, it follows that $i\partial_t V=(H(t)-\mAE[t])V=\agp_\mAE(t)V$ with $V(t_0)=\identity$, i.e., $V=W$. 

The Wilson line $W$ thus carries the geometric part of the dynamics, whereas $e^{-i(t-t_0)\mAE[t_0]}$ carries the dynamical part. 
Evaluated after one full period, the Wilson line reduces to
\begin{equation}
  \label{app:eq:HK_stroboscopic}
  W(t_0+T,t_0) = e^{i\Gamma[t_0]},
  \qquad
  \UF[t_0] = e^{i\Gamma[t_0]}\, e^{-iT\mAE[t_0]},
\end{equation}
with $\Gamma[t_0]=\sum_n\gamma_n\Pi_n[t_0]$ and the geometric phases of the Floquet states are given by
\begin{equation}
  \label{app:eq:berry_phase}
  \gamma_n = \int_{t_0}^{t_0+T}\!\dd t\, \braket{n_\F[t]}{i\partial_tn_\F[t]} ,
\end{equation}
in agreement with Eq.~\eqref{app:eq:geometric_phase} for $\lambda=t$. 
Since the integrand is $T$-periodic, $\gamma_n$ does not depend on $t_0$. A periodic gauge transformation $\chi_n=\ell_n\omega t$ shifts $\gamma_n$ by $-2\pi\ell_n$. 
Equation~\eqref{app:eq:berry_phase} follows from the Schr\"odinger equation for $W(t,t_0)\ket{n_\F[t_0]}=e^{i\theta_n(t)}\ket{n_\F[t]}$, which gives $\dot\theta_n=i\braket{n_\F[t]}{\partial_tn_\F[t]}-\bra{n_\F[t]}\agp_\mAE(t)\ket{n_\F[t]}$; by Eq.~\eqref{app:eq:AE_conditions}, the second term integrates to zero over one period. 

Note that $\mAE[t_0]$ and $\Gamma[t_0]$ commute, as both are diagonal in the Floquet states. 
Comparing the eigenvalues of Eqs.~\eqref{app:eq:floquet_stroboscopic} and~\eqref{app:eq:HK_stroboscopic}, we arrive at the geometric decomposition of the quasienergy~\cite{fainshtein1978some,grifoni1998contemporary,le2022missing,schindler2025geometric}
\begin{equation}
  \label{app:eq:eps_ae_gamma}
  \epsilon_{\F,n} = \Bigl(\mae_n - \frac{\gamma_n}{T}\Bigr) \mod \omega .
\end{equation}
Unlike the micromotion operator, $P(t_0+T,t_0)=\identity$, the one-period Wilson line $W(t_0+T,t_0)$ generally differs from the identity operator. 
Hence, the average-energy decomposition~\eqref{app:eq:AE_evolution} separates the evolution into a dynamical contribution generated by $\mAE[t]$, which plays the role of energy in a static system, and a geometric contribution accumulated by $W$, which encodes the inherently nonequilibrium content of the drive~\cite{schindler2025geometric}.
 
\subsubsection{Degenerate average-energy manifolds.}
 
Several Floquet states can share the same average energy while having different quasienergies, as for the symmetry-broken doublet of the GS-DTC in Sec.~\ref{sec:MFIM}. 

Let $\Pi_0[t]$ denote the projector onto such an eigenspace of $\mAE[t]$ with eigenvalue $\mae_0$, and let $r=\mathrm{rank}\,\Pi_0$ denote its dimension. 
Since $\mAE[t_0]$ acts as $\mae_0$ on $\Pi_0[t_0]$, Eq.~\eqref{app:eq:AE_evolution} gives $\Pi_0[t]=W(t,t_0)\Pi_0[t_0]W^\dagger(t,t_0)$: the Wilson line transports the entire Floquet eigenspace. 
The one-period Wilson line commutes with $\Pi_0[t_0]$, and its restriction
\begin{equation}
  \label{app:eq:holonomy}
  \begin{aligned}
  W_0[t_0] &= \Pi_0[t_0]\,W(t_0+T,t_0)\,\Pi_0[t_0]
  \\
  &= e^{iT\mae_0}\,\Pi_0[t_0]\,\UF[t_0]\,\Pi_0[t_0]
  \end{aligned}
\end{equation}
is a unitary operator on the $r$-dimensional iamge of $\Pi_0[t_0]$, analogous to the Wilczek--Zee holonomy of degenerate adiabatic evolution~\cite{wilczek1984appearance}. 

In a basis of the eigenspace which does not consist of Floquet states, e.g., symmetry-broken states, $W_0[t_0]$ is represented by a general $r\times r$ unitary matrix, whose eigenvalues $e^{i\gamma_n}$ are basis-independent and coincide with the geometric phases~\eqref{app:eq:berry_phase} of the Floquet states in the eigenspace. 
By Eq.~\eqref{app:eq:eps_ae_gamma}, the quasienergy splittings inside the eigenspace are then set by the geometric phases alone, $\epsilon_{\F,n}-\epsilon_{\F,m}=-(\gamma_n-\gamma_m)/T\mod\omega$. 
For a doublet with $\Delta\gamma=\gamma_1-\gamma_0=\pi$, the splitting equals $\omega/2$; this is precisely the $\pi$-pairing of the GS-DTC, which is invisible in its degenerate average-energy spectrum. 

The second line of Eq.~\eqref{app:eq:holonomy} also provides a direct numerical procedure: we evolve a basis of $\Pi_0[t_0]$ over one period, compute the overlap matrix with the initial basis, and multiply by $e^{iT\mae_0}$, as done for Fig.~\ref{fig:Ising_GS-DTC}(f).

\subsection{Geometric vs.~conventional Floquet theory}
\label{app:sec:comparison}
 
\begin{table*}[t]
  \label{app:tab:decompositions}
  \renewcommand{\arraystretch}{1.5}
  \begin{ruledtabular}
  \begin{tabular}{lccc}
    & Floquet & Kato & average energy \\
    \hline
    defining decomposition & $H(t) = \HF[t] + A_\F(t)$ & $H(t) = \HK(t) + \AK(t)$ & $H(t) = \mAE[t] + A_\mAE(t)$ \\
    CD Hamiltonian & $H(t) $ & $H(t)$ & $H(t)$ \\
    control Hamiltonian & $\HF[t]$ & $\HK(t)$ & $\mAE[t]$ \\
    gauge potential & $\AF(t)$ & $\AK(t)$ & $\agp_\mAE(t)$ \\
    generalized force & $F_\F(t)\equiv 0$ & $F_\K(t)=\sum_n\dot E_{\K,n}(t)\Pi_n[t]$ & $F_\mAE(t)\equiv 0$ \\
    stationary states/eigenstates & $\Pi_n[t]=\ketbra{n_\F[t]}$ & $\Pi_n[t]=\ketbra{n_\F[t]}$ & $\Pi_n[t]=\ketbra{n_\F[t]}$ \\
    eigenvalues & $\epsilon_{\F,n}$ & $E_{\K,n}(t)$ & $\mae_n$ \\
    dynamical invariant & yes & no & yes \\
    unique & up to $\omega\ell_n$ & yes & yes \\
    gauge condition 
      & $P(t_0{+}T,t_0)=\identity$ 
      & $\bra{n_\F[t]}\AK(t)\ket{n_\F[t]}=0\ \ \forall t$ 
      & $\int_0^T\!\dd t\,\bra{n_\F[t]}\agp_\mAE(t)\ket{n_\F[t]}=0$ \\
    gauge potential holonomy 
      & \makecell{$P(t_0{+}T,t_0) = $ \\ $\mathcal{T}\exp\bigl(-i\int_{t_0}^{t_0+T}\!\dd t\,\AF(t)\bigr){=}\identity$}
      & \makecell{$W(t_0+T,t_0)=$ \\ 
      $\mathcal{T}\exp\bigl(-i\int_{t_0}^{t_0+T}\!\dd t\,\AK(t)\bigr)$}
      & \makecell{$W(t_0+T,t_0)=$ \\
      $\mathcal{T}\exp\bigl(-i\int_{t_0}^{t_0+T}\!\dd t\,\agp_\mAE(t)\bigr)$} \\
  \end{tabular}
  \end{ruledtabular}
  \caption{Floquet, Kato, and average-energy decompositions, $H(t)=\Hctrl(t)+\agp(t)$, with control Hamiltonian $\Hctrl(t)=\sum_n E_n(t)\Pi_n[t]$ diagonal in the Floquet projectors. The generalized force $F_\ctrl$ is defined in Eq.~\eqref{app:eq:gen_force}, where the subscript "$\text{ctr}$" stands for one of F, K, \AE. The last row lists the holonomy acquired by the Floquet states under the one-period evolution generated by the respective gauge potential alone.
  We use the average energy decomposition [rightmost column] to define the the Floquet ground state and its variational principle.
  }
\end{table*}
 
Table~\ref{app:tab:decompositions} summarizes the three decompositions of the lab-frame Hamiltonian $H(t)$ discussed in Sec.~\ref{app:sec:gft}. 
All three share the eigenprojectors $\Pi_n[t]$, and differ only in how the diagonal part of $H(t)$ w.r.t.~the Floquet states is split between the control Hamiltonian and the AGP.
 
The Floquet decomposition~\eqref{app:eq:floquet_decomposition} assigns no geometric phase to the AGP: the micromotion $P(t,t_0)$ returns to the identity operator after one period, and the stroboscopic evolution is generated by $\HF[t_0]$ alone. 
The price to pay is the folding ambiguity~\eqref{app:eq:floquet_gauge}, which precludes a natural ordering of the Floquet states; in large systems, the $D$ quasienergies fill the Floquet zone of width $\omega$ densely, cf.~Fig.~\ref{fig:GFT_ED}(b), and no quasienergy gap above a distinguished state can be identified.
 
The Kato decomposition~\eqref{app:eq:Kato_decomp} fixes the gauge at every instant of time and is free of folding, but its eigenvalues $E_{\K,n}(t)$ change in time. 
It therefore provides an ordering of the Floquet states at each time $t$, which may change during the drive period. 
Its gauge potential accumulates the same phase factors $e^{i\gamma_n}$ over one period as the average-energy gauge potential, see the last row of Table~\ref{app:tab:decompositions}.
 
Finally, the average-energy decomposition~\eqref{app:eq:AE_decomposition} fixes the gauge on average over one period. 
Its eigenvalues are time-independent, so that the ordering of the Floquet states is the same at all times, and the notions of a ground state and of an average-energy gap are well defined. 
Compared with the Floquet decomposition, the information contained in the quasienergy is redistributed according to Eq.~\eqref{app:eq:eps_ae_gamma}: the average energy $\mae_n$ is unambiguously sorted (and not folded), whereas the folding ambiguity is carried entirely by the geometric phase $\gamma_n$, defined modulo $2\pi$. 
In the infinite-frequency limit, $\gamma_n\to0$, and the average energies reduce to the eigenvalues of the period-averaged Hamiltonian $H^{(0)}$.
 
For a degenerate average-energy eigenspace, the phase factors $e^{i\gamma_n}$ are the eigenvalues of the holonomy~\eqref{app:eq:holonomy} operators shown in the last row of Table~\ref{app:tab:decompositions}. 
The Floquet decomposition encodes the same information in the quasienergy splittings within the eigenspace, whereas the average-energy decomposition separates it from the dynamical phase $\mae_0T$ common to all states in the eigenspace. 
For instance, both the F-AFM and the GS-DTC of Sec.~\ref{sec:MFIM} feature a twofold degenerate average-energy ground-state manifold; they differ in their holonomy, which is trivial for the F-AFM and has $\Delta\gamma=\pi$ for the GS-DTC.

\subsection{Operator expressions for the average energy}
\label{app:sec:AE_operator}
 
The definition $\mAE[t]=\sum_n\mae_n\Pi_n[t]$ requires knowledge of the Floquet states $\Pi_n[t]$, which are not accessible for large nonintegrable systems. 
We now derive expressions that involve only the lab-frame Hamiltonian $H(t)$.
 
\subsubsection{Generalized-force equation}
\label{app:sec:AE_force}
 
Consider any decomposition $H(t)=\Hctrl(t)+\agp(t)$ in which the control Hamiltonian is diagonal in the Floquet projectors, $\Hctrl(t)=\sum_n E_n(t)\Pi_n[t]$; the Floquet, Kato, and average-energy decompositions are all of this form. 
Since the Floquet projectors evolve following $i\partial_t\Pi_n[t]=\comm{H(t)}{\Pi_n[t]}=\comm{\agp(t)}{\Pi_n[t]}$, we have $i\sum_n E_n\partial_t\Pi_n=\comm{\agp}{\Hctrl}$, and therefore
\begin{equation}
  \label{app:eq:gen_force}
  \begin{aligned}
  \comm{\Hctrl(t)}{\agp(t)} + i\partial_t\Hctrl(t) &= iF_\ctrl(t),
  \\
  F_\ctrl(t) &= \sum_n \dot{E}_n(t)\,\Pi_n[t].
  \end{aligned}
\end{equation}
Equation~\eqref{app:eq:gen_force} is the Floquet counterpart~\cite{schindler2024counterdiabatic} of the defining equation of the AGP in static systems~\cite{Kolodrubetz2017Geometry,Sels2017Minimizing}, where the generalized force $F_\ctrl$ collects the time derivatives of the eigenvalues of the control Hamiltonian. 
Eliminating $\Hctrl=H-\agp$ from Eq.~\eqref{app:eq:gen_force} gives
\begin{equation}
  \label{app:eq:AGP_force_eq}
  \mathcal{L}(\agp) = -i\partial_t H + iF_\ctrl .
\end{equation}

In the Kato gauge, $F_\K=\sum_n\dot E_{\K,n}\Pi_n$ is generically nonzero. 
By contrast, in the Floquet and in the average-energy gauge the control Hamiltonian is a dynamical invariant; hence, the generalized force vanishes, and the AGP solves the simpler linear equation
\begin{equation}
  \label{app:eq:AGP_linear}
  \mathcal{L}(\agp) = -i\partial_t H = \mathcal{L}(H),
\end{equation}
whose right-hand side is the known derivative of the Hamiltonian, $\partial_t H(t)$. In the last step above we used $\comm{H}{H}=0$ and the definition of $\mathcal{L}$.
 
Howwever, equation~\eqref{app:eq:AGP_linear} alone does not determine $\agp$ uniquely: if $\agp$ is a solution, so is $\agp-I$ for any dynamical invariant $I\in\ker\mathcal{L}$, and each solution defines a invariant $H-\agp$. 
For instance, in the Floquet gauge $\AF=\agp_\mAE-I$ with $I=\sum_n(\epsilon_{\F,n}-\mae_n)\Pi_n$, cf.~Eq.~\eqref{app:eq:eps_ae_gamma}. 
The kernel of $\mathcal{L}$ is thus the basis-independent form of the gauge freedom~\eqref{app:eq:floquet_gauge}, enlarged from integer to real shifts of the diagonal; indeed, it is the periodicity of the micromotion, $P(t_0+T,t_0)=\identity$, which restricts the Floquet gauge to integer shifts. 
Hence, an additional condition is required to select a unique solution; for the average-energy gauge this condition is linear, as we show in Sec.~\ref{app:sec:AE_projector}.
 
\subsubsection{Vectorized form}
\label{app:sec:AE_vectorized}
 
In Sambe superspace, Eq.~\eqref{app:eq:AGP_linear} becomes a linear system for the vector $\floquetket{\agp}$. 
Since $\mathcal{C}_H\floquetket{H}=0$, we have $\mathcal{L}\floquetket{H}=\omega\mathcal{N}\floquetket{H}$, and Eq.~\eqref{app:eq:liouvillian_vec} gives
\begin{equation}
  \label{app:eq:AGP_vectorized}
  \bigl(\mathcal{C}_H+\omega\mathcal{N}\bigr)\floquetket{\agp}
  = \omega\mathcal{N}\floquetket{H}
  = \floquetket{-i\partial_t H}.
\end{equation}
In components, Eq.~\eqref{app:eq:AGP_vectorized} reads as
\begin{equation}
    \sum_m\mathcal{C}_H^{(n-m)}\floquetket{\agp^{(m)}}+n\omega\floquetket{\agp^{(n)}}=n\omega\floquetket{H^{(n)}},
\end{equation}
where the right-hand side contains the Fourier harmonics of $\omega\mathcal{N}\floquetket{H}$ and, in particular, has no zero-harmonic component. In Appendices~\ref{app:sec:varl} and~\ref{app:sec:mps_algo}, we denote this right-hand side by $\floquetket{\partial H}$, Eq.~\eqref{app:eq:dH_def}. 
Imposing a finite Fourier cutoff $|n|\leq n_c$, Eq.~\eqref{app:eq:AGP_vectorized} constitutes a linear system of $(2n_c+1)D^2$ linear equations. Appendix~\ref{app:sec:varl} formulates variational principles for it; in Appendix~\ref{app:sec:mps_algo} we discuss how to solve it using a matrix product state representation of $\floquetket{\agp}$.
 
\subsubsection{Projector form and uniqueness}
\label{app:sec:AE_projector}

Since $\mathcal{L}$ is self-adjoint and has closed range (Sec.~\ref{app:sec:varl:superspace}), its pseudo-inverse commutes with it, and
\begin{equation}
  \label{app:eq:L_projectors}
  \mathcal{L}^+\mathcal{L} = \mathcal{L}\mathcal{L}^+ = \mathcal{P}_{\im\mathcal{L}},
  \qquad
  \identity-\mathcal{L}^+\mathcal{L} = \mathcal{P}_{\ker\mathcal{L}},
\end{equation}
are the orthogonal projectors onto the image and the kernel of $\mathcal{L}$, respectively, with $\hilbert{S}_S=\ker\mathcal{L}\oplus\im\mathcal{L}$. 
These are the properties of $\mathcal{L}$ used in Eq.~\eqref{eq:AE_ker_proj} of the main text and in Appendix~\ref{app:sec:varl}. 

The pseudo-inverse also encodes a minimal-norm property which underlies the variational principle of Appendix~\ref{app:sec:varl}: for a consistent linear equation $\mathcal{L}X=b$ with $b\in\im\mathcal{L}$, the general solution reads $X=\mathcal{L}^+b+(\identity-\mathcal{L}^+\mathcal{L})Z$ with arbitrary $Z\in\hilbert{S}_S$, and since the two terms are orthogonal, $\norm{X}^2=\norm{\mathcal{L}^+b}^2+\norm{\mathcal{P}_{\ker\mathcal{L}}Z}^2$, the pseudo-inverse solution $\mathcal{L}^+b$ is the unique solution of minimal norm, i.e., the unique solution in $\im\mathcal{L}$~\cite{groetsch1977generalized}. When applied to the gauge-potential equation~\eqref{app:eq:AGP_linear}, whose right-hand side $\mathcal{L}(H)$ lies in $\im\mathcal{L}$, this identifies $\agp_\mAE=\mathcal{L}^+\mathcal{L}(H)$ as the minimal-norm solution among all gauge potentials $\agp_\mAE-I$, $I\in\ker\mathcal{L}$, of Sec.~\ref{app:sec:AE_force}. 

The average-energy decomposition~\eqref{app:eq:AE_decomposition} is the corresponding orthogonal decomposition of $H(t)$:
\begin{equation}
  \label{app:eq:orthogonal_decomp}
  \begin{aligned}
    \mAE[t] &= \bigl(\identity - \mathcal{L}^+\mathcal{L}\bigr)(H(t)) &&\in \ker\mathcal{L},
    \\
    \agp_\mAE(t) &= \mathcal{L}^+\mathcal{L}(H(t)) &&\in \im\mathcal{L},
  \end{aligned}
\end{equation}
where, in vectorized form, $\floquetket{\agp_\mAE}=\omega\mathcal{L}^+\mathcal{N}\floquetket{H}$, cf.~Eq.~\eqref{app:eq:AGP_vectorized}. 

To show that Eq.~\eqref{app:eq:orthogonal_decomp} reproduces $\mAE[t]=\sum_n\mae_n\Pi_n[t]$ and the conditions~\eqref{app:eq:AE_conditions}, we first characterize the kernel of $\mathcal{L}$. 
The condition $\mathcal{L}(I)=0$ is equivalent to $I(t)=U(t,t_0)I(t_0)U^\dagger(t,t_0)$, and periodicity, $I(t_0+T)=I(t_0)$, requires $\comm{\UF[t_0]}{I(t_0)}=0$; hence,
\begin{equation}
  \label{app:eq:kernel}
  \ker\mathcal{L} = \bigl\{\,U(t,t_0)\,X\,U^\dagger(t,t_0) \;:\; \comm{X}{\UF[t_0]}=0\,\bigr\}
\end{equation}
consists of all operators $X$ that commute with the FLoquet unitary $U_\F[t_0]$.
If the quasienergies are nondegenerate modulo $\omega$, the commutant of $\UF[t_0]$ is spanned by its eigenprojectors, and $\ker\mathcal{L}=\mathrm{span}\{\Pi_n[t]\}_n$ with orthonormal basis elements, $\floquetbraket{\Pi_m}{\Pi_n}=\delta_{mn}$. 
Quasienergy degeneracies modulo $\omega$ add off-diagonal invariants to the kernel; however, a degeneracy in the average energy alone does not; for instance, the quasienergies of the GS-DTC doublet differ by $\omega/2$ whereas the average energies are degenerate.
 
Using this characterization, the proof proceeds in four steps. 
Note first that, by Eq.~\eqref{app:eq:L_projectors}, $(\identity-\mathcal{L}^+\mathcal{L})(H)=\mathcal{P}_{\ker\mathcal{L}}(H)$ lies in $\ker\mathcal{L}$, while $\mathcal{L}^+\mathcal{L}(H)=H-\mathcal{P}_{\ker\mathcal{L}}(H)$ solves Eq.~\eqref{app:eq:AGP_linear}. 
Second, for quasienergies that are nondegenerate modulo $\omega$, the orthonormal basis $\{\Pi_n[t]\}$ of the kernel gives
\begin{equation}
  \label{app:eq:AE_projection}
  \mathcal{P}_{\ker\mathcal{L}}(H) = \sum_n \floquetbraket{\Pi_n}{H}\,\Pi_n[t] = \sum_n \mae_n\,\Pi_n[t],
\end{equation}
since $\floquetbraket{\Pi_n}{H}=T^{-1}\int_0^T\dd t\,\bra{n_\F[t]}H(t)\ket{n_\F[t]}=\mae_n$ by Eq.~\eqref{app:eq:average-energies}. 
Third, the averaged parallel-transport condition in Eq.~\eqref{app:eq:AE_conditions} reads $\floquetbraket{\Pi_n}{\agp_\mAE}=0$ for all $n$, i.e., $\agp_\mAE\perp\ker\mathcal{L}$, which holds since $\agp_\mAE\in\im\mathcal{L}=(\ker\mathcal{L})^\perp$. 
And finally, fourth, suppose $H=\mAE'+\agp'$ with $\mAE'\in\ker\mathcal{L}$ and $\agp'$ satisfying the averaged parallel-transport condition, i.e., $\agp'\perp\ker\mathcal{L}$; then $\mAE'-\mAE=\agp_\mAE-\agp'$ lies in both $\ker\mathcal{L}$ and $(\ker\mathcal{L})^\perp$, and therefore vanishes. 
Equation~\eqref{app:eq:orthogonal_decomp} thus defines $\mAE$ and $\agp_\mAE$ uniquely as solutions to linear equations in Sambe superspace, without reference to the Floquet states. 

If some quasienergies are degenerate modulo $\omega$, Eq.~\eqref{app:eq:orthogonal_decomp} remains the definition, and the conditions~\eqref{app:eq:AE_conditions} are replaced by $\agp_\mAE\perp\ker\mathcal{L}$, i.e., $\floquetbraket{I}{\agp_\mAE}=0$ for all dynamical invariants $I$ in the kernel~\eqref{app:eq:kernel}, including the off-diagonal ones.
 
\subsubsection{The average-energy gauge}
\label{app:sec:AE_choice}
 
In principle, each of the three decompositions in Table~\ref{app:tab:decompositions} provides access to the Floquet states; let us therefore explain why we formulate the variational principle in the average-energy gauge.
 
First, $\mAE$ is uniquely defined and its spectrum is time-independent and not folded; the average energies thus endow the Floquet states with a physically meaningful ordering~\cite{schindler2025geometric}, which defines the Floquet ground state and the average-energy gap in the main text. 
By contrast, the Floquet Hamiltonian $\HF[t]$ lacks such an ordering due to the folding ambiguity. 
Similarly, the Kato Hamiltonian $\HK(t)$, although free of folding, is not suitable either: because the Kato energies $E_{\K,n}(t)$ depend on time, the lowest-energy state of $\HK(t)$ may change abruptly with $t$ whenever two Kato energies cross; hence, the lowest-average Kato eigenstate need not correspond to the same Floquet state throughout the drive period.
 
Second, the condition that selects the average-energy gauge, $\mAE\in\ker\mathcal{L}$ together with $\agp_\mAE\perp\ker\mathcal{L}$, is linear: Eq.~\eqref{app:eq:orthogonal_decomp} is the minimal-norm solution of the linear equation~\eqref{app:eq:AGP_linear}. 
This form is compatible with any given variational family, since restricting the AGP to an ansatz turns Eq.~\eqref{app:eq:AGP_linear} into a least-squares problem, see Appendix~\ref{app:sec:varl}. 
The Floquet gauge, in contrast, is selected by the stroboscopic condition $P(t_0+T,t_0)=\mathcal{T}\exp\bigl(-i\int_{t_0}^{t_0+T}\dd t\,\agp(t)\bigr)=\identity$, which is nonlinear in the gauge potential $\agp=\agp_\F$. 
Should an alternative gauge condition be cast in a form compatible with a given variational family, the subsequent step -- the ground-state search in Appendix~\ref{app:sec:mps_algo} and the Krylov refinement in Appendix~\ref{app:sec:krylov} -- apply without modification.


\section{Variational principles for the average-energy gauge potential}
\label{app:sec:varl}

Recall from Appendix~\ref{app:sec:AE_operator} that the average-energy decomposition $H=\mAE+\agp_\mAE$ admits two equivalent characterizations. 
On the one hand, $\mAE=(\identity-\mathcal{L}^+\mathcal{L})H$ and $\agp_\mAE=\mathcal{L}^+\mathcal{L}H$ are the orthogonal projections of the lab-frame Hamiltonian onto the kernel and the image of the Floquet Liouvillian, Eq.~\eqref{app:eq:orthogonal_decomp}, see Fig.~\ref{fig:GFT_DMRG}(a). 
On the other hand, $\agp_\mAE$ is the unique psuedoinverse solution of the linear equation~\eqref{app:eq:AGP_linear} that lies in $\im\mathcal{L}$. 

Throughout this and the following appendix, we denote the right-hand side of Eq.~\eqref{app:eq:AGP_linear} by
\begin{equation}
  \label{app:eq:dH_def}
  \partial H = -i\partial_t H(t) = \mathcal{L}(H),
  \qquad
  \floquetket{\partial H} = \omega\mathcal{N}\floquetket{H},
\end{equation}
whose Fourier harmonics are $(\partial H)^{(n)}=n\omega H^{(n)}$, see Eq.~\eqref{app:eq:AGP_vectorized}; note that the factor $-i$ is absorbed into the definition of $\partial H$, which is therefore anti-hermitian. 
Since $H(t)$ is known, knowing either $\mAE$ or $\agp_\mAE$ is sufficient; for the reasons laid out in Sec.~\ref{app:sec:AE_choice} above, we target the gauge potential $\agp_\mAE$.

Notice that none of the two characterizations can be computed directly for a many-body system, because the pseudo-inverse $\mathcal{L}^+$ of $\mathcal{L}$ acts on a space of dimension $(2n_c+1)D^2$, with $D=\dim\hilbert{H}$ growing exponentially with the system size $L$. 
We therefore seek to formulate a variational ansatz (e.g., a Floquet matrix product state of fixed bond dimension, see Appendix~\ref{app:sec:mps_algo}), and derive the target equation that the restricted variational approximant should satisfy. 

In this appendix, we give an overview of the variational formulations we considered. We explain the core problem of fixing the gauge of the AGP within the restricted variational family. 
We emphasize that all numerical data presented in this work use only the \emph{normal equation} $\mathcal{L}^2X=\partial H$ introduced in Sec.~\ref{app:sec:varl:formulations}(v) and implemented using matrix product states in Appendix~\ref{app:sec:mps_algo}; the remaining formulations serve to put this choice into context, and to document some unsuccessful approaches. 
Throughout, $X$ denotes the variational unknown of the formulation at hand, and $X_\ast$ its solution.

\subsection{Target equations and the gauge problem}
\label{app:sec:varl:gauge}

The average-energy gauge potential $\agp_\mAE$ is singled out by the two conditions:
\begin{equation}
  \label{app:eq:varl_target}
  \mathcal{L}X = \partial H
  \qquad\text{and}\qquad
  X\perp\ker\mathcal{L},
\end{equation}
cf.~Secs.~\ref{app:sec:AE_force} and~\ref{app:sec:AE_projector}. 
Equivalently, $\agp_\mAE$ is the \emph{minimal-norm} solution of the above linear equation,
\begin{equation}
  \label{app:eq:AGP_minnorm}
  \agp_\mAE = \underset{X:\;\mathcal{L}X=\partial H}{\mathrm{argmin}}\;\norm{X}^2 ,
\end{equation}
since any other solution $X=\agp_\mAE-I$ with $I\in\ker\mathcal{L}$ some dynamical invariant has $\norm{X}^2=\norm{\agp_\mAE}^2+\norm{I}^2$ by orthogonality.

Observe that the two conditions in Eq.~\eqref{app:eq:varl_target} are of different nature. 
The first is a linear equation whose data, $\mathcal{L}$ and $\partial H$, are known a priori. 
The second refers to the space of dynamical invariants, $\ker\mathcal{L}$, which is not known: by Eq.~\eqref{app:eq:kernel}, knowing $\ker\mathcal{L}$ amounts to knowing the Floquet projectors, i.e., to having solved the Floquet problem. 
Therefore, within a restricted variational family the constraint $X\perp\ker\mathcal{L}$ cannot be imposed directly; moreover, a solver that controls only the residual of $\mathcal{L}X=\partial H$ is blind to the kernel since adding any dynamical invariant to $X$ leaves the residual unchanged. 
This is the essential issue that has to be overcome to design are working variational principle. 

To understand it better, observe that every solution $X$ of the linear equation defines a dynamical invariant $H-X=\mAE+I$, which shares its eigenvectors with $\mAE$ but whose eigenvalues are shifted by the (arbitrary, real) eigenvalues of $I$; the Floquet Hamiltonian $\HF$ and the average energy operator $\mAE$ are two concrete examples, cf.~Sec.~\ref{app:sec:AE_force}. 
Shifting $X$ by a kernel component thus leaves the Floquet eigenvectors unchanged; however, it destroys the ordering of the Floquet states by their average energy, and with it the notion of the Floquet ground state. 
We therefore refer to the task of selecting the solution with $X\perp\ker\mathcal{L}$ as the \emph{gauge problem}, and to $\agp_\mAE$ as the minimal-norm gauge. 

Note that global Krylov methods, such as GMRES or MINRES~\cite{saad1986gmres,paige1975minres}, initialized from a vector $X_0\in\im\mathcal{L}$, solve the gauge problem automatically, since the initial residual $\partial H-\mathcal{L}X_0$ and all Krylov vectors generated from it by applying $\mathcal{L}$ lie in $\im\mathcal{L}$. The difficulty arises once the unknown is restricted to a nonlinear variational family, such as matrix product states,\footnote{An MPS $\{M_1,\dots,M_L\}\mapsto\sum_{\boldsymbol\sigma}M_1^{\sigma_1}\cdots M_L^{\sigma_L}|\boldsymbol\sigma\rangle$ is multilinear in the tensors $M^\sigma_j(\theta)$ and represents a polynomial of degree $L$ in the variational parameters $\theta$.} whose local updates need not preserve the image of the Liouvillian.

\subsection{Formulations}
\label{app:sec:varl:formulations}

Let us now list the variational formulations of Eq.~\eqref{app:eq:varl_target} we considered, in increasing order of suitability for a restricted ansatz; Table~\ref{app:tab:formulations} summarizes their properties and indicates which of them we explored numerically.

\paragraph*{(i) Image-constrained least squares.}
The most direct formulation minimizes the residual of the linear equation subject to the image constraint,
\begin{equation}
  \label{app:eq:AGP_variational}
  \agp_\mAE = \underset{X\in\im\mathcal{L}}{\mathrm{argmin}}\,\norm{\mathcal{L}X-\partial H}^2 .
\end{equation}
This is the formulation solved exactly by global Krylov methods. 
For a restricted family it serves as a reference only, because to the best of our knowledge the constraint $X\in\im\mathcal{L}$ cannot be embedded in the variational ansatz.

\paragraph*{(ii) Constrained minimum norm.}
Alternatively, one can enforce the minimal norm explicitly, cf.~Eq.~\eqref{app:eq:AGP_minnorm},
\begin{equation}
  \label{app:eq:AGP_constrained}
  \min_X \norm{X}^2 \quad \text{subject to} \quad \mathcal{L}X - \partial H = 0 .
\end{equation}
Introducing a time-periodic Lagrange multiplier $\lambda$, the stationarity conditions are $X=\mathcal{L}\lambda$ and $\mathcal{L}X=\partial H$, or, as a single linear system,
\begin{equation}
  \label{app:eq:AGP_lagrange}
  \begin{pmatrix}
    -\identity     & \mathcal{L}\\
    \mathcal{L}    & 0
  \end{pmatrix}
  \begin{pmatrix}
    X \\
    \lambda
  \end{pmatrix}
  =
  \begin{pmatrix}
    0\\
    \partial H
  \end{pmatrix}.
\end{equation}
The first row expresses the image constraint, $X=\mathcal{L}\lambda\in\im\mathcal{L}$, through the multiplier, which plays the role of a pre-image of $X$. 
The price to pay is a doubling of the unknowns and an indefinite (saddle-point) system; indeed, Eq.~\eqref{app:eq:AGP_lagrange} does not derive from a cost function to be minimized. Within a restricted family, both rows hold only approximately.

A cost function can be constructed by solving Eq.~\eqref{app:eq:AGP_lagrange} as a least-squares problem, i.e., by minimizing
\begin{equation}
  \label{app:eq:AGP_lagrange_normal}
  \norm{X-\mathcal{L}\lambda}^2+\norm{\mathcal{L}X-\partial H}^2
\end{equation}
jointly over $(X,\lambda)$; this variant, which we label (ii$'$) in Table~\ref{app:tab:formulations}, retains the direct control over the residual of the linear equation and enforces the image constraint through the multiplier, at the price of normal equations which again involve $\mathcal{L}^2$. 
We comment on our experience with both variants in Sec.~\ref{app:sec:mps_algo:alternatives}.

\paragraph*{(iii) Tikhonov regularization.}
The hard constraint in Eq.~\eqref{app:eq:AGP_constrained} can be replaced by a soft quadratic penalty,
\begin{equation}
  \label{app:eq:AGP_tikhonov}
  X_\mu = \underset{X}{\mathrm{argmin}}\bqty{\norm{X}^2 + \mu \norm{\mathcal{L}X - \partial H}^2},
\end{equation}
with $\agp_\mAE=\lim_{\mu\to\infty}X_\mu$. 
The functional is unconstrained and strictly convex for any $\mu>0$, with normal equation
\begin{equation}
  \label{app:eq:AGP_tikhonov_normal}
  \pqty{\identity + \mu\,\mathcal{L}^2} X_\mu = \mu\,\mathcal{L}\,\partial H ,
\end{equation}
which interpolates between $X_0=0$ and the Moore--Penrose solution. 
Equation~\eqref{app:eq:AGP_tikhonov} is particularly amenable to gradient-based optimization of nonlinear ans\"atze $X=X(\theta)$; we have not employed it in the present work.

\paragraph*{(iv) Petrov--Galerkin projection.}
One can avoid $\mathcal{L}^2$ altogether by requiring the residual of the linear equation to be orthogonal to all variations $\delta X$ within the family,
\begin{equation}
  \label{app:eq:AGP_galerkin}
  \floquetbraket{\delta X}{\mathcal{L}X-\partial H} = 0 \qquad \forall\,\delta X .
\end{equation}
For a linear family this is a Galerkin projection of Eq.~\eqref{app:eq:AGP_linear}; for the Krylov family it reproduces the global Krylov methods. 
Equation~\eqref{app:eq:AGP_galerkin} is the stationarity condition of the cost function $\floquetbra{X}\mathcal{L}\floquetket{X}-2\,\mathrm{Re}\floquetbraket{X}{\partial H}$, which has a saddle point rather than a minimum at the solution because $\mathcal{L}$ is indefinite. Moreover, the condition $X\in\im\mathcal{L}$ is not enforced by Eq.~\eqref{app:eq:AGP_galerkin}.

\paragraph*{\textbf{(v) Normal equation.}}
The formulation we adopt in this work makes the pre-image of the Lagrange-multiplier formulation the primary unknown. 
Observe that the two steps $X=\mathcal{L}\lambda$ and $\mathcal{L}X=\partial H$ of Eq.~\eqref{app:eq:AGP_lagrange} can be combined into a single equation for the pre-image, which we rename $X$: 
\begin{equation}
  \label{app:eq:normal_eq}
  \mathcal{L}^2 X = \partial H = \mathcal{L}H
  \quad\Longleftrightarrow\quad
  X_\ast = \underset{X}{\mathrm{argmin}}\,\norm{\mathcal{L}X-H}^2 ,
\end{equation}
i.e., $X_\ast$ is a least-squares solution of $\mathcal{L}X=H$, and Eq.~\eqref{app:eq:normal_eq} is its normal equation. 

The gauge potential is then obtained in a second step,
\begin{equation}
  \label{app:eq:Xstar_solution}
  \agp_\mAE = \mathcal{L}X_\ast .
\end{equation}
To see this, note that $\mathcal{L}X_\ast$ lies in $\im\mathcal{L}$ by construction, and satisfies $\mathcal{L}(\mathcal{L}X_\ast)=\mathcal{L}^2X_\ast=\partial H$; by the uniqueness established in Sec.~\ref{app:sec:AE_projector}, $\mathcal{L}X_\ast=\agp_\mAE$. 
In terms of the pseudo-inverse, Eq.~\eqref{app:eq:Xstar_solution} gives $\agp_\mAE=\mathcal{L}\mathcal{L}^+H=\mathcal{L}^+\mathcal{L}H$, where we used that $\mathcal{L}$ and $\mathcal{L}^+$ commute, Eq.~\eqref{app:eq:L_projectors}. 

The following three properties make the normal equation formulation suitable for a restricted family. 
First, the gauge problem is solved exactly: whatever kernel component $X_\ast$ may acquire\footnote{The solution of Eq.~\eqref{app:eq:normal_eq} is unique only modulo $\ker\mathcal{L}$.} is annihilated by the final application of $\mathcal{L}$, cf.~Fig.~\ref{fig:local_vs_global-problem}(a). 
Second, Eq.~\eqref{app:eq:normal_eq} can be derived from a cost function, $\norm{\mathcal{L}X-H}^2$, whose operator $\mathcal{L}^2$ is Hermitian and positive semidefinite; hence, local (DMRG-type) minimization is well founded. 
Last, the right-hand side $\partial H$ lies in $\im\mathcal{L}^2=\im\mathcal{L}$, which ensures that the global problem is consistent. 

The price to pay is the conditioning of the linear operator: the nonzero eigenvalues of $\mathcal{L}$ are differences of quasienergies modulo $\omega$, whose smallest value for generic integrable systems is bounded by the quasienergy spacing $\omega/D$, and decays exponentially. In the normal-equation formulation, $\mathcal{L}^2$ squares the resulting condition number. 
Note also that the gauge potential $\mathcal{L}X_\ast$ obtained from a restricted family does \emph{not} minimize $\norm{\mathcal{L}\agp-\partial H}$ over that family; instead, the quantity controlled by the variational principle is the residual of the normal equation, see Sec.~\ref{app:sec:varl:diagnostics}.

To see the latter, observe that since $A=\mathcal{L}X\in\im\mathcal{L}$ is orthogonal to $\mAE\in\ker\mathcal{L}$, the cost function decomposes as
\begin{equation}
  \label{app:eq:normal_eq_cost}
  \norm{\mathcal{L}X-H}^2 = \norm{\mAE}^2 + \norm{\mathcal{L}X-\agp_\mAE}^2
  = \norm{\mAE}^2 + \norm{\tilde\mAE-\mAE}^2 ,
\end{equation}
with $\tilde\mAE=H-\mathcal{L}X$. Therefore, up to the constant $\norm{\mAE}^2$, the normal equation minimizes the Hilbert-Schmidt distance between the variational and the exact average-energy operators over the family. 
In particular, it does \emph{not} minimize the residual $\norm{\mathcal{L}A-\partial H}=\norm{\mathcal{L}(\tilde\mAE-\mAE)}$ of the linear equation, which weighs the same error by the eigenvalues of $\mathcal{L}$ and therefore emphasizes the far-off-resonant components of $A$. 
Within a restricted variational family, the two minimizers differ; only at the exact solution do they both vanish. 
Since $\norm{\tilde\mAE-\mAE}$ is not computable, we monitor the residual instead, see Sec.~\ref{app:sec:varl:diagnostics}.

We emphasize that the advantages listed above are precisely those which formulations (i)--(iv) lack in a restricted variational family: (i) cannot impose its constraint, (ii) and (iv) lead to saddle-point problems without a cost function, (ii$'$) doubles the unknowns while still involving $\mathcal{L}^2$, and (iii) recovers the gauge only in a limit. 
For these reasons we choose the normal equation (v); its DMRG implementation is presented in Appendix~\ref{app:sec:mps_algo}.

\begin{table*}[t]
  \label{app:tab:formulations}
  \renewcommand{\arraystretch}{1.35}\footnotesize
  \begin{ruledtabular}
  \begin{tabular}{llccccl}
    formulation & equation & unknowns & cost function & $X\in\im\mathcal{L}$ & operator & status \\
    \hline
    (i) image-constrained least squares\hphantom{x} & \eqref{app:eq:AGP_variational} & $X$ & yes & by constraint & $\mathcal{L}$ & reference only \\
    (ii) constrained minimum norm & \eqref{app:eq:AGP_lagrange} & $(X,\lambda)$ & no (saddle point) & $X=\mathcal{L}\lambda$, approximate & $\mathcal{L}$ & explored, Sec.~\ref{app:sec:mps_algo:alternatives} \\
    (ii$'$) constrained minimum norm, least squares & \eqref{app:eq:AGP_lagrange_normal} & $(X,\lambda)$ & yes & $X=\mathcal{L}\lambda$, approximate & $\mathcal{L}^2$ & explored ($L\le5$), Sec.~\ref{app:sec:mps_algo:alternatives} \\
    (iii) Tikhonov regularization & \eqref{app:eq:AGP_tikhonov_normal} & $X$ & yes, strictly convex & for $\mu\to\infty$ & $\identity+\mu\mathcal{L}^2$ & not explored \\
    (iv) Petrov--Galerkin projection & \eqref{app:eq:AGP_galerkin} & $X$ & no (saddle point) & not enforced & $\mathcal{L}$ & explored, Sec.~\ref{app:sec:mps_algo:alternatives} \\
    \hline
    \textbf{(v) normal equation} & \textbf{\eqref{app:eq:normal_eq}} & $\boldsymbol{X}$ \textbf{(pre-image)} & \textbf{yes} & \textbf{exact,} $\boldsymbol{\agp_\mAE=\mathcal{L}X_\ast}$ & $\boldsymbol{\mathcal{L}^2}$ & \textbf{this work, Sec.~\ref{app:sec:dmrg}} \\
  \end{tabular}
  \end{ruledtabular}
  \caption{Variational formulations of the target equations~\eqref{app:eq:varl_target}. The column ``$X\in\im\mathcal{L}$'' states how the gauge problem of Sec.~\ref{app:sec:varl:gauge} is handled; the column ``operator'' lists the operator whose condition number governs the iterative solution. The ``status'' column refers to the DMRG implementations of Appendix~\ref{app:sec:mps_algo}. The last row (bold) is the formulation we used for all results presented in this work.}
\end{table*}

\subsection{Diagnostics: normal-equation residual}
\label{app:sec:varl:diagnostics}

Within the normal-equation formulation, the natural measure of convergence is the relative residual of Eq.~\eqref{app:eq:normal_eq},
\begin{equation}
  \label{app:eq:residual}
  \mathcal{R} = \frac{\norm{\mathcal{L}^2\floquetket{X}-\floquetket{\partial H}}}{\norm{\floquetket{\partial H}}} ,
\end{equation}
evaluated in the inner product~\eqref{eq:HS_inner_product}; this is the quantity shown in Figs.~\ref{fig:XY}(e) and~\ref{app:fig:residual}.

This residual has a direct physical interpretation. 
Using the approximate average-energy operator $\tilde\mAE= H-\mathcal{L}X$ obtained from a variational $X$, we have $\mathcal{L}^2X-\partial H=\mathcal{L}(\mathcal{L}X-H)=-\mathcal{L}(\tilde\mAE)$, and hence
\begin{equation}
  \label{app:eq:residual_invariant}
  \mathcal{R} = \frac{\norm{\mathcal{L}(\tilde\mAE)}}{\norm{\mathcal{L}(H)}} .
\end{equation}
Thus, $\mathcal{R}$ measures the extent to which $\tilde\mAE$ fails to be a dynamical invariant of $H(t)$, Eq.~\eqref{app:eq:IOM}, relative to the lab-frame Hamiltonian $H$ itself. 
Moreover, whereas the normal equation minimizes $\norm{\tilde\mAE-\mAE}$ [see Eq.~\eqref{app:eq:normal_eq_cost}], the residual $\mathcal{R}$ measures $\norm{\mathcal{L}(\tilde\mAE-\mAE)}$.
Since $\mathcal{R}$ is insensitive to the presence of any kernel components in $X$, it does not detect a violation of the gauge condition; by Eq.~\eqref{app:eq:Xstar_solution}, however, kernel components of $X$ do not affect $\tilde\mAE$; therefore, the residual $\mathcal{R}$ controls the quality of the approximation we are after.

By contrast, notice that the variance of $\tilde\mAE$ in its approximate ground state $|\tilde{n}^\F_\text{GS}\rangle$, $\langle \tilde{n}^\F_\text{GS} |\tilde\mAE^2|\tilde{n}^\F_\text{GS}\rangle-\langle \tilde{n}^\F_\text{GS} |\tilde\mAE|\tilde{n}^\F_\text{GS}\rangle^2$, is not a reliable diagnostic of the Floquet ground state. 
A small variance only certifies that $|\tilde{n}^\F_\text{GS}\rangle$ is close to an eigenstate of $\tilde\mAE$, i.e., it tests the ground-state search for the approximate operator, but not the deviation $\tilde\mAE-\mAE$, which is unknown; a small variance can therefore still be compatible with a sizable error in $\tilde\mAE$. 
For this reason, we use $\mathcal{R}$ to assess the operator and assess the state using the stroboscopic infidelity, which involves only the exact dynamics generated by $H$, see Sec.~\ref{subsec:varl}.

\section{Matrix product state algorithm for the average-energy operator}
\label{app:sec:mps_algo}

In this appendix, we describe the tensor-network implementation of the normal-equation formulation of Appendix~\ref{app:sec:varl}, which produces the Floquet ground states reported in the main text. 
Before turning to the technical details, we summarize the conceptual points behind the algorithm.
\begin{enumerate}
    \item Time-periodic operators are represented as Floquet matrix product states (FMPS) in Sambe superspace, see Sec.~\ref{app:sec:varl:superspace}: a matrix product state on $L$ sites of local dimension $d^2$ carrying one open photon index. The Floquet Liouvillian $\mathcal{L}=\mathcal{C}+\omega\mathcal{N}$ acts on such states as a convolution over the photon index; we store only the harmonics $\mathcal{C}^{(n)}$ and $(\mathcal{C}^2)^{(n)}$ as matrix product operators (Sec.~\ref{app:sec:mps_algo:tn}).
    \item The normal equation $\mathcal{L}^2X=\partial H$ is solved by two-site DMRG sweeps~\cite{White1992,Schollwock2011,kim2024variational,mckeever2024towards}, i.e., by minimizing the cost function $\norm{\mathcal{L}X-H}^2$ one bond at a time. The local two-site problems are consistent, i.e., they always admit a solution, provided the effective operator is obtained by reducing $\mathcal{L}^2$ to the bond; reducing $\mathcal{L}$ and squaring the reduced operator yields inconsistent local problems (Sec.~\ref{app:sec:dmrg:consistency} and Fig.~\ref{fig:local_vs_global-problem}).
    \item The gauge problem of Sec.~\ref{app:sec:varl:gauge} reappears at the level of the local updates: local Krylov solutions do not stay in $\im\mathcal{L}$ globally. The two-step procedure $\agp_\mAE=\mathcal{L}X_\ast$ removes this issue, since any kernel component acquired by $X$ during the sweeps is annihilated by the final application of $\mathcal{L}$.
    \item The computational bottleneck is the repeated application of the effective $\mathcal{L}^2$ inside the Krylov solver of each local problem. The contraction order of Sec.~\ref{app:sec:dmrg:cost} and the evaluation on a GPU reduce the run time by about two orders of magnitude for the largest bond dimensions used (Fig.~\ref{app:fig:CPU_GPU}).
    \item The approximate average-energy operator $\tilde\mAE=H-\mathcal{L}X_\ast$ is a matrix product operator; its ground state is obtained with standard DMRG (Sec.~\ref{app:sec:mps_algo:gs}).
\end{enumerate}
Sections~\ref{app:sec:mps_algo:alternatives} and~\ref{app:sec:mps_algo:outlook} report on the alternative formulations of Table~\ref{app:tab:formulations} which we explored, and on directions we have not pursued but which could lead to more efficient implementations.

\subsection{Tensor networks in Sambe superspace}
\label{app:sec:mps_algo:tn}

Sambe space maps a time-periodic problem to a time-independent one by augmenting the physical Hilbert space with the photon (Fourier) index, Sec.~\ref{app:sec:varl:sambe}. 
We now describe the tensor network representation of states, operators, and their action in this enlarged space~\cite{cheng2022matrix}.

\paragraph*{Floquet matrix product states (FMPSs).}
A state in Sambe space is expanded over photon sectors, $\timeket{\psi}=\sum_m\ket{\psi^{(m)}}\otimes\ket{m}_\floquethilbert$, see Eq.~\eqref{app:eq:sambe_state}; we represent each sector as a matrix product state (MPS) with one extra open photon index $m$,
\begin{equation}\label{app:eq:fmps}
  \ket{\psi^{(m)}}
  \;=\;
  \sum_{\boldsymbol{\sigma}}\;
  \prod_{j=1}^{L} M_{j}^{\sigma_j\,(m)}\,
  \ket{\boldsymbol{\sigma}} ,
\end{equation}
where $M_j^{(m)}$ are the local tensors, $\{\sigma_j\}$ the physical indices, and the bond indices $\{a_j\}$ are suppressed. 
Graphically,

\begin{equation}
\tag{\ref{app:eq:fmps}$'$}\label{app:eq:diag_fmps}
\vcenter{\hbox{\resizebox{0.78\columnwidth}{!}{%
\begin{tikzpicture}
\def\rad{0.4}
\def\sp{2.0}
\def\pl{0.5}
\def\hl{0.5}
\def\stub{0.5}
\def\roff{5.2}

\foreach \i in {1,2}{
  \pgfmathsetmacro{\cx}{(\i-1)*\sp}
  \filldraw[fill=blue!40, draw=black, line width=1.4pt] (\cx,0) circle (\rad);
  \node at (\cx,0) {$M_{\i}$};
  \draw[vleg] (\cx,\rad) -- (\cx,\rad+\pl);
  \node[above, green!60!black] at (\cx,\rad+\pl) {$\sigma_{\i}$};
  \draw[photon]
    (\cx+\rad*0.707,\rad*0.707) -- (\cx+\rad*0.707+\hl,\rad*0.707+\hl);
  \node[red] at (\cx+\rad*0.707+\hl+0.20,\rad*0.707+\hl+0.08) {$m$};
}
\draw (\rad,0) -- (\sp-\rad,0);
\node[below] at (\sp/2,0) {$a_1$};
\draw (\sp+\rad,0) -- (\sp+\rad+\stub,0);
\node[below] at (\sp+\rad+\stub,0) {$a_2$};
\node[font=\huge] at (\sp+\rad+\stub+0.6,0) {$\cdots$};

\draw (\roff-\rad-\stub,0) -- (\roff-\rad,0);
\node[below] at (\roff-\rad-\stub,0) {$a_{L-2}$};

\foreach \i/\idx/\siglbl in {
  0/{L\!-\!1}/{\sigma_{L-1}},
  1/L/{\sigma_L}}{
  \pgfmathsetmacro{\cx}{\roff+\i*\sp}
  \filldraw[fill=blue!40, draw=black, line width=1.4pt] (\cx,0) circle (\rad);
  \node at (\cx,0) {$M_{\idx}$};
  \draw[vleg] (\cx,\rad) -- (\cx,\rad+\pl);
  \node[above, green!60!black] at (\cx,\rad+\pl) {$\siglbl$};
  \draw[photon]
    (\cx+\rad*0.707,\rad*0.707) -- (\cx+\rad*0.707+\hl,\rad*0.707+\hl);
  \node[red] at (\cx+\rad*0.707+\hl+0.20,\rad*0.707+\hl+0.08) {$m$};
}
\draw (\roff+\rad,0) -- (\roff+\sp-\rad,0);
\node[below] at (\roff+\sp/2,0) {$a_{L-1}$};
\end{tikzpicture}%
}}}
\end{equation}

\noindent
where the red wiggly leg $m$ is the open photon index labelling the Fourier sector. Note that there is a single shared photon index $m$ for all local tensors, unlike the physical legs $\sigma_j$ which are site-dependent. 
We refer to such an object as a Floquet matrix product state (FMPS).

\paragraph*{Floquet matrix product operators (FMPOs).}
An operator $\sambe{O}$ on Sambe space is represented as a matrix product operator (MPO) with two open photon indices,
\begin{equation}\label{app:eq:fmpo}
  \sambe{O}^{(n,m)}
  \;=\;
  \sum_{\boldsymbol{\sigma},\boldsymbol{\sigma}'}\;
  \prod_{j=1}^{L} W_{j}^{\sigma_j\sigma_j'\,(n,m)}\,
  \ketbra{\boldsymbol{\sigma}}{\boldsymbol{\sigma}'} .
\end{equation}
Gaphically,

\begin{equation}
\tag{\ref{app:eq:fmpo}$'$}\label{app:eq:diag_fmpo}
\vcenter{\hbox{\resizebox{0.78\columnwidth}{!}{%
\begin{tikzpicture}
\def\bs{0.4}
\def\sp{2.0}
\def\pl{0.5}
\def\hl{0.5}
\def\stub{0.5}
\def\roff{5.2}

\foreach \i in {1,2}{
  \pgfmathsetmacro{\cx}{(\i-1)*\sp}
  \filldraw[fill=green!40, draw=black, rounded corners=1pt, line width=1.4pt]
    (\cx-\bs,-\bs) rectangle (\cx+\bs,\bs);
  \node at (\cx,0) {$W_{\i}$};
  \draw[vleg] (\cx,\bs) -- (\cx,\bs+\pl);
  \node[above, green!60!black] at (\cx,\bs+\pl) {$\sigma_{\i}'$};
  \draw[vleg] (\cx,-\bs) -- (\cx,-\bs-\pl);
  \node[below, green!60!black] at (\cx,-\bs-\pl) {$\sigma_{\i}$};
  \draw[photon] (\cx+\bs,\bs) -- (\cx+\bs+\hl,\bs+\hl);
  \node[red] at (\cx+\bs+\hl+0.22,\bs+\hl+0.08) {$n$};
  \draw[photon] (\cx+\bs,-\bs) -- (\cx+\bs+\hl,-\bs-\hl);
  \node[red] at (\cx+\bs+\hl+0.20,-\bs-\hl-0.14) {$m$};
}
\draw[bleg] (\bs,0) -- (\sp-\bs,0);
\node[below, blue] at (\sp/2,0) {$c_1$};
\pgfmathsetmacro{\stubRx}{\sp+\bs+\stub}
\draw[bleg] (\sp+\bs,0) -- (\stubRx,0);
\node[below, blue] at (\sp+\bs+\stub,0) {$c_2$};
\node[font=\huge] at (\stubRx+0.6,0) {$\cdots$};

\draw[bleg] (\roff-\bs-\stub,0) -- (\roff-\bs,0);
\node[below, blue] at (\roff-\bs-\stub,0) {$c_{L-2}$};

\foreach \i/\idx/\sup/\sdown in {
  0/{L\!-\!1}/{\sigma_{L-1}'}/{\sigma_{L-1}},
  1/L/{\sigma_L'}/{\sigma_L}}{
  \pgfmathsetmacro{\cx}{\roff+\i*\sp}
  \filldraw[fill=green!40, draw=black, rounded corners=1pt, line width=1.4pt]
    (\cx-\bs,-\bs) rectangle (\cx+\bs,\bs);
  \node at (\cx,0) {$W_{\idx}$};
  \draw[vleg] (\cx,\bs) -- (\cx,\bs+\pl);
  \node[above, green!60!black] at (\cx,\bs+\pl) {$\sup$};
  \draw[vleg] (\cx,-\bs) -- (\cx,-\bs-\pl);
  \node[below, green!60!black] at (\cx,-\bs-\pl) {$\sdown$};
  \draw[photon] (\cx+\bs,\bs) -- (\cx+\bs+\hl,\bs+\hl);
  \node[red] at (\cx+\bs+\hl+0.22,\bs+\hl+0.08) {$n$};
  \draw[photon] (\cx+\bs,-\bs) -- (\cx+\bs+\hl,-\bs-\hl);
  \node[red] at (\cx+\bs+\hl+0.20,-\bs-\hl-0.14) {$m$};
}
\draw[bleg] (\roff+\bs,0) -- (\roff+\sp-\bs,0);
\node[below, blue] at (\roff+\sp/2,0) {$c_{L-1}$};
\end{tikzpicture}%
}}}
\end{equation}

\noindent
Each local tensor $W_j^{(n,m)}$ carries the two open photon indices $n$ and $m$ (red wiggly legs) in addition to the physical indices $\sigma_j,\sigma_j'$ and the bond indices $\{c_j\}$. 
For the operators of interest, which are multiplication operators by a time-periodic $O(t)=\sum_n e^{in\omega t}O^{(n)}$, the FMPO is block-Toeplitz, i.e., $\sambe{O}^{(n,m)}=O^{(n-m)}$, cf.~Eq.~\eqref{app:eq:sambe_mats}. The same holds for the superoperators $\mathcal{C}$ and $\mathcal{C}^2$ on Sambe superspace, cf.~q.~\eqref{app:eq:mats}. 
Hence, due to the time-periodic structure, only the harmonics $W_j^{(n)}$ need to be stored.

\paragraph*{FMPO acting on an FMPS.}
The action $\sambe{O}\timeket{\psi}$ is obtained by contracting the FMPO onto the FMPS: the physical leg $\sigma_j$ between $W_j$ and $M_j$ is contracted locally, the shared photon index $m$ is summed over, and the open photon index $n$ labels the sector of the resulting state,
\begin{equation}\label{app:eq:fmpo-fmps}
  \bigl(\sambe{O}\,\timeket{\psi}\bigr)^{(n)}
  \;=\;
  \sum_{m}\sum_{\boldsymbol{\sigma}}\;
  \prod_{j=1}^{L}
  W_{j}^{\sigma_j'\sigma_j\,(n,m)}\,M_{j}^{\sigma_j\,(m)}\,
  \ket{\boldsymbol{\sigma}'} .
\end{equation}
Graphically,

\begin{equation}
\tag{\ref{app:eq:fmpo-fmps}$'$}\label{app:eq:diag_fmpo_fmps}
\vcenter{\hbox{\resizebox{0.78\columnwidth}{!}{%
\begin{tikzpicture}
\def\bs{0.4}
\def\rad{0.4}
\def\sp{2.0}
\def\pl{0.5}
\def\hl{0.5}
\def\stub{0.5}
\def\vsep{2.0}
\def\roff{5.2}

\foreach \i in {1,2}{
  \pgfmathsetmacro{\cx}{(\i-1)*\sp}
 
  \filldraw[fill=green!40, draw=black, rounded corners=1pt, line width=1.4pt]
    (\cx-\bs,\vsep-\bs) rectangle (\cx+\bs,\vsep+\bs);
  \node at (\cx,\vsep) {$W_{\i}$};
  \draw[vleg] (\cx,\vsep+\bs) -- (\cx,\vsep+\bs+\pl);
  \node[above, green!60!black] at (\cx,\vsep+\bs+\pl) {$\sigma_{\i}'$};
  \draw[photon] (\cx+\bs,\vsep+\bs) -- (\cx+\bs+\hl,\vsep+\bs+\hl);
  \node[red] at (\cx+\bs+\hl+0.22,\vsep+\bs+\hl+0.08) {$n$};
 
  \filldraw[fill=blue!40, draw=black, line width=1.4pt] (\cx,0) circle (\rad);
  \node at (\cx,0) {$M_{\i}$};
 
  \draw[vleg] (\cx,\vsep-\bs) -- (\cx,\rad);
  \node[right, green!60!black] at (\cx,\bs+\pl) {$\sigma_{\i}$};
 
  \draw[photon]
    (\cx+\bs, \vsep-\bs)
    .. controls (\cx+1.1, \vsep-\bs-0.7) and (\cx+1.2, 1.0) ..
    (\cx+\rad*0.707, \rad*0.707);
  \node[red] at (\cx+1.28,\bs+\pl) {$m$};
}
 
\draw[bleg] (\bs,\vsep) -- (\sp-\bs,\vsep);
\node[below, blue] at (\sp/2,\vsep) {$c_1$};
\draw[bleg] (\sp+\bs,\vsep) -- (\sp+\bs+\stub,\vsep);
\node[below, blue] at (\sp+\bs+\stub,\vsep) {$c_2$};
\node[font=\huge] at (\sp+\bs+\stub+0.6,\vsep) {$\cdots$};
 
\draw (\rad,0) -- (\sp-\rad,0);
\node[below] at (\sp/2,0) {$a_1$};
\draw (\sp+\rad,0) -- (\sp+\rad+\stub,0);
\node[below] at (\sp+\rad+\stub,0) {$a_2$};
\node[font=\huge] at (\sp+\rad+\stub+0.6,0) {$\cdots$};
 
\draw[bleg] (\roff-\bs-\stub,\vsep) -- (\roff-\bs,\vsep);
\node[below, blue] at (\roff-\bs-\stub,\vsep) {$c_{L-2}$};
\draw (\roff-\rad-\stub,0) -- (\roff-\rad,0);
\node[below] at (\roff-\rad-\stub,0) {$a_{L-2}$};
 
\foreach \i/\ws/\sup/\sdn in {
  0/{L\!-\!1}/{\sigma_{L-1}'}/{\sigma_{L-1}},
  1/L/{\sigma_L'}/{\sigma_L}}{
  \pgfmathsetmacro{\cx}{\roff+\i*\sp}
 
  \filldraw[fill=green!40, draw=black, rounded corners=1pt, line width=1.4pt]
    (\cx-\bs,\vsep-\bs) rectangle (\cx+\bs,\vsep+\bs);
  \node at (\cx,\vsep) {$W_{\ws}$};
  \draw[vleg] (\cx,\vsep+\bs) -- (\cx,\vsep+\bs+\pl);
  \node[above, green!60!black] at (\cx,\vsep+\bs+\pl) {$\sup$};
  \draw[photon] (\cx+\bs,\vsep+\bs) -- (\cx+\bs+\hl,\vsep+\bs+\hl);
  \node[red] at (\cx+\bs+\hl+0.22,\vsep+\bs+\hl+0.08) {$n$};
 
  \filldraw[fill=blue!40, draw=black, line width=1.4pt] (\cx,0) circle (\rad);
  \node at (\cx,0) {$M_{\ws}$};
 
  \draw[vleg] (\cx,\vsep-\bs) -- (\cx,\rad);
  \node[right, green!60!black] at (\cx,\bs+\pl) {\scriptsize$\sdn$};
 
  \draw[photon]
    (\cx+\bs, \vsep-\bs)
    .. controls (\cx+1.1, \vsep-\bs-0.7) and (\cx+1.2, 1.0) ..
    (\cx+\rad*0.707, \rad*0.707);
  \node[red] at (\cx+1.28,\bs+\pl) {$m$};
}
 
\draw[bleg] (\roff+\bs,\vsep) -- (\roff+\sp-\bs,\vsep);
\node[below, blue] at (\roff+\sp/2,\vsep) {$c_{L-1}$};
\draw (\roff+\rad,0) -- (\roff+\sp-\rad,0);
\node[below] at (\roff+\sp/2,0) {$a_{L-1}$};
\end{tikzpicture}%
}}}
\end{equation}

\noindent
For a block-Toeplitz FMPO the sum over $m$ is a convolution over the harmonics,
\begin{equation}
    \bigl(\sambe{O}\,\timeket{\psi}\bigr)^{(n)} = \sum_m O^{(n-m)}\, \ket{\psi^{(m)}} ,
    \label{app:eq:conv1}
\end{equation}
see the convolution rule of Sec.~\ref{app:sec:varl:physical}.

In practice, we impose a Fourier cutoff $|n|\leq n_c$ on all FMPS and evaluate the convolution~\eqref{app:eq:conv1} for the $2n_c+1$ retained harmonics; for the Hamiltonians considered in this work, $H$ has harmonics $n\in\{-1,0,1\}$ only, and we find $n_c=1$ sufficient, cf.~Appendix~\ref{app:sec:spin_models}.

\paragraph*{Operators as FMPS.}
The unknown of the normal equation, $X(t)$, is itself a time-periodic operator on $\hilbert{H}$, which we vectorize as in Eq.~\eqref{app:eq:superspace_vector}: the row and column indices of $X^{(m)}$ at site $j$, $\sigma_j$ and $\sigma_j'$, are combined into a single physical index of dimension $d^2$, and $\floquetket{X}$ is an FMPS on $L$ sites of local dimension $d^2$, with tensors $M_{X,j}^{(m)}$. 
Superoperators such as $\mathcal{C}^{(n)}$ are then ordinary FMPOs on this chain, with the structure of Eq.~\eqref{app:eq:liouvillian_vec}. 
In the implementation, we split each site of local dimension $d^2$ into two sites of local dimension $d$ by a singular value decomposition~\cite{kim2024variational}, so the DMRG code operates on a chain of $2L$ sites. 

We also tested an expansion of $X^{(m)}$ in the local Pauli basis rather than in the computational basis, which replaces the combined index $(\sigma_j,\sigma_j')$ by a Pauli label; this formulation uses a local dimension of $d^2=4$ and an $L$-site chain.
We found no difference in performance compared to the computational basis with $2L$ sites. However, we chose the computational basis, since the splitting respects the product structure of $\identity\otimes H^{(n)}$.

\subsection{Representing $\mathcal{L}$ and $\mathcal{L}^2$}
\label{app:sec:mps_algo:L}

In Sambe superspace, the Floquet Liouvillian takes the vectorized form of Eq.~\eqref{app:eq:liouvillian_vec},
\begin{equation}\label{app:eq:liouvillian-sambe}
    \mathcal{L} = \mathcal{C} + \omega \mathcal{N},
\end{equation}
where $\mathcal{N}$ is the photon-number superoperator, diagonal in the harmonics, and $\mathcal{C}=\mathcal{C}_H$ is block-Toeplitz commutator with the time-periodic Hamiltonian with harmonics $\mathcal{C}^{(n)}=\identity\otimes H^{(n)}-H^{(n)T}\otimes\identity$. 
Each harmonic $\mathcal{C}^{(n)}$ is an MPO in the doubled Hilbert space, since the commutator contains the two terms $\identity\otimes H^{(n)}$ and $H^{(n)T}\otimes\identity$, which we implement on the $2L$-site chain~\cite{kim2024variational}.

To form $\mathcal{L}^2$ we additionally need
\begin{equation}
 (\mathcal{C}^2)^{(n)} = \sum_m \mathcal{C}^{(n-m)}\, \mathcal{C}^{(m)} ,
 \label{app:eq:conv2}
\end{equation}
the harmonics of the squared commutator superoperator, which we precompute once as MPOs by multiplying the MPOs $\mathcal{C}^{(n-m)}$ and $\mathcal{C}^{(m)}$ site by site.
Note that the order of the factors in Eq.~\eqref{app:eq:conv2} matters, since $\mathcal{C}^{(n)}$ and $\mathcal{C}^{(m)}$ do not commute in general. 
Due to the block-Toeplitz structure, Eq.~\eqref{app:eq:mats}, neither $\mathcal{C}$ nor $\mathcal{C}^2$ is stored as a full matrix in the photon indices $(n,m)$.

With these definitions, the action of $\mathcal{L}$ on an FMPS $\floquetket{X}$ reads as
\begin{equation}\label{app:eq:Lx}
  \bigl(\mathcal{L}\floquetket{X}\bigr)^{(n)}
  \;=\; \sum_m \mathcal{C}^{(n-m)}\,\floquetket{X^{(m)}}
  \;+\; n\omega\,\floquetket{X^{(n)}} ,
\end{equation}
and, using $\mathcal{L}^2=\mathcal{C}^2+\omega(\mathcal{C}\mathcal{N}+\mathcal{N}\mathcal{C})+\omega^2\mathcal{N}^2$, we have
\begin{multline}\label{app:eq:L2x}
  \bigl(\mathcal{L}^2\floquetket{X}\bigr)^{(n)} \\
  = \sum_m \Bigl( (\mathcal{C}^2)^{(n-m)} + \omega(n+m)\,\mathcal{C}^{(n-m)} \Bigr)\,
        \floquetket{X^{(m)}} \\
  +\; \omega^2 n^2\,\floquetket{X^{(n)}} .
\end{multline}
Both operations are convolutions in the photon index and MPO-MPS contractions in the physical indices. 

Finally, the right-hand side of the normal equation is the FMPS
\begin{equation}\label{app:eq:dH_fmps}
  \floquetket{\partial H} = \omega\mathcal{N}\floquetket{H},
  \qquad
  \floquetket{(\partial H)^{(n)}} = n\omega\,\floquetket{H^{(n)}} ,
\end{equation}
cf.~Eq.~\eqref{app:eq:dH_def}, obtained from the MPO of $H^{(n)}$ by combining its physical legs; it has no zero-harmonic component.

\subsection{Overall procedure}
\label{app:subsec:AE}

The complete procedure for computing the average-energy operator $\mAE[t]$ and its ground state is summarized in Algorithm~\ref{alg:AE}.
\begin{enumerate}
    \item Starting from the harmonics $H^{(n)}$ of the lab-frame Hamiltonian, we construct the MPOs $\mathcal{C}^{(n)}$ and $(\mathcal{C}^2)^{(n)}$ from Eq.~\eqref{app:eq:conv2}, and the FMPS $\floquetket{\partial H}$ from Eq.~\eqref{app:eq:dH_fmps}.
    \item We choose the initial guess $\floquetket{X_0}=\mathcal{L}\floquetket{\partial H}$. A Krylov solver for the normal equation started from a vector in $\im\mathcal{L}$ only adds vectors from $\im\mathcal{L}^2\subseteq\im\mathcal{L}$ and therefore targets the minimal-norm pre-image $X_\ast\perp\ker\mathcal{L}$ at the global level; since $\partial H=\mathcal{L}H$ lies in $\im\mathcal{L}$ itself, the right-hand side would serve this purpose as well, and so does $\mathcal{L}\floquetket{\partial H}$.
    In practice, we observe that starting from $\mathcal{L}\floquetket{\partial H}$ leads to a faster convergence of the DMRG sweeps.
    \item The key step is solving the normal equation~\eqref{app:eq:normal_eq} $\mathcal{L}^2\floquetket{X}=\floquetket{\partial H}$ for the FMPS $\floquetket{X}$ by the DMRG sweeps of Sec.~\ref{app:sec:dmrg}.
    \item From the solution we obtain the gauge potential~\eqref{app:eq:Xstar_solution} $\floquetket{\agp_\mAE}=\mathcal{L}\floquetket{X_\ast}$ by one application of Eq.~\eqref{app:eq:Lx}. This step is essential conceptually as it removes any residual kernel components in $X_\ast$ accumulated during the sweeps, and enforces the minimum-norm gauge, cf.~Sec.~\ref{app:sec:dmrg:gauge}.
    \item Finally, we 
    (i) contract $\floquetket{\agp_\mAE}$ back into an MPO, (ii) form the MPO $\tilde\mAE^{(n)}=H^{(n)}-\agp_\mAE^{(n)}$ harmonic by harmonic, and (iii) return to the time domain using Eq.~\eqref{app:eq:fourier}:
    \begin{equation}
    \tilde\mAE[t] = \sum_n e^{in\omega t}\, \tilde\mAE^{(n)} ,
    \label{app:eq:IFT}
    \end{equation} 
    (iv) A standard DMRG run~\cite{White1992,Schollwock2005,Schollwock2011} applied to $\tilde\mAE[t]$ at the desired time $t$ (we use $t{=}0$) then gives the approximate Floquet ground state $|\tilde n^\F_\text{GS}[t]\rangle$ and its average energy $\mae_0$, Sec.~\ref{app:sec:mps_algo:gs}.
\end{enumerate}

\begin{algorithm}[H]
\caption{Calculation of the average-energy operator $\mAE$}
\label{alg:AE}
\begin{algorithmic}[1]
\REQUIRE harmonics $H^{(n)}$ of $H(t)$ as MPOs, $|n|\leq n_c$
\STATE $\mathcal{C}^{(n)} \leftarrow \identity \otimes H^{(n)} - H^{(n)T} \otimes \identity$ \hfill $\triangleright$ Eq.~\eqref{app:eq:liouvillian_vec}
\STATE $(\mathcal{C}^2)^{(n)} \leftarrow \sum_m \mathcal{C}^{(n-m)}\mathcal{C}^{(m)}$ \hfill $\triangleright$ Eq.~\eqref{app:eq:conv2}
\STATE $\floquetket{\partial H} \leftarrow \omega\mathcal{N}\floquetket{H}$ \hfill $\triangleright$ Eq.~\eqref{app:eq:dH_fmps}
\STATE $\floquetket{X_0} \leftarrow \mathcal{L}\floquetket{\partial H}$ \hfill $\triangleright$ Initial guess in $\im\mathcal{L}$
\STATE \textbf{Solve} $\mathcal{L}^2\floquetket{X} = \floquetket{\partial H}$ by DMRG \hfill $\triangleright$ Alg.~\ref{alg:dmrg}, Eq.~\eqref{app:eq:normal_eq}
\STATE $\floquetket{\agp_\mAE} \leftarrow \mathcal{L}\floquetket{X_\ast}$ \hfill $\triangleright$ Eqs.~\eqref{app:eq:Xstar_solution}, \eqref{app:eq:Lx}
\STATE $\tilde\mAE^{(n)} \leftarrow H^{(n)} - \agp_\mAE^{(n)}$ \hfill $\triangleright$ Reshape to MPO, Eq.~\eqref{app:eq:AE_decomposition}
\STATE $\tilde\mAE[t] \leftarrow \sum_n e^{in\omega t}\tilde\mAE^{(n)}$ \hfill $\triangleright$ Eq.~\eqref{app:eq:IFT}
\STATE \textbf{DMRG} ground state of $\tilde\mAE[t]$ \hfill $\triangleright$ Sec.~\ref{app:sec:mps_algo:gs}
\end{algorithmic}
\end{algorithm}

\subsection{Two-site DMRG for the normal equation}
\label{app:sec:dmrg}

We now describe the DMRG algorithm used to solve the normal equation $\mathcal{L}^2\floquetket{X}=\floquetket{\partial H}$. 
The procedure, summarized in Algorithm~\ref{alg:dmrg}, is a standard two-site variational sweep which minimizes the cost function $\norm{\mathcal{L}X-H}^2$ of Eq.~\eqref{app:eq:normal_eq} one bond at a time, in the spirit of DMRG linear solvers for static problems~\cite{kim2024variational,mckeever2024towards}, but using tensor-network methods in Sambe super space~\cite{cheng2022matrix,keliri2026sambe,perez2026floquet}.

\subsubsection{Gauges}
\label{app:sec:dmrg:gauge}

The algorithm has two different gauge choices, which we now briefly discuss. 

The first is the gauge of the target gauge potential, cf.~Sec.~\ref{app:sec:varl:gauge}. The Moore--Penrose solution $\agp_\mAE$ is the minimal-norm solution of $\mathcal{L}X=\partial H$: global Krylov methods started in $\im\mathcal{L}$ are guaranteed to find that gauge. 
A \textit{local} DMRG update, however, solves a reduced linear problem on two sites with a local Krylov solver; hence, the resulting two-site tensor is in general not the restriction of any vector in the global kernel $\im\mathcal{L}$ to the two-site Hilbert space. Accordingly, we observed that sweeps of local minimal-norm solutions of $\mathcal{L}X=\partial H$ do not converge to the global minimal-norm solution. 
This is the practical reason for solving the normal equation~\eqref{app:eq:normal_eq} instead and for the two-step formulation where $\agp_\mAE=\mathcal{L}X_\ast$; the second step makes the kernel component that $X$ accumulates during the sweeps harmless, cf.~Fig.~\ref{fig:local_vs_global-problem}(a). Hence, we do not fix this gauge explicitly during the DMRG sweeps. 

The second gauge is that of the FMPS itself, which we fix. 
We define $\floquetket{X}$ in the mixed-canonical form with the orthogonality center at the bond $(j,j+1)$ currently being optimized. The reduced problem is expressed in an orthonormal basis of the two-site subspace, and we move the orthogonality center along with the DMRG sweep.

\subsubsection{Sweeps and environments}
\label{app:sec:dmrg:sweeps}

The algorithm works with three objects besides the unknown $\floquetket{X}$: the FMPOs $\mathcal{C}$ and $\mathcal{C}^2$, and the FMPS $\floquetket{\partial H}$. 
For each of them we store left and right environments, listed in Table~\ref{app:tab:environments}, which contain the contraction of all sites to the left (right) of the current bond with the bra and ket tensors of $\floquetket{X}$. 
The FMPO environments carry two photon indices $(n,m)$, since the bra and the ket of $\floquetket{X}$ may be in different harmonics; the FMPS environments carry a single photon index. 

\begin{table}[t]
  \label{app:tab:environments}
  \renewcommand{\arraystretch}{1.3}
  \begin{ruledtabular}
  \begin{tabular}{lccl}
    object & left env. & right env. & photon indices \\
    \hline
    $\mathcal{C}$ & $L_{\mathcal{C},j}^{(n,m)}$ & $R_{\mathcal{C},j+1}^{(n,m)}$ & two \\
    $\mathcal{C}^2$ & $L_{\mathcal{C}^2,j}^{(n,m)}$ & $R_{\mathcal{C}^2,j+1}^{(n,m)}$ & two \\
    $\floquetket{\partial H}$ & $L_{\partial H,j}^{(m)}$ & $R_{\partial H,j+1}^{(m)}$ & one \\
  \end{tabular}
  \end{ruledtabular}
  \caption{Environments used in the two-site DMRG sweep at bond $(j,j+1)$. All environments are contractions with the FMPS $\floquetket{X}$.
  }
\end{table}

We first initialize all environments from the initial guess $\floquetket{X_0}$. 
The algorithm then alternates between right sweeps ($j=1,\dots,2L-1$) and left sweeps ($j=2L-1,\dots,1$). 
At each bond $(j,j+1)$, the two-site step of Algorithm~\ref{alg:two_site} updates the two local tensors of $\floquetket{X}$, after which the corresponding environment is extended by one site for the next bond. 
The bond dimension $\chi_X$ of $\floquetket{X}$ is increased in stages: for the results of the main text, we perform two full sweeps at each $\chi_X\in\{16,32,64,96,128\}$, using the converged FMPS as the starting point for the next stage; an example for the convergence of the residual~\eqref{app:eq:residual} with $\chi_X$ is shown in Fig.~\ref{app:fig:residual}.

\begin{algorithm}[H]
\caption{DMRG sweeps for the normal equation $\mathcal{L}^2\floquetket{X}=\floquetket{\partial H}$}
\label{alg:dmrg}
\begin{algorithmic}[1]
\REQUIRE MPOs $\mathcal{C}^{(n)}$, $(\mathcal{C}^2)^{(n)}$; FMPSs $\floquetket{\partial H}$, $\floquetket{X_0}$; frequency $\omega$; schedule for bond dimension $\chi_X$
\STATE Initialize the environments of Table~\ref{app:tab:environments} from $\floquetket{X_0}$
\FOR{each target bond dimension $\chi_X$ in the schedule}
\FOR{$s = 1$ to number of sweeps}
  \STATE \textbf{Right sweep:}
  \FOR{$j = 1$ to $2L-1$}
    \STATE Two-site step at bond $(j,j+1)$ \hfill $\triangleright$ Alg.~\ref{alg:two_site}
    \STATE Extend the left environments to site $j+1$ \hfill $\triangleright$ Eq.~\eqref{app:eq:env-update}
  \ENDFOR
  \STATE \textbf{Left sweep:}
  \FOR{$j = 2L-1$ downto $1$}
    \STATE Two-site step at bond $(j,j+1)$ \hfill $\triangleright$ Alg.~\ref{alg:two_site}
    \STATE Extend the right environments to site $j$
  \ENDFOR
\ENDFOR
\ENDFOR
\RETURN $\floquetket{X_\ast}=\floquetket{X}$
\end{algorithmic}
\end{algorithm}

\subsubsection{Two-site step}
\label{app:sec:dmrg:twosite}

The two-site step, Algorithm~\ref{alg:two_site}, updates the local tensors of $\floquetket{X}$ at the adjacent sites $(j,j+1)$ which define the bond. 
It solves the reduced linear problem
\begin{equation}
\label{app:eq:local}
  (\mathcal{L}^2)_{j,j+1}\, X_{j, j+1} \;=\; \partial H_{j,j+1},
\end{equation}
where $(\mathcal{L}^2)_{j,j+1}$ and $\partial H_{j,j+1}$ are the effective operator $\mathcal{L}^2$ and the effective right-hand side $\floquetket{\partial H}$ at the bond $(j,j+1)$, respectively. 
We show in Sec.~\ref{app:sec:dmrg:consistency} that Eq.~\eqref{app:eq:local} always admits a solution.
In the following, we discuss the steps of Algorithm~\ref{alg:two_site} in detail. 

\textbf{Steps 1--4: local tensors.} 
The initial guess for the two-site tensor is the contraction of the current local tensors,
\begin{equation}\label{app:eq:Xguess}
  X_{j,j+1}^{(m)} \;=\; M_{X,j}^{(m)} \, M_{X,j+1}^{(m)} ,
\end{equation}
which is the exact two-site tensor of the current FMPS in the mixed-canonical gauge. 
The effective operators $\mathcal{C}_{j,j+1}$ and $(\mathcal{C}^2)_{j,j+1}$ are obtained by contracting the local FMPO tensors with the corresponding environments,
\begin{align}\label{app:eq:Aeff}
  \mathcal{C}_{j,j+1}^{(n,m)}
    &= L_{\mathcal{C},j}^{(n,m)} \, W_{\mathcal{C},j}^{(n,m)} \, W_{\mathcal{C},j+1}^{(n,m)} \,R_{\mathcal{C},j+1}^{(n,m)}, \nonumber \\
  (\mathcal{C}^2)^{(n,m)}_{j,j+1}
    &= L_{\mathcal{C}^2,j}^{(n,m)} \, W_{\mathcal{C}^2,j}^{(n,m)} \, W_{\mathcal{C}^2,j+1}^{(n,m)} \, R_{\mathcal{C}^2,j+1}^{(n,m)},
\end{align}
and the effective right-hand side is given by
\begin{equation}\label{app:eq:beff}
  \partial H_{j,j+1}^{(m)}
  = L_{\partial H,j}^{(m)} \, M_{\partial H,j}^{(m)} \, M_{\partial H,j+1}^{(m)} \, R_{\partial H,j+1}^{(m)} .
\end{equation}
Graphically, Eq.~\eqref{app:eq:Aeff} reads as
\begin{equation}
\tag{\ref{app:eq:Aeff}$'$}\label{app:eq:diag_eff_op}
\vcenter{\hbox{\resizebox{0.85\columnwidth}{!}{%
\begin{tikzpicture}[line width=1.4pt]
\node at (-1.9,0) {$O_{j,j+1}^{({\color{red}m,n})}\;=$};
\def\bs{0.55}
\def\rad{0.55}
\def\sp{2.0}
\def\pl{0.5}
\def\ybra{1.6}
\def\ympo{0}
\def\yket{-1.6}
\def\xL{0}
\def\xWj{2.2}
\def\xWjj{4.4}
\def\xR{6.6}
\def\envW{0.55}
\def\envH{2.2}
\pgfmathsetmacro{\envYmid}{\ympo}

\filldraw[fill=teal!40, rounded corners=3pt, draw=black]
  (\xL-\envW, \envYmid-\envH) rectangle (\xL+\envW, \envYmid+\envH);
\node at (\xL, \envYmid) {$L_{O,j}^{({\color{red}m,n})}$};

\filldraw[fill=teal!40, rounded corners=3pt, draw=black]
  (\xR-\envW, \envYmid-\envH) rectangle (\xR+\envW, \envYmid+\envH);
\node at (\xR, \envYmid) {$R_{O,j+1}^{({\color{red}m,n})}$};

\foreach \cx/\lbl in {\xWj/{W_{O,j}^{({\color{red}m,n})}}, \xWjj/{W_{O,j+1}^{({\color{red}m,n})}}}{
  \filldraw[fill=green!40, rounded corners=1pt, draw=black]
    (\cx-\bs,\ympo-\bs) rectangle (\cx+\bs,\ympo+\bs);
  \node at (\cx,\ympo) {$\lbl$};
}

\draw[bleg] (\xL+\envW, \ympo) -- (\xWj-\bs, \ympo)
      node[midway, below, blue] {$c_{j-1}$};
\draw[bleg] (\xWj+\bs, \ympo) -- (\xWjj-\bs, \ympo)
      node[midway, below, blue] {$c_{j}$};
\draw[bleg] (\xWjj+\bs, \ympo) -- (\xR-\envW, \ympo)
      node[midway, below, blue] {$c_{j+1}$};

\draw[vleg] (\xWj,  \ympo+\bs) -- (\xWj,  \ympo+\bs+\pl);
\node[above, green!60!black] at (\xWj,  \ympo+\bs+\pl) {$\sigma_{j}'$};
\draw[vleg] (\xWjj, \ympo+\bs) -- (\xWjj, \ympo+\bs+\pl);
\node[above, green!60!black] at (\xWjj, \ympo+\bs+\pl) {$\sigma_{j+1}'$};

\draw (\xL+\envW, \ybra) -- (\xL+\envW+\pl, \ybra) node[right]{$a_{j-1}'$};
\draw (\xR-\envW-\pl, \ybra) -- (\xR-\envW, \ybra)  node[pos=0,left]{$a_{j+1}'$};

\draw[vleg] (\xWj,  \ympo-\bs) -- (\xWj,  \ympo-\bs-\pl);
\node[below, green!60!black] at (\xWj,  \ympo-\bs-\pl) {$\sigma_{j}$};
\draw[vleg] (\xWjj, \ympo-\bs) -- (\xWjj, \ympo-\bs-\pl);
\node[below, green!60!black] at (\xWjj, \ympo-\bs-\pl) {$\sigma_{j+1}$};

\draw (\xL+\envW, \yket) -- (\xL+\envW+\pl, \yket) node[right]{$a_{j-1}$};
\draw (\xR-\envW-\pl, \yket) -- (\xR-\envW, \yket)  node[pos=0,left]{$a_{j+1}$};
\end{tikzpicture}%
}}}
\end{equation}
\noindent
with $O=\mathcal{C}$ and $O=\mathcal{C}^2$, where the environments $L_{O,j}$ and $R_{O,j+1}$ carry the open bond indices of the bra ($a'$) and of the ket ($a$) of $\floquetket{X}$. 
In turn, Eq.~\eqref{app:eq:beff} reads graphically as
\begin{equation}
\tag{\ref{app:eq:beff}$'$}\label{app:eq:diag_eff_rhs}
\vcenter{\hbox{\resizebox{0.85\columnwidth}{!}{%
\begin{tikzpicture}[line width=1.4pt]
\def\rad{0.55}
\def\pl{0.5}
\def\ybra{1.6}
\def\yket{-1.6}
\def\xL{0}
\def\xWj{2.2}
\def\xWjj{4.4}
\def\xR{6.6}
\def\envW{0.55}
\def\envH{2.2}
\node at (-2.1,0) {$\partial H_{j,j+1}^{({\color{red}m})}=$};
\filldraw[fill=teal!40, rounded corners=3pt, draw=black]
  (\xL-\envW, -\envH) rectangle (\xL+\envW, \envH);
\node at (\xL, 0) {$L_{\partial H,j}^{({\color{red}m})}$};
\filldraw[fill=teal!40, rounded corners=3pt, draw=black]
  (\xR-\envW, -\envH) rectangle (\xR+\envW, \envH);
\node at (\xR, 0) {$R_{\partial H,j+1}^{({\color{red}m})}$};
\draw (\xL+\envW, \ybra) -- (\xL+\envW+\pl, \ybra) node[right]{$a_{j-1}'$};
\draw (\xR-\envW-\pl, \ybra) -- (\xR-\envW, \ybra)  node[pos=0,left]{$a_{j+1}'$};
\foreach \cx/\lbl in {\xWj/{M_{\partial H,j}^{({\color{red}m})}}, \xWjj/{M_{\partial H,j+1}^{({\color{red}m})}}}{
  \filldraw[fill=blue!40, draw=black] (\cx,\yket) circle (\rad);
  \node at (\cx,\yket) {$\lbl$};
}
\draw[vleg] (\xWj,  \yket+\rad) -- (\xWj,  \yket+\rad+1.2);
\node[above, green!60!black] at (\xWj,  \yket+\rad+1.2) {$\sigma_{j}$};
\draw[vleg] (\xWjj, \yket+\rad) -- (\xWjj, \yket+\rad+1.2);
\node[above, green!60!black] at (\xWjj, \yket+\rad+1.2) {$\sigma_{j+1}$};
\draw (\xL+\envW, \yket) -- (\xWj-\rad, \yket)   node[midway,below]{$b_{j-1}$};
\draw (\xWj+\rad, \yket) -- (\xWjj-\rad,\yket)   node[midway,below]{$b_{j}$};
\draw (\xWjj+\rad,\yket) -- (\xR-\envW, \yket)   node[midway,below]{$b_{j+1}$};
\end{tikzpicture}%
}}}
\end{equation}
\noindent
where the environments of $\floquetket{\partial H}$ are contractions of the tensors $M_{\partial H}$ with the conjugate tensors of $\floquetket{X}$, so that the result carries the bra indices $a'_{j-1},a'_{j+1}$ and the physical indices $\sigma_j,\sigma_{j+1}$ of a two-site tensor; $b_j$ denote the bond indices of $\floquetket{\partial H}$. 
In practice, the effective operators are never formed as matrices; only their action on a two-site tensor is required, see below.

\textbf{Step 5: linear solve.} 
We use the generalized minimal residual method (GMRES)~\cite{saad1986gmres}, based on a Krylov-subspace iteration, to solve Eq.~\eqref{app:eq:local}, since the exact pseudo-inverse of $(\mathcal{L}^2)_{j,j+1}$ is out of reach for the bond dimensions of interest; the initial guess is Eq.~\eqref{app:eq:Xguess}. 
Each GMRES iteration applies $(\mathcal{L}^2)_{j,j+1}$ once to a two-site tensor. 
Following Eq.~\eqref{app:eq:L2x}, this application reads
\begin{multline}\label{app:eq:localLHS}
  \bigl((\mathcal{L}^2)_{j,j+1}\,X_{j,j+1}\bigr)^{(n)} \\
  = \sum_m \Bigl( (\mathcal{C}^2)^{(n,m)}_{j,j+1} + \omega(n+m)\,\mathcal{C}^{(n,m)}_{j,j+1} \Bigr)\,
        X^{(m)}_{j,j+1} \\
  +\; \omega^2 n^2\,X^{(n)}_{j,j+1} .
\end{multline}
The last term is a scalar multiplication and is treated separately. 
The first two terms involve the contraction of the two tensor networks $(\mathcal{C}^2)^{(n,m)}_{j,j+1}X^{(m)}_{j,j+1}$ and $\mathcal{C}^{(n,m)}_{j,j+1}X^{(m)}_{j,j+1}$, which share the graphical structure of Eq.~\eqref{app:eq:diag_local_op},

\begin{equation}
\tag{\ref{app:eq:localLHS}$'$}\label{app:eq:diag_local_op}
\vcenter{\hbox{\resizebox{0.78\columnwidth}{!}{%
\begin{tikzpicture}[line width=1.4pt]
\def\bs{0.55}
\def\rad{0.55}
\def\sp{2.0}
\def\pl{0.5}
\def\ybra{1.6}
\def\ympo{0}
\def\yket{-1.6}
\def\xL{0}
\def\xWj{2.2}
\def\xWjj{4.4}
\def\xR{6.6}
\def\envW{0.55}
\def\envH{2.2}
\pgfmathsetmacro{\envYmid}{\ympo}

\filldraw[fill=teal!40, rounded corners=3pt, draw=black]
  (\xL-\envW, \envYmid-\envH) rectangle (\xL+\envW, \envYmid+\envH);
\node at (\xL, \envYmid) {$L_{O,j}^{({\color{red}m,n})}$};

\filldraw[fill=teal!40, rounded corners=3pt, draw=black]
  (\xR-\envW, \envYmid-\envH) rectangle (\xR+\envW, \envYmid+\envH);
\node at (\xR, \envYmid) {$R_{O,j+1}^{({\color{red}m,n})}$};

\foreach \cx/\lbl in {\xWj/{W_{O,j}^{({\color{red}m,n})}}, \xWjj/{W_{O,j+1}^{({\color{red}m,n})}}}{
  \filldraw[fill=green!40, rounded corners=1pt, draw=black]
    (\cx-\bs,\ympo-\bs) rectangle (\cx+\bs,\ympo+\bs);
  \node at (\cx,\ympo) {$\lbl$};
}

\draw[bleg] (\xL+\envW, \ympo) -- (\xWj-\bs, \ympo)
      node[midway, below, blue] {$c_{j-1}$};
\draw[bleg] (\xWj+\bs, \ympo) -- (\xWjj-\bs, \ympo)
      node[midway, below, blue] {$c_{j}$};
\draw[bleg] (\xWjj+\bs, \ympo) -- (\xR-\envW, \ympo)
      node[midway, below, blue] {$c_{j+1}$};

\draw[vleg] (\xWj,  \ympo+\bs) -- (\xWj,  \ympo+\bs+\pl);
\node[above, green!60!black] at (\xWj,  \ympo+\bs+\pl) {$\sigma_{j}'$};
\draw[vleg] (\xWjj, \ympo+\bs) -- (\xWjj, \ympo+\bs+\pl);
\node[above, green!60!black] at (\xWjj, \ympo+\bs+\pl) {$\sigma_{j+1}'$};

\draw (\xL+\envW, \ybra) -- (\xL+\envW+\pl, \ybra) node[right]{$a_{j-1}'$};
\draw (\xR-\envW-\pl, \ybra) -- (\xR-\envW, \ybra)  node[pos=0,left]{$a_{j+1}'$};

\foreach \cx/\lbl in {\xWj/{M_{X,j}^{({\color{red}m})}}, \xWjj/{M_{X,j+1}^{({\color{red}m})}}}{
  \filldraw[fill=blue!40, draw=black] (\cx,\yket) circle (\rad);
  \node at (\cx,\yket) {$\lbl$};
}

\draw[vleg] (\xWj,  \ympo-\bs) -- (\xWj,  \yket+\rad);
\node[right, green!60!black] at (\xWj,  {(\ympo-\bs+\yket+\rad)/2}) {$\sigma_{j}$};
\draw[vleg] (\xWjj, \ympo-\bs) -- (\xWjj, \yket+\rad);
\node[right, green!60!black] at (\xWjj, {(\ympo-\bs+\yket+\rad)/2}) {$\sigma_{j+1}$};

\draw (\xL+\envW, \yket) -- (\xWj-\rad, \yket)   node[midway,below]{$a_{j-1}$};
\draw (\xWj+\rad, \yket) -- (\xWjj-\rad,\yket)   node[midway,below]{$a_{j}$};
\draw (\xWjj+\rad,\yket) -- (\xR-\envW, \yket)   node[midway,below]{$a_{j+1}$};
\end{tikzpicture}%
}}}
\end{equation}

\noindent
with $O=\mathcal{C}$ and $O=\mathcal{C}^2$, respectively, and are followed by the sum over $m$. 

The choice of GMRES is dictated by consistency: minimal-residual Krylov solvers such as GMRES and MINRES~\cite{paige1975minres} are stable only for solvable systems, $b\in\im A$, and for the reduced problem~\eqref{app:eq:local} this is guaranteed by the construction of Sec.~\ref{app:sec:dmrg:consistency}. 
Since $(\mathcal{L}^2)_{j,j+1}$ is Hermitian, MINRES is applicable as well. In our tests, both GMRES and MINRES performed similarly in terms of Krylov dimensions and run time.

\textbf{Steps 6--7: truncation and update.} 
The solution of Eq.~\eqref{app:eq:local} is split back into two local tensors by a singular value decomposition,
\begin{equation}\label{app:eq:svd}
  X_{j,j+1}^{(m)} \;\xrightarrow{\text{SVD}}\;
  \tilde{M}_{X,j}^{(m)},\;\tilde{M}_{X,j+1}^{(m)} ,
\end{equation}
truncated to the current target bond dimension $\chi_X$, and the local tensors are replaced, $M_{X,j}^{(m)}\leftarrow\tilde M_{X,j}^{(m)}$ and $M_{X,j+1}^{(m)}\leftarrow\tilde M_{X,j+1}^{(m)}$; the orthogonality center is placed on the site towards which the sweep proceeds.

\begin{algorithm}[H]
\caption{Two-site step at bond $(j,j+1)$ for the normal equation $\mathcal{L}^2\floquetket{X}=\floquetket{\partial H}$}
\label{alg:two_site}
\begin{algorithmic}[1]
\REQUIRE environments of Table~\ref{app:tab:environments} at bond $(j,j+1)$
\REQUIRE local tensors $M_{X,j}^{(m)}, M_{X,j+1}^{(m)}$, $M_{\partial H,j}^{(m)}, M_{\partial H,j+1}^{(m)}$, $W_{\mathcal{C},j}^{(n,m)}, W_{\mathcal{C},j+1}^{(n,m)}$, $W_{\mathcal{C}^2,j}^{(n,m)}, W_{\mathcal{C}^2,j+1}^{(n,m)}$
\STATE $X_{j,j+1}^{(m)} \leftarrow M_{X,j}^{(m)} \, M_{X,j+1}^{(m)}$ \hfill $\triangleright$ Initial guess, Eq.~\eqref{app:eq:Xguess}
\STATE $\mathcal{C}_{j,j+1}^{(n,m)} \leftarrow L_{\mathcal{C},j}^{(n,m)} \, W_{\mathcal{C},j}^{(n,m)} \,W_{\mathcal{C},j+1}^{(n,m)} \, R_{\mathcal{C},j+1}^{(n,m)}$ \hfill $\triangleright$ Eq.~\eqref{app:eq:Aeff}
\STATE $(\mathcal{C}^2)^{(n,m)}_{j,j+1} \leftarrow L_{\mathcal{C}^2,j}^{(n,m)} \, W_{\mathcal{C}^2,j}^{(n,m)} \, W_{\mathcal{C}^2,j+1}^{(n,m)} \, R_{\mathcal{C}^2,j+1}^{(n,m)}$ \hfill $\triangleright$ Eq.~\eqref{app:eq:Aeff}
\STATE $\partial H_{j,j+1}^{(m)} \leftarrow L_{\partial H,j}^{(m)} \, M_{\partial H,j}^{(m)} \, M_{\partial H,j+1}^{(m)} \, R_{\partial H,j+1}^{(m)}$ \hfill $\triangleright$ Eq.~\eqref{app:eq:beff}
\STATE \textbf{Solve} $(\mathcal{L}^2)_{j,j+1}\, X_{j,j+1} = \partial H_{j,j+1}$ by GMRES \hfill $\triangleright$ Eqs.~\eqref{app:eq:local}, \eqref{app:eq:localLHS}
\STATE $\tilde{M}_{X,j}^{(m)}, \tilde{M}_{X,j+1}^{(m)} \leftarrow$ SVD$\bigl(X_{j,j+1}^{(m)}\bigr)$ \hfill $\triangleright$ Truncate to $\chi_X$, Eq.~\eqref{app:eq:svd}
\STATE $M_{X,j}^{(m)}, M_{X,j+1}^{(m)} \leftarrow \tilde{M}_{X,j}^{(m)}, \tilde{M}_{X,j+1}^{(m)}$ \hfill $\triangleright$ Update local tensors
\STATE Extend the environments \hfill $\triangleright$ Eq.~\eqref{app:eq:env-update}
\end{algorithmic}
\end{algorithm}

\textbf{Step 8: environment update.} 
After the update at bond $(j,j+1)$, the environments are extended so the algorithm can proceed to the adjacent bond. 
During a right sweep, the left environments at site $j+1$ follow from those at site $j$ by absorbing the updated tensor $\tilde M_{X,j}$, its conjugate, and the FMPO or FMPS tensor at site $j$,
\begin{align}\label{app:eq:env-update}
  L_{\mathcal{C}^2,j+1}^{(n,m)}
    &= L_{\mathcal{C}^2,j}^{(n,m)} \, W_{\mathcal{C}^2,j}^{(n,m)}
       \, \tilde{M}_{X,j}^{(m)} \, \tilde{M}_{X,j}^{(n)*}, \nonumber \\
  L_{\mathcal{C},j+1}^{(n,m)}
    &= L_{\mathcal{C},j}^{(n,m)} \, W_{\mathcal{C},j}^{(n,m)}
       \, \tilde{M}_{X,j}^{(m)} \, \tilde{M}_{X,j}^{(n)*}, \nonumber \\
  L_{\partial H,j+1}^{(m)}
    &= L_{\partial H,j}^{(m)} \, M_{\partial H,j}^{(m)} \, \tilde{M}_{X,j}^{(m)} .
\end{align}
Graphically, the first two updates read as

\begin{equation}
\tag{\ref{app:eq:env-update}$'$}\label{app:eq:diag_env_update}
\vcenter{\hbox{\resizebox{0.78\columnwidth}{!}{%
\begin{tikzpicture}

\def\bs{0.55}
\def\rad{0.55}
\def\pl{0.55}
\def\envW{0.55}
\def\envH{2.2}

\def\ybra{ 1.6}
\def\ympo{ 0.0}
\def\yket{-1.6}

\filldraw[fill=teal!40, rounded corners=3pt, draw=black]
  (-\envW, \ympo-\envH) rectangle (\envW, \ympo+\envH);
\node at (0,\ympo) {$L_{O,j+1}^{({\color{red}m,n})}$};

\draw (\envW, \ybra) -- (\envW+\pl, \ybra) node[above]{$\tilde a_j^{\prime}$};
\draw[bleg] (\envW, \ympo) -- (\envW+\pl, \ympo) node[below, blue]{$c_j$};
\draw (\envW, \yket) -- (\envW+\pl, \yket) node[below]{$\tilde a_j$};

\node[font=\Large] at (1.7, \ympo) {\textbf{=}};

\def\xLj{3.0}
\def\xSite{5.2}

\filldraw[fill=teal!40, rounded corners=3pt, draw=black]
  (\xLj-\envW, \ympo-\envH) rectangle (\xLj+\envW, \ympo+\envH);
\node at (\xLj,\ympo) {$L_{O,j}^{({\color{red}m,n})}$};

\draw (\xLj+\envW, \ybra) -- (\xSite-\rad, \ybra)
      node[midway,above]{$a_{j-1}^{\prime}$};
\draw[bleg] (\xLj+\envW, \ympo) -- (\xSite-\bs, \ympo)
      node[midway, below, blue]{$c_{j-1}$};
\draw (\xLj+\envW, \yket) -- (\xSite-\rad, \yket)
      node[midway,below]{$a_{j-1}$};

\filldraw[fill=green!35, rounded corners=1pt, draw=black]
  (\xSite-\bs,\ympo-\bs) rectangle (\xSite+\bs,\ympo+\bs);
\node at (\xSite,\ympo){$W_{O,j}^{({\color{red}m,n})}$};

\filldraw[fill=orange!40, draw=black] (\xSite,\ybra) circle (\rad);
\node at (\xSite,\ybra){$\tilde{M}_{X,j}^{({\color{red}n})*}$};

\filldraw[fill=blue!40, draw=black] (\xSite,\yket) circle (\rad);
\node at (\xSite,\yket){$\tilde{M}_{X,j}^{({\color{red}m})}$};

\draw[vleg] (\xSite,\ympo+\bs) -- (\xSite,\ybra-\rad);
\node[right, green!60!black] at (\xSite,{(\ympo+\bs+\ybra-\rad)/2}) {$\sigma_j'$};
\draw[vleg] (\xSite,\ympo-\bs) -- (\xSite,\yket+\rad);
\node[right, green!60!black] at (\xSite,{(\ympo-\bs+\yket+\rad)/2}) {$\sigma_j$};

\draw (\xSite+\rad,\ybra) -- (\xSite+\rad+\pl,\ybra) node[above]{$\tilde{a}_j^{\prime}$};
\draw[bleg] (\xSite+\bs, \ympo) -- (\xSite+\bs+\pl, \ympo) node[below, blue]{$c_j$};
\draw (\xSite+\rad,\yket) -- (\xSite+\rad+\pl,\yket) node[below]{$\tilde{a}_j$};

\end{tikzpicture}%
}}}
\end{equation}

\noindent
with $O=\mathcal{C}$ and $O=\mathcal{C}^2$. 
The right environments are updated analogously during left sweeps. 
The same contractions initialize the environments from $\floquetket{X_0}$ in step~1 of Algorithm~\ref{alg:dmrg}.

\begin{figure*}[t!]
\centering
\tikzset{
  f7pt/.style={circle,fill,inner sep=1.7pt},
  f7open/.style={circle,draw,fill=white,inner sep=1.5pt,line width=0.8pt},
  f7plane/.style={draw=blue!70!black,fill=blue!8,line width=0.9pt},
  f7ker/.style={->,draw=red!80!black,line width=0.9pt,>=stealth},
  f7L/.style={->,draw=green!55!black,line width=0.9pt,>=stealth,shorten >=2.5pt,shorten <=2.5pt},
  f7Llab/.style={font=\footnotesize,green!55!black},
  f7Lp/.style={->,draw=magenta!80!black,line width=0.9pt,>=stealth,shorten >=2.5pt,shorten <=2.5pt},
  f7Lplab/.style={font=\footnotesize,magenta!80!black},
  f7solve/.style={->,draw=black,line width=1.1pt,>=stealth,shorten >=2.5pt,shorten <=2.5pt},
  f7dash/.style={densely dashed,draw=gray!80!black,line width=0.6pt},
  f7sol/.style={draw=orange!90!black,line width=1.0pt},
  f7lab/.style={font=\footnotesize},
  f7title/.style={font=\footnotesize,anchor=west},
}
\resizebox{0.32\textwidth}{!}{\begin{tikzpicture}[scale=0.7,baseline]
  \useasboundingbox (-2.0,-2.4) rectangle (6.9,3.6);
  \node[f7title] at (-2.0,3.3) {(a) global problem: $\mathcal{L}^2X=\partial H$};
  \node[f7title] at (-2.0,-2.1) {consistent: $\partial H\in\im\mathcal{L}^2=\im\mathcal{L}$};
  \fill[f7plane] (-1.6,-1.6) -- (0.5,-0.7) -- (0.5,2.5) -- (-1.6,1.6) -- cycle;
  \node[blue!70!black,f7lab] at (0.3,2.85) {$\im\mathcal{L}$};
  \draw[f7ker] (0.0,-0.9) -- (5.2,-0.9) node[right,f7lab,red!80!black] {$\ker\mathcal{L}$};
  \node[f7pt] at (0.0,-0.9) {}; \node[below,f7lab] at (0.0,-0.9) {$0$};
  \coordinate (O) at (0.0,-0.9);
  \coordinate (HF) at (1.4,-0.9);
  \coordinate (AE) at (3.2,-0.9);
  \coordinate (H) at (3.2,1.3);
  \coordinate (A) at (0.0,1.3);
  \coordinate (AF) at (1.8,1.3);
  \coordinate (dH) at (-1.0,-0.5);
  \coordinate (Xs) at (-0.8,1.8);
  \coordinate (X) at (4.4,1.8);
  \node[f7pt,red!80!black] at (HF) {}; \node[below,f7lab] at (HF) {$\HF$};
  \node[f7pt,red!80!black] at (AE) {}; \node[below,f7lab] at (AE) {$\mAE$};
  \draw[f7dash] (AE) -- (H) -- (A);
  \draw[f7dash] (HF) -- (AF);
  \node[f7pt,violet] at (H) {}; \node[below right,f7lab] at (H) {$H$};
  \node[f7pt,blue!70!black] at (A) {}; \node[above right,f7lab] at (A) {$\agp_\mAE$};
  \node[f7pt,violet] at (AF) {}; \node[below left,f7lab] at (AF) {$\AF$};
  \node[f7pt,blue!70!black] at (dH) {}; \node[below,f7lab] at (dH) {$\partial H$};
  \draw[f7sol] (Xs) -- (5.2,1.8);
  \node[f7open] at (Xs) {}; \node[above left,f7lab] at (Xs) {$X_\ast$};
  \node[f7pt,orange!90!black] at (X) {}; \node[above,f7lab] at (X) {$X$};
  \draw[f7solve] (H) to[bend right=25] node[below right,f7lab,pos=0.6] {DMRG} (X);
  \draw[f7L] (A) to[bend right=30] node[left,f7Llab,pos=0.5] {$\mathcal{L}$} (dH);
  \draw[f7Lp] (dH) to[bend right=30] node[right,f7Lplab,pos=0.4] {$\mathcal{L}^+$} (A);
  \draw[f7L] (Xs) to[bend right=25] node[left,f7Llab,pos=0.5] {$\mathcal{L}$} (A);
  \draw[f7L] (H) to[bend left=12] node[below,f7Llab,pos=0.3] {$\mathcal{L}$} (dH);
  \draw[f7Lp] (H) to[bend right=10] node[above,f7Lplab,pos=0.55] {$\mathcal{L}^+$} (Xs);
  \draw[f7L] (X) to[right=30] node[above,f7Llab,pos=0.5] {$\mathcal{L}$} (A);
\end{tikzpicture}}
\hfill
\resizebox{0.32\textwidth}{!}{\begin{tikzpicture}[scale=0.7,baseline]
  \useasboundingbox (-2.0,-2.4) rectangle (6.9,3.6);
  \node[f7title] at (-2.0,3.3) {(b) local problem: $(\mathcal{L}^2)_{j,j+1}X_{j,j+1}=\partial H_{j,j+1}$};
  \node[f7title] at (-2.0,-2.1) {consistent: $\partial H_{j,j+1}\in\im(\mathcal{L}^2)_{j,j+1}$};
  \fill[f7plane] (-1.6,-1.6) -- (0.5,-0.7) -- (0.5,2.5) -- (-1.6,1.6) -- cycle;
  \node[blue!70!black,f7lab] at (0.3,2.85) {$\im(\mathcal{L}^2)_{j,j+1}$};
  \draw[f7ker] (0.0,-0.9) -- (5.2,-0.9) node[right,f7lab,red!80!black] {$\ker(\mathcal{L}^2)_{j,j+1}$};
  \node[f7pt] at (0.0,-0.9) {}; \node[below,f7lab] at (0.0,-0.9) {$0$};
  \coordinate (b) at (-1.0,1.1);
  \coordinate (Xs) at (-0.6,0.25);
  \coordinate (Xg) at (2.2,-0.4);
  \coordinate (Xsol) at (3.72,0.25);
  \node[f7pt,blue!70!black] at (b) {}; \node[left,f7lab] at (b) {$\partial H_{j,j+1}$};
  \draw[f7sol] (Xs) -- (5.2,0.25);
  \node[f7open] at (Xs) {}; \node[below,f7lab] at (Xs) {$X_{\ast,j,j+1}$};
  \draw[f7L] (Xs) to[bend left=30] node[left,f7Llab,pos=0.5] {$(\mathcal{L}^2)_{j,j+1}$} (b);
  \draw[f7L] (Xsol) to[bend right=25] node[above,f7Llab,pos=0.5] {$(\mathcal{L}^2)_{j,j+1}$} (b);
  \node[f7pt,violet] at (Xg) {}; \node[ left,f7lab] at (Xg) {$X_{j,j+1}^{\rm guess}$};
  \draw[f7solve] (Xg) -- node[below right,f7lab,pos=0.55] {GMRES} (Xsol);
  \node[f7pt,orange!90!black] at (Xsol) {}; \node[above right,f7lab] at (Xsol) {$X_{j,j+1}$};
\end{tikzpicture}}
\hfill
\resizebox{0.32\textwidth}{!}{\begin{tikzpicture}[scale=0.7,baseline]
  \useasboundingbox (-2.0,-2.4) rectangle (6.9,3.6);
  \node[f7title] at (-2.0,3.3) {(c) local problem: $(\mathcal{L}_{j,j+1})^2X_{j,j+1}=\partial H_{j,j+1}$};
  \node[f7title] at (-2.0,-2.1) {inconsistent: $\partial H_{j,j+1}\notin\im\mathcal{L}_{j,j+1}$};
  \fill[f7plane] (-1.6,-1.6) -- (0.5,-0.7) -- (0.5,2.5) -- (-1.6,1.6) -- cycle;
  \node[blue!70!black,f7lab] at (0.3,2.85) {$\im\mathcal{L}_{j,j+1}$};
  \draw[f7ker] (0.0,-0.9) -- (5.2,-0.9) node[right,f7lab,red!80!black] {$\ker\mathcal{L}_{j,j+1}$};
  \node[f7pt] at (0.0,-0.9) {}; \node[below,f7lab] at (0.0,-0.9) {$0$};
  \coordinate (b) at (3.0,1.5);
  \coordinate (bp) at (-0.6,1.5);
  \coordinate (Xs) at (-1.0,0.0);
  \coordinate (Xg) at (2.2,-0.4);
  \coordinate (LXg) at (-0.7,-0.9);
  \node[f7pt,violet] at (b) {}; \node[right,f7lab] at (b) {$\partial H_{j,j+1}$};
  \draw[f7dash] (b) -- (bp);
  \node[f7open] at (bp) {};
  \draw[decorate,decoration={brace,amplitude=4pt,raise=3pt},orange!90!black,line width=0.8pt] (bp) -- (b)
     node[midway,above=8pt,f7lab,orange!90!black] {$r_{\min}$};
  \node[f7pt,orange!90!black] at (Xs) {}; \node[below,f7lab] at (Xs) {$X_{j,j+1}$};
  \draw[f7L] (Xs) to[bend left=30] node[left,f7Llab,pos=0.6] {$(\mathcal{L}_{j,j+1})^2$} (bp);
  \draw[f7Lp] (b) to[bend left=15] node[right,f7Lplab,pos=0.4] {$[(\mathcal{L}_{j,j+1})^2]^{+}$} (Xs);
  \node[f7pt,violet] at (Xg) {}; \node[right,f7lab] at (Xg) {$X_{j,j+1}^{\rm guess}$};
  \draw[f7L] (Xg) to[bend left=12] node[above,f7Llab,pos=0.35] {$(\mathcal{L}_{j,j+1})^2$} (LXg);
  \node[f7pt,blue!70!black] at (LXg) {};
\end{tikzpicture}}
\caption{
\textbf{Schematic representation of global and local linear problems in Sambe superspace.}
Blue planes: image of the respective operator; red axes: its kernel; green arrows: application of the operator; magenta arrows: its pseudo-inverse; orange: solution spaces.
(a) Global problem, $\mathcal{L}^2X=\partial H$. The lab-frame Hamiltonian $H$ decomposes orthogonally into its kernel component $\mAE$ and its image component $\agp_\mAE$ (dashed parallelogram); the Floquet decomposition $H=\HF+\AF$ differs from it by a trade-off kernel component. The normal equation is solved by any point $X$ on the line $X_\ast+\ker\mathcal{L}$ (orange), where $X_\ast=\mathcal{L}^+H$ is the minimal-norm solution; the DMRG sweeps converge to one such point. We obtain the AGP via $\agp_\mAE=\mathcal{L}X$ irrespective of its kernel component.
(b) Local two-site problem obtained by reducing $\mathcal{L}^2$ (Sec.~\ref{app:sec:dmrg:consistency}). The reduced right-hand side lies in the image of the reduced operator and the linear system is consistent; its solutions form the line $X_{\ast,j,j+1}+\ker(\mathcal{L}^2)_{j,j+1}$. GMRES moves the initial guess only along the image (black arrow, parallel to the blue image plane); the kernel component of the guess is preserved.
(c) Local two-site problem obtained by reducing $\mathcal{L}$ and squaring afterwards. The reduced right-hand side generically has a component outside the image of the reduced operator; the linear system is inconsistent: a minimal-residual solver returns the minimum-residual solution $X_{j,j+1}=[(\mathcal{L}_{j,j+1})^2]^+\partial H_{j,j+1}$, whose residual $r_{\min}$ cannot be removed by further iterations.
}
\label{fig:local_vs_global-problem}
\end{figure*}
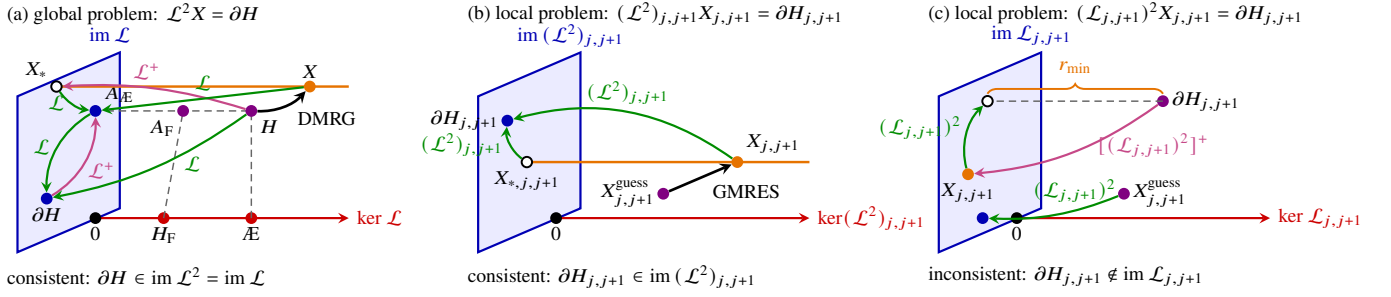

\subsubsection{Consistency of the reduced problem}
\label{app:sec:dmrg:consistency}

To see that the reduced problem~\eqref{app:eq:local} always has a solution, let $\Phi_{j,j+1}$ denote the linear map $X_{j,j+1}\mapsto\floquetket{X}$ which embeds the two-site tensor into the FMPS $\floquetket{X}$, all other tensors being held fixed; its Hermitian conjugate $\Phi_{j,j+1}^\dagger$ reduces the global problem to the bond $(j,j+1)$. 

By construction, cf.~Eqs.~\eqref{app:eq:Aeff}--\eqref{app:eq:localLHS}, the effective operator is the reduction of the \emph{squared} Liouvillian, $(\mathcal{L}^2)_{j,j+1}=\Phi_{j,j+1}^\dagger\mathcal{L}^2\Phi_{j,j+1}$, and the right-hand side $\partial H_{j,j+1}=\Phi_{j,j+1}^\dagger\floquetket{\partial H}$. 
Using $\mathcal{L}=\mathcal{L}^\dagger$ and $\floquetket{\partial H}=\mathcal{L}\floquetket{H}$, Eq.~\eqref{app:eq:local} reads as
\begin{equation}
\label{app:eq:local2}
  \Phi_{j,j+1}^\dagger\mathcal{L}^\dagger\mathcal{L}\,\Phi_{j,j+1}X_{j,j+1}
  = \Phi_{j,j+1}^\dagger\mathcal{L}^\dagger\floquetket{H}.
\end{equation}
It can be further recast as $K^\dagger K X_{j,j+1}=K^\dagger\floquetket{H}$ with $K=\mathcal{L}\Phi_{j,j+1}$. 
The right-hand side lies in $\im(K^\dagger)$. 
Since $\norm{Kx}^2=\langle x,K^\dagger Kx\rangle$, we have $\ker(K^\dagger K)=\ker(K)$, and taking orthogonal complements in the finite-dimensional space of two-site tensors gives $\im(K^\dagger K)=\im(K^\dagger)$. 
Hence $\partial H_{j,j+1}\in\im((\mathcal{L}^2)_{j,j+1})$, and Eq.~\eqref{app:eq:local} has a solution for every bond and at every stage of the sweep; the canonical form of the FMPS is not required for this argument. 

Equation~\eqref{app:eq:local2} also identifies the reduced problem as the normal equation of the restricted least-squares problem 
\begin{equation}
    \min_{X_{j,j+1}}\norm{\mathcal{L}\floquetket{X}-\floquetket{H}}^2,
\end{equation}
i.e., of the global cost function~\eqref{app:eq:normal_eq} with all tensors but $X_{j,j+1}$ held fixed. It is this step that makes the sweep a proper local minimization. 
The solution is unique modulo $\ker(K)$, the space of two-site tensors whose embedding lies in $\ker\mathcal{L}$; these directions leave the residual unchanged and are the local remnant of the gauge freedom of Sec.~\ref{app:sec:varl:gauge}. 
GMRES started from the guess~\eqref{app:eq:Xguess} modifies the two-site tensor only along $\im(K^\dagger)=\ker(K)^\perp$, so that the kernel component of the guess is retained, see Fig.~\ref{fig:local_vs_global-problem}(b).

By Eq.~\eqref{app:eq:normal_eq_cost}, the sweeps minimize $\norm{\tilde\mAE-\mAE}$, whereas the residual  $\mathcal{R}$ measures $\norm{\mathcal{L}(\tilde\mAE-\mAE)}$; the residual therefore bounds the error of the operator only up to the smallest nonzero eigenvalue of $\mathcal{L}$, which closes with the system size.

The order of the squaring and the reducing operations matters. 
Reducing $\mathcal{L}$ to the two-site subspace first and squaring afterwards gives $(\Phi_{j,j+1}^\dagger\mathcal{L}\Phi_{j,j+1})^2=K^\dagger\Phi_{j,j+1}\Phi_{j,j+1}^\dagger K$, whose image generically does not contain $\partial H_{j,j+1}$, since $\Phi_{j,j+1}\Phi_{j,j+1}^\dagger$ projects the intermediate vector onto the two-site subspace and discards the part of $\mathcal{L}\floquetket{H}$ outside it. 
The reduced problem then becomes inconsistent, see Fig.~\ref{fig:local_vs_global-problem}(c), and a minimal-residual solver returns a minimum-residual rather than an exact solution. 
We verified that sweeps built on such inconsistent local problems, even when solved with a minimal-norm solver such as MINRES-QLP~\cite{choi2011minresqlp}, produce a wrong global solution. 
The precomputation of $(\mathcal{C}^2)^{(n)}$ in Sec.~\ref{app:sec:mps_algo:L}, at the cost of an MPO bond dimension $\chi_{\mathcal{C^2}} > \chi_{\mathcal{C}}$, is the price to pay for having a consistent local problem.

\subsubsection{Contraction order and cost}
\label{app:sec:dmrg:cost}

The linear operation~\eqref{app:eq:localLHS} is performed multiple times inside GMRES at every bond; hence, its implementation determines the run time. 
Let the bond dimensions of $\floquetket{X}$ and of the FMPO $O$ be $\chi_X$ and $\chi_O$, respectively, and let $d$ be the local dimension of the operator chain of length $2L$. 
We evaluate the network in Eq.~\eqref{app:eq:diag_local_op} in the order
\begin{equation}
  \label{app:eq:contraction_order}
  \Bigl[\bigl\{\bigl(L_{O,j} W_{O,j}\bigr)\bigl(M_{X,j}M_{X,j+1}\bigr)\bigr\}\bigl\{W_{O,j+1}R_{O,j+1}\bigr\}\Bigr]
\end{equation}
for every pair of photon indices: i.e., we first absorb the FMPO tensors into the environments, then contract the left block with the two-site tensor, and finally contract the result with the right block over the shared MPO bond $c_j$ and the shared indices $\sigma_{j+1}$, $a_{j+1}$ of site $j+1$. 

The two contractions involving the two-site tensor dominate the cost.
In the first, the left block $L_{O,j}W_{O,j}$, viewed as a $\chi_X\chi_Od\times\chi_Xd$ matrix, is applied to the two-site tensor $X_{j,j+1}$ of size $\chi_Xd\times d\chi_X$; in the second, the resulting $\chi_Xd\times\chi_O d\chi_X$ object is contracted with the right block of size $\chi_Od\chi_X\times d\chi_X$. 
Each step costs $\chi_X^3\chi_Od^3$ operations, and the largest intermediate has $\chi_X^2\chi_Od^2$ entries; note that $\chi_O$ enters linearly, since the MPO bond $c_j$ is summed over and the result carries no MPO index. 
Instead, contracting the two blocks $L_{O,j}W_{O,j}$ and $W_{O,j+1}R_{O,j+1}$ with each other first would produce the effective operator as a $\chi_X^2d^2\times\chi_X^2d^2$ matrix, at a cost of $\chi_X^4\chi_Od^4$ operations for its construction and $\chi_X^4d^4$ for each application. The chosen order is thus cheaper by a factor $\chi_Xd/\chi_O$.

\begin{table}[t]
  \renewcommand{\arraystretch}{1.3}
  \begin{ruledtabular}
  \begin{tabular}{lc}
    chain length & $2L$ \\
    local dimension & $d$ \\
    two-site tensor $X_{j,j+1}^{(m)}$ & $\chi_X^2d^2$ \\
    one application of $(\mathcal{L}^2)_{j,j+1}$ & $\chi_X^3\chi_Od^3$ \\
    largest intermediate & $\chi_X^2\chi_Od^2$ \\
    effective operator as a matrix & $\chi_X^4\chi_Od^4$ \\
    two-site steps per sweep in one direction & $2L-1$ \\
  \end{tabular}
  \end{ruledtabular}
  \caption{Cost of the two-site step of Algorithm~\ref{alg:two_site} per pair of photon indices. $\chi_X$ and $\chi_O$ denote the bond dimensions of $\floquetket{X}$ and of the FMPO $O\in\{\mathcal{C},\mathcal{C}^2\}$, and $d$ the local Hilbert space dimension.
  }
  \label{app:tab:cost}
\end{table}

\begin{figure}[t]
  \centering
  \includegraphics[width=\columnwidth]{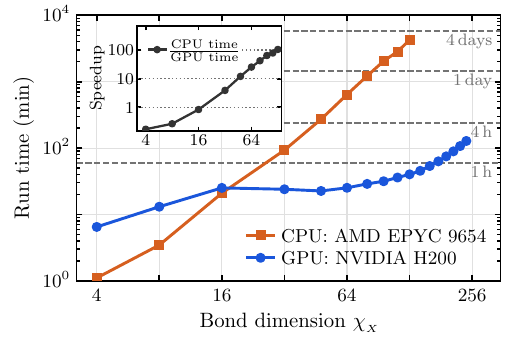}
  \caption{
Run time of the DMRG sweeps against the bond dimension $\chi_X$, on a CPU (orange squares) and on a GPU (blue circles). For the CPU run we use a single core of an AMD EPYC 9654 (Zen 4) processor, for the GPU run -- an NVIDIA H200 NVL (Hopper). 
Inset: ratio of the two run times. Timings refer to two complete sweeps at each $\chi_X$, whose output serves as the starting point for the next larger $\chi_X$. Data for the mixed-field Ising chain~\eqref{eq:Ising_chain} with $L=20$, $h_x/J=1.2$, $h_z/J=0.3$, and $TJ=0.5$.
The bond dimension $\chi_X$ is the maximum over all bonds of the $2L$-site chain, see Table~\ref{app:tab:cost}.
  }
  \label{app:fig:CPU_GPU}
\end{figure}

Since the same two-site tensor is contracted with both $O=\mathcal{C}$ and $O=\mathcal{C}^2$ and for all photon-index pairs, these operations are performed in batches on a GPU. 
Figure~\ref{app:fig:CPU_GPU} shows the resulting run times: the GPU evaluation is slower for small $\chi_X$, where the overhead dominates, but its advantage improves with increasing bond dimension and reaches a speed-up of about $105\approx10^2$ at $\chi_X=128$.

\subsection{From the gauge potential to the Floquet ground state}
\label{app:sec:mps_algo:gs}

Once the DMRG sweeps have converged, the gauge potential is obtained as $\floquetket{\agp_\mAE}=\mathcal{L}\floquetket{X_\ast}$ by a single application of Eq.~\eqref{app:eq:Lx}; the result is an FMPS of bond dimension more than $\chi_X$, which we compress using singular value decomposition. 

Splitting the combined physical index of each site restores the operator form. The harmonics $\tilde\mAE^{(n)}=H^{(n)}-\agp_\mAE^{(n)}$ define the approximate average-energy operator as an MPO at any time $t$ through Eq.~\eqref{app:eq:IFT}. 
Since $\tilde\mAE$ is Hermitian only up to the truncation and round-off errors of the sweeps, we symmetrize it, $\tilde\mAE\leftarrow(\tilde\mAE+\tilde\mAE^\dagger)/2$, before proceeding. 

We then run a standard two-site DMRG ground-state search~\cite{White1992,Schollwock2011} on $\tilde\mAE[0]$, which yields the approximate Floquet ground state $|\tilde n^\F_\text{GS}[0]\rangle$ and its average energy $\mae_0$; the average-energy gaps reported in the main text require, in addition, the lowest excited states of $\tilde\mAE[0]$, which we obtain from the same DMRG code by targeting excited states. 
Since $\tilde\mAE$ is only approximately a dynamical invariant, cf.~Sec.~\ref{app:sec:varl:diagnostics}, we measure the quality of $|\tilde{n}^\F_\text{GS}[0]\rangle$ using the stroboscopic infidelity under the exact dynamics generated by $H(t)$. To achieve close-to-machine-precision stroboscopic infidelity, the state is further improved using the Krylov--Arnoldi procedure of Appendix~\ref{app:sec:krylov}. All gaps, correlators, order parameters, entanglement data, and holonomies reported in the main text are evaluated on Krylov-converged states using the exact lab-frame generator $H(t)$.
For the exactly solvable XY chain of Sec.~\ref{sec:XY}, the variational results are in addition compared with the exact free-fermion solution for open boundary conditions, Sec.~\ref{app:subsec:XY_obc}.

\subsection{Alternative formulations explored}
\label{app:sec:mps_algo:alternatives}

The algorithm of Sec.~\ref{app:sec:dmrg} is the most stable and efficient implementation we found; it is, however, not the only one we tried. 
Here, we report on the behavior of the DMRG implementations of the other formulations in Table~\ref{app:tab:formulations} which we explored originally, together with our observations; whether the issues encountered can be overcome is currently an open question.

\paragraph*{Petrov--Galerkin projection.}
The main drawback of the normal equation is the conditioning of the squared Liouvillian $\mathcal{L}^2$, whose gap above zero closes exponentially with the system size for nonintegrable systems, cf.~Sec.~\ref{app:sec:varl:formulations}. 
Hence, it would be preferable to work with a single application of $\mathcal{L}$, i.e., to solve $\mathcal{L}X=\partial H$ directly. 
The DMRG adaptation of the Petrov--Galerkin condition~\eqref{app:eq:AGP_galerkin} is obtained by forming the environments with $\mathcal{L}$ instead of $\mathcal{L}^2$, so that the reduced problem at each bond reads $\mathcal{L}_{j,j+1}X_{j,j+1}=\partial H_{j,j+1}$ with $\mathcal{L}_{j,j+1}=\Phi_{j,j+1}^\dagger\mathcal{L}\Phi_{j,j+1}$. 

We encountered two issues. 
First, the reduced equation is the stationarity condition of an indefinite quadratic form rather than of a cost function, so that the sweep searches for a saddle point; to our knowledge, there is no convergence guarantee for DMRG in this setting. 
Second, the local Krylov updates do not preserve $X\in\im\mathcal{L}$, which global Krylov methods guarantee. 
We observed both failure modes: in some runs the sweeps did not converge, and when they did, the solution generically had a component in $\ker\mathcal{L}$. 
The second issue can be circumvented as discussed in Sec.~\ref{app:sec:varl:formulations}(v), by solving for the least-squares pre-image of $H$; globally, MINRES-QLP~\cite{choi2011minresqlp} handles such singular Hermitian problems efficiently, and is designed to return the unique minimal-norm least-squares solution. 

We emphasize that this property holds at the local two-site level as well: at every bond, MINRES-QLP returns the unique minimal-norm solution of the reduced problem $\mathcal{L}_{j,j+1}X_{j,j+1}=\partial H_{j,j+1}$, even when the latter is inconsistent, see Sec.~\ref{app:sec:dmrg:consistency}. 
Nevertheless, we found that the sweeps built from these local minimal-norm solutions do not converge to the global solution, presumably because no cost function is being minimized and the local minimal-norm condition does not translate into the global one. 

Modified local updates, including a different gauge for the FMPS, or additional constraints on the two-site tensor, could restore convergence; constructing such an update, or showing that none exists, remains open.

\paragraph*{Lagrange multiplier formalism.}
The constrained formulation~\eqref{app:eq:AGP_lagrange} enforces $X\in\im\mathcal{L}$ through the multiplier $\lambda$ at the expense of doubling the unknowns. 
We implemented it with two independent FMPSs for $X$ and $\lambda$, and with the Petrov--Galerkin projection of the linear system, which requires four types of environments, corresponding to $\floquetbraket{\lambda}{X}$, $\floquetbra{\lambda}\mathcal{L}\floquetket{X}$, $\floquetbra{X}\mathcal{L}\floquetket{X}$, and $\floquetbraket{X}{\partial H}$. 
The sweeps again did not find the saddle point of the system: in almost all runs the local updates quickly converged to an $X$ solving $\mathcal{L}X=\partial H$, but the constraint $X=\mathcal{L}\lambda$ was not satisfied at convergence, i.e., the gauge problem persisted. 
We observed this behavior already for small chains of $L\lesssim 5$ spins in the exactly solvable driven XY model, see Sec.~\ref{app:subsec:XY}. 

We also implemented the least-squares variant (ii$'$), Eq.~\eqref{app:eq:AGP_lagrange_normal}, which does derive from a cost function. 
It works in principle, but the reduced problems are considerably less well-behaved than those of Sec.~\ref{app:sec:dmrg}, requiring a few orders of magnitude more Krylov iterations per bond; within reasonable exploration times, we could not go beyond $L=5$ sites; hence, we did not pursue this direction for larger systems. 
The normal equation is expected to be the more stable of the two variants (ii) and (ii') [see Sec.~\ref{app:sec:varl:formulations}], and the direct control over $\norm{\mathcal{L}X-\partial H}$ may yield smaller residuals. The problem with the conditioning of $\mathcal{L}^2$, however, persists.

\subsection{Directions not explored}
\label{app:sec:mps_algo:outlook}

We close with directions which we consider promising but have not pursued.
\begin{enumerate}
  \item \emph{Solver.} Since $(\mathcal{L}^2)_{j,j+1}$ is Hermitian and positive semidefinite, MINRES~\cite{paige1975minres} could replace GMRES in the two-site step at lower memory cost; we have not benchmarked the two. Moreover, the reduced problems need not be solved to full convergence in the early sweeps and at small $\chi_X$, as is common practice in DMRG eigensolvers; an inexact Krylov strategy may reduce the run time substantially.
  \item \emph{Convolutions.} The sums over harmonics in Eqs.~\eqref{app:eq:conv2} and~\eqref{app:eq:localLHS} are convolutions and could be evaluated with fast Fourier transforms, scaling as $n_c\log n_c$ instead of $n_c^2$; for the small cutoffs used here the overhead is likely to dominate, but the option becomes relevant for drives with many harmonics.
  \item \emph{Tikhonov regularization.} The unconstrained functional~\eqref{app:eq:AGP_tikhonov} is suited to gradient-based optimization of nonlinear ans\"atze beyond matrix product states. This is particularly interesting in higher dimensions.
  \item \emph{Kernel projector.} Instead of computing the gauge potential, one may target the kernel projector $\identity-\mathcal{L}^+\mathcal{L}$ directly, e.g., with the projector algorithms of Ref.~\cite{kuo2022computing}, and apply it to the lab frame Hamiltonian $H$.
  \item \emph{Interleaved sweeps.} The linear solve for $X$ and the ground-state search for $\tilde\mAE$ could be interleaved, updating the ground state as $X$ converges; the DMRG implementation for sums of operators, $\tilde\mAE=H-\mathcal{L}X$, as available in ITensor~\cite{itensor, itensor-r0.3}, already provides the required machinery.
  \item \emph{Ground-state-only variational principle.} The gauge potential $\agp_\mAE$ contains more information than needed for the Floquet ground state, since it encodes the transitions among all Floquet states. For nonintegrable many-body systems, it is a nonlocal operator which leads to large residual $\mathcal{R}$ when the bond dimension is insufficient; for small mixed-field Ising chains, we observed that the residual is kept to zero only when the bond dimension $\chi_X$ is close to its maximal possible value.
  A variational principle targeting the ground state of $\mAE$ directly with the operator defined only implicitly through Eq.~\eqref{app:eq:orthogonal_decomp}, or a quasilocal ansatz for the average-energy ground state only, would avoid this redundancy; formulating useful cost functions and variational principles with these properties is an open problem.
\end{enumerate}

\section{Krylov-Arnoldi optimization of Floquet eigenstates}
\label{app:sec:krylov}
In the previous section, we described how we approximate $\mAE[t]$ using the FDMRG method. Given $\mAE$, its low-energy eigenstates can be calculated via the DMRG method used for ground state search. In this section, we discuss how we can further optimize these states using an iterative Krylov-Arnoldi method to improve the numerical accuracy.

\subsection{Fidelity and spectral transformation}
\label{app:sec:krylov-shift}
Given a normalized trial state $\ket{\psi[t]}$, we define the stroboscopic Floquet overlap
\begin{equation}
  z = 
  \melem{\psi[t]}{U_F[t]}{\psi[t]},
  \label{app:eq:overlap}
\end{equation}
where $U_F[t] = U(t+T;t)$, and from it, the fidelity and quasienergy phase
\begin{equation}
  F(T) = \lvert z\rvert^{2},
  \qquad
  \varphi = \arg z.
  \label{app:eq:fidelity}
\end{equation}
$F(T)=1$ if and only if $\ket{\psi[t]}$ is an exact eigenstate of $U_F[t]$; the infidelity $1-F(T)$ therefore serves both as the error measure and as the convergence criterion. The phase $\varphi$ estimates the quasienergy $\epsilon_\F$ of the
target state through $\varphi = T\epsilon_\F \bmod 2\pi$.

Because $U_F[t]$ is unitary, its spectrum lies on the unit circle. That means no eigenvalue is dominant in modulus; a power iteration does not converge to a nonexistent extremal value. Targeting must therefore be defined with respect to a reference point on the circle. We supply the reference point by the phase shift
\begin{equation}
  \tilde{U}_F[t] = e^{-\mathrm{i}\varphi}\,U_F[t] ,
  \label{app:eq:shift}
\end{equation}
which moves the eigenvalue associated with the current trial state to the neighborhood of $+1$, so that the target becomes the eigenvalue of $U_F[t]$ closest to $+1$.

\subsection{Arnoldi iteration}
\label{app:sec:krylov-arnoldi}

Starting from $\ket{v_1}=\ket{\psi[t]}/\lVert\ket{\psi[t]}\rVert$, we build the Krylov subspace
\begin{equation}
  \mathcal{K}_{m}(\tilde{U}_F,v_1) =
  \begin{multlined}[t]
    \bigl\{\ket{v_1},\, \tilde{U}_F[t]\ket{v_1},\, \tilde{U}_F^{2}[t]\ket{v_1},\\
    \dots,\, \tilde{U}_F^{m-1}[t]\ket{v_1}\bigr\}
  \end{multlined}
  \label{app:eq:krylov}
\end{equation}
by the Arnoldi recursion with Gram--Schmidt orthogonalization. At step $k$ we form $\ket{w}=\tilde{U}_F[t]\ket{v_k}$ and then, for $j=1,\dots,k$,
\begin{align}
  H_{jk} &= \braket{v_j}{w},
  \label{app:eq:arn-h}\\
  \ket{w} &\leftarrow \ket{w} - H_{jk}\ket{v_j},
  \label{app:eq:arn-mgs}
\end{align}
followed by
\begin{equation}
  H_{k+1,k} = \lVert \ket{w} \rVert,
  \qquad
  \ket{v_{k+1}} = \ket{w}/H_{k+1,k},
  \label{app:eq:arn-norm}
\end{equation}
for $k=1,\dots,m-1$. 

The
iteration terminates early if $H_{k+1,k} < \epsilon_{\text{tol}}$, where $\epsilon_{\text{tol}}$ is a given tolerance. This signals that the Krylov subspace $\mathcal{K}_k$ has become invariant under $\tilde{U}_F$. After $n$ completed steps, the recursion realizes the Arnoldi factorization
\begin{equation}
  \tilde{U}_F[t]V_n = V_n H_n + H_{n+1,n}\ket{v_{n+1}}e_n^{\mathsf{T}},
  \label{app:eq:arnoldi-fact}
\end{equation}
with $V_n=[\ket{v_1},\dots,\ket{v_n}]$, $H_n\in\mathbb{C}^{n\times n}$ an upper
Hessenberg matrix and \(e_{n}^{T}\) is the transpose of the $n$-th standard basis vector.

\subsection{Eigenvalue selection rule}
\label{app:sec:krylov-ritz}
The Hessenberg matrix built during the recursion is
\begin{equation}
  V_n^{\dagger} \tilde{U}_F[t] V_n = H_n.
  \label{app:eq:rayleigh-quotient}
\end{equation}
To find the target eigenstate, we solve the eigenvalue problem
\begin{equation}
  H_n\,y^{(i)} = \lambda_i\,y^{(i)},
  \qquad i=1,\dots,n .
  \label{app:eq:ritz}
\end{equation}
Eq.~\eqref{app:eq:ritz} is solved by exact diagonalization on the $n$-dimensional space $\mathbb{C}^n$.

Following the spectral transformation~\eqref{app:eq:shift}, the target Ritz value is the one whose eigenvalue lies closest to $+1$,
\begin{equation}
  i_{\star} = \arg\min_{i}\,\lvert \lambda_i - 1\rvert ,
  \label{app:eq:select}
\end{equation}
and the refined state is the corresponding Ritz vector, created using a linear combination of the Krylov basis,
\begin{equation}
  \ket{\psi_\star[t]} = \sum_{j=1}^{n} y^{(i_\star)}_j \ket{v_j}.
  \label{app:eq:ritzvec}
\end{equation}
Equation~\eqref{app:eq:select} targets the eigenstate closest to the input state rather than an arbitrary state of the spectrum. The initial state supplied by the ground state of the average-energy operator found using DMRG in Sec.~\ref{app:sec:mps_algo} is therefore instrumental, especially in the nonperturbative `low'-frequency DTC regime.

The method described above can be numerically simulated using the following algorithm.
\begin{algorithm}[H]
\caption{Krylov-Arnoldi refinement of a Floquet eigenstate}
\label{alg:krylov}
\begin{algorithmic}[1]
\REQUIRE initial Floquet GS guess $\ket{\psi[t]}$, Floquet unitary $U_F[t]$, tolerance $\epsilon_{\text{tol}}$, maximum Krylov dimension $m$
\ENSURE refined state $\ket{\psi_\star[t]}$ and its fidelity $F_\star$
  \STATE $z \gets \melem{\psi[t]}{U_F[t]}{\psi[t]}$
  \STATE $F \gets \lvert z\rvert^{2}$; \quad $\varphi \gets \arg z$
  \STATE $U_F[t] \gets e^{-\mathrm{i}\varphi}U_F[t]$
         \hfill $\triangleright$ {target eigenvalue rotated to $+1$}
  \STATE $\ket{v_1} \gets \ket{\psi}/\lVert\ket{\psi}\rVert$
  \FOR{$k = 1,\dots,m-1$}
     \STATE $\ket{w} \gets U_F[t]\ket{v_k}$
     \FOR{$j = 1,\dots,k$}
        \STATE $H_{jk} \gets \braket{v_j}{w}$;
        \STATE $\ket{w} \gets {\ket{w} - H_{jk}\ket{v_j}}$
        \hfill $\triangleright$ {Gram--Schmidt orthogonalization}
     \ENDFOR
     \STATE $H_{k+1,k} \gets \lVert \ket{w}\rVert$
     \IF{$H_{k+1,k} < \epsilon_{\text{tol}}$} \STATE \textbf{break}
        \hfill $\triangleright$ {invariant subspace reached} 
     \ENDIF
     \STATE $\ket{v_{k+1}} \gets \ket{w}/H_{k+1,k}$
  \ENDFOR
  \STATE solve $H y^{(i)} = \lambda_i y^{(i)}$ \hfill $\triangleright$ {$i$ is the eigenstate index}
  \STATE $\ket{\psi_\star[t]} \gets {\sum_{j} y^{(i_\star)}_j \ket{v_j}}$\\
  \hfill $\triangleright$ {$i_\star$ is the index of Ritz value closest to $+1$}
\STATE $F_\star$ = evaluate $F$ for the final state
\RETURN $\ket{\psi_\star[t]}, F_\star$
\end{algorithmic}
\end{algorithm}

\begin{figure}
  \centering
  \includegraphics[width=\columnwidth]{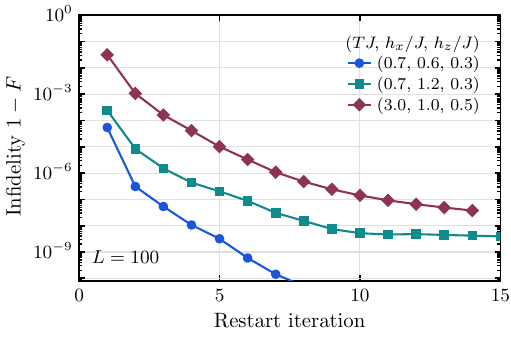}
  \caption{
    Convergence of the Krylov-Arnoldi optimization of the Floquet ground state, for the three parameter sets in the mixed-field Ising model~\eqref{eq:Ising_chain} of $L=100$ sites. The infidelity $1-F$ of the state entering each restart, with $F=F(T)=\bigl|\langle\psi[0]|{U}_F[0]|\psi[0]\rangle\bigr|^{2}$ for the two parameter sets with $TJ=0.7$ (F-AFM/F-PM regimes) and $F=F(2T) = \bigl|\langle\psi[0]|{U}^2_F[0]|\psi(0)\rangle\bigr|^{2}$ for the parameter set with $TJ=3.0$ (GS-FDTC regime),  measuring how far $\ket{\psi[t]}$ is from being an eigenstate of the Floquet unitary. Small values of $1-F$ indicate convergence. Each restart builds a Krylov space of maximum dimension $m=30$ , and the refined state is kept at bond dimension $\chi_{\psi} =70$. All three converge monotonically, but at rates that differ in each parameter regime. The first point for each parameter set corresponds to the raw state that comes from the ground state DMRG applied to $\tilde\mAE$, calculated using Floquet DMRG with $\chi_X =128$. The y-axis is cut to $10^{-10}$; the blue curve goes below $10^{-12}$ in the further iterations.
  }
  \label{app:fig:mfi_infidelity}
\end{figure}

In our simulations, we perform Algorithm~\ref{alg:krylov} several times for a single state to reach good numerical precision. In each iteration, we use the output of the previous iteration as trial $\ket{\psi[t]}$. Figure~\ref{app:fig:mfi_infidelity} shows the performance of this algorithm to refine the ground state of the mixed-field Ising model~\eqref{eq:Ising_chain} for a system of $L=100$ in different parameter regimes. In the main text we consider maximum Krylov dimension, maximum restart iteration and maximum bond dimension $\chi_\psi$ as $30$, $20$, and $70$ respectively.


\section{Additional data for the driven XY and Ising spin chains}
\label{app:sec:spin_models}

\subsection{Continuously driven XY chain}
\label{app:subsec:XY}

The circularly driven XY chain~\eqref{eq:XY_chain} plays, in Floquet ground-state phases, a similar role to the paradigmatic transverse-field Ising and XY chains for equilibrium quantum phase transitions. It is exactly solvable, exhibits ordered and critical Floquet ground states, and admits a closed-form analytical solution which enables us to compute exactly the Floquet-ground state phase diagram. 

\begin{figure*}[t]
  \centering
  \includegraphics[width=\textwidth]{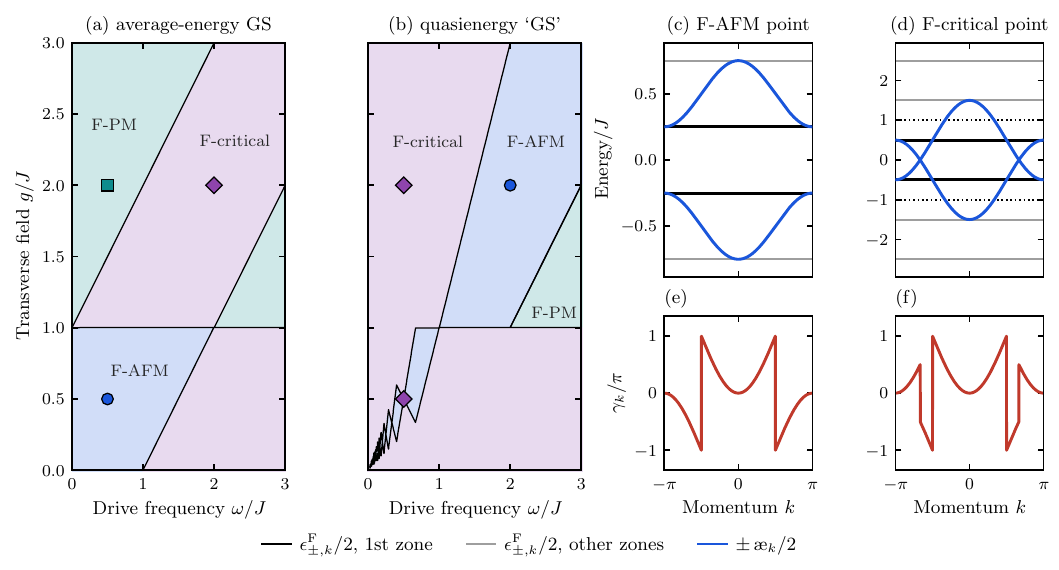}
  \caption{
  \textbf{Average-energy versus quasienergy ground state of the driven XY chain~\eqref{eq:gft:circXY}} at $A/J=1/2$.
  (a) Floquet ground-state phase diagram in the $(\omega/J, g/J)$ plane obtained from the average-energy operator, i.e., from the gap-closing condition~\eqref{eq:gft:circXY_gapclosing} and the winding criterion $|g-\omega|<J$ of Sec.~\ref{app:subsec:XY_phases}; same as Fig.~\ref{fig:XY}(f), reproduced for comparison. Symbols mark the F-AFM, F-PM, and F-critical parameter points of Fig.~\ref{fig:XY}.
  (b) Same, for the `ground state' obtained by filling the single-particle quasienergies in the first Floquet zone $(-\omega/2,\omega/2]$: critical wherever Eq.~\eqref{eq:gft:circXY_quasigap} holds for some $k$, and otherwise F-AFM (F-PM) for $|g-\omega|<J$ ($|g-\omega|>J$). The two diagrams agree at high frequency but differ in the adiabatic regime $\omega\ll J, g \gtrsim J$, where (b) shows a spurious critical region, and in the resonance regime $\omega\approx g$, where (b) shows order while the average-energy ground state is critical; the narrow slivers at $\omega/J\lesssim0.5$ in (b) are the gapped windows between successive photon resonances.
  (c), (d) Single-particle quasienergies $\epsilon^\F_{\pm, k}/2=\pm(\epsilon_k-\omega)/2$ (black: first Floquet zone, bounded by the dotted lines; gray: shifted by multiples of $\omega$) and average energies $\pm\mae_k/2$ (blue) at the F-AFM point $(g/J,\omega/J)=(0.5,0.5)$ (c) and at the F-critical point $(g/J,\omega/J)=(2.0,2.0)$ (d). At the F-AFM point, $\epsilon_k=2\omega$ for all momenta $k$: the two quasienergy bands coincide at the zone boundary and the folded spectrum is gapless, whereas the average-energy gap $\min_k|\mae_k|/2=0.25J$ stays open; at the F-critical point the average-energy bands cross at the photon resonance while the quasienergy bands remain separated.
  (e), (f) Geometric phases $\gamma_k$ of the lower average-energy band, Eq.~\eqref{app:eq:eps_ae_gamma}, folded into $(-\pi,\pi]$, for the same two points; $|\gamma_k|$ reaches $\pi$ where the quasienergy bands touch the zone boundary.
  Throughout the caption, the phase diagram points are denoted and referred to according to the average energy.}
  \label{fig:XY_app}
\end{figure*}

Here, we use this model with a two-fold purpose: 
(i) as a benchmark for the variational algorithm of Appendix~\ref{app:sec:mps_algo}, and 
(ii) to illustrate how the average-energy ground state differs from a na\"ive `ground state' defined through the quasienergies. 
Below, we recall the exact solution used in Figs.~\ref{fig:XY} and~\ref{fig:XY_app}; a more detailed study is presented in Ref.~\cite{schindler2026thesis}.

\subsubsection{Free-fermion solution for periodic boundary conditions}
\label{app:subsec:XY_pbc}

We consider the Hamiltonian~\eqref{eq:XY_chain} with periodic boundary conditions, 
\begin{equation}
    \label{eq:gft:circXY}
    H(t) = 
            \frac{1}{2}\sum_{n=1}^L \left[ 
            \left( \frac{J}{2} S^+_{n+1} S^-_{n} + A i e^{-i\omega t} S^+_{n+1} S^+_{n} + \mathrm{h.c.}\right) 
            + g S_n^z
            \right]\,,
\end{equation}
with spin-$1{/}2$ matrices $S_n^\alpha = \sigma^\alpha_n/2$, $\alpha=x,y,z$. 
The Jordan-Wigner transformation
\begin{equation}
\label{eq:JordanWigner}
    f_n = \prod_{j=1}^{n-1}\pqty{-\sigma^z_j} \sigma_n^{-}\,,
    \qquad
    f^\dagger_n f_n = S^z_n + \frac{1}{2} \,,
\end{equation}
followed by the Fourier transform $f_n^{\dagger}=L^{-1/2}\sum_k e^{ikn}f_k^\dagger$, maps the chain to free fermions~\cite{schindler2026thesis}, 
\begin{equation}
    \label{eq:gft:circXY_fermion}
    \begin{aligned}
        H(t)    & \hat{=} \sum_{k>0} \boldsymbol{\psi}_k^\dagger \boldsymbol{h}(k,t) \boldsymbol{\psi}_k\, ,\\
        \boldsymbol{h}(k,t)  &=
            \Delta_k\tau^z + A_k\left[ \cos(\omega t) \tau^x + \sin(\omega t) \tau^y \right]\, ,
    \end{aligned}
\end{equation}
where we introduced the momentum-dependent level splitting $\Delta_k=g + J \cos k$ and drive amplitude $A_k=2A\sin k$, the Nambu spinor $\boldsymbol{\psi}_k^\dagger=(f_k^\dagger, f_{-k})$, and the pseudo-spin operators $\tau^\alpha=\sigma^\alpha/2$; since $\boldsymbol\psi_k$ and $\boldsymbol\psi_{-k}$ are not independent, the sum runs over half the Brillouin zone $k>0$. 
With periodic boundary conditions on the spin chain, the mapping holds within each sector of the fermion parity $\prod_n(-\sigma^z_n)$, resulting in periodic (antiperiodic) boundary conditions for the fermion chain in the odd (even) sector.

For each $k>0$, the pair of modes $(k,-k)$ spans a four-dimensional space, in which the even-parity subspace $\{\ket{0_k0_{-k}},\ket{1_k1_{-k}}\}$ is a pseudo-spin on which $\boldsymbol\psi_k^\dagger\boldsymbol\tau\boldsymbol\psi_k$ acts as $\boldsymbol\tau$. 
The many-body problem thus reduces to a collection of circularly driven two-level systems, Eq.~\eqref{eq:gft:circXY_fermion}, one per $k>0$, for which Floquet theory and GFT can be worked out in closed form~\cite{schindler2025geometric,schindler2026thesis}. 
Passing to the rotating frame, $e^{i\omega t\tau^z}$, removes the time dependence from $\boldsymbol h(k,t)$ completely, leading to the static Hamiltonian $(\Delta_k-\omega)\tau^z+A_k\tau^x$, with eigenvalues $\pm\epsilon_k/2$, where $\epsilon_k^2=(\Delta_k-\omega)^2+A_k^2$. 
The time-periodic Floquet states are the rotated eigenvectors, $|n^\F_{\pm,k}[t]\rangle=e^{-i\omega t(\tau^z-1/2)}|n^\F_{\pm,k}[0]\rangle$.

Hence, all three decompositions of Sec.~\ref{app:sec:gft} are diagonal in the time-dependent unit vector
\begin{equation}
    \label{eq:gft:circXY_nk}
    \hat{\boldsymbol n}_k(t)=\frac{1}{\epsilon_k}\bigl(A_k\cos(\omega t),\,A_k\sin(\omega t),\,\Delta_k-\omega\bigr)\,,
\end{equation}
which precesses about the $z$-axis at the drive frequency $\omega$. 
Explicitly, the Floquet Hamiltonian and its AGP read as
\begin{equation}
\label{eq:gft:circXY_HF}
    \begin{aligned}
        \boldsymbol{h}_{\F,k}[t] &= \pqty{\epsilon_k - \omega}\,\hat{\boldsymbol n}_k(t)\cdot\boldsymbol\tau \, ,\\
        \boldsymbol{a}_{\F,k}(t) &= \omega\bqty{ \tau^z + \hat{\boldsymbol n}_k(t)\cdot\boldsymbol\tau }\, ,
    \end{aligned}
\end{equation}
and the average-energy operator and its AGP~\cite{schindler2025geometric}
\begin{equation}
\label{eq:gft:circXY_AE}
    \begin{aligned}
        \textbf{\ae}_k[t] &= \mae_k\,\hat{\boldsymbol n}_k(t)\cdot\boldsymbol\tau \, ,
        \quad
        \mae_k=\frac{\Delta_k(\Delta_k-\omega) + A_k^2}{\epsilon_k}\,,\\
        \boldsymbol{a}_{\text{\ae},k}(t) &= \frac{A_k\omega}{\epsilon_k}\,\hat{\boldsymbol m}_k(t)\cdot\boldsymbol\tau \, ,
    \end{aligned}
\end{equation}
where we denote by $\hat{\boldsymbol m}_k(t)=\epsilon_k^{-1}\bigl(-(\Delta_k-\omega)\cos(\omega t),\,-(\Delta_k-\omega)\sin(\omega t),\,A_k\bigr)$ the unit vector in the plane of $\hat{\boldsymbol n}_k$ and $\hat z$ perpendicular to $\hat{\boldsymbol n}_k$. 
The single-particle quasienergies $\epsilon^\F_{\pm,k}/2=\pm(\epsilon_k-\omega)/2 \mod \omega$ decompose into the average energies $\pm\mae_k/2$ and the corresponding geometric phases $\gamma_k/2$ by Eq.~\eqref{app:eq:eps_ae_gamma}. 
Note that $\mae_k$, in contrast to the quasienergy, carries a sign: it is the time-averaged energy of the Floquet state, Eq.~\eqref{app:eq:average-energies}, and tells which of the two Floquet states of the pseudo-spin $k$ is the lower one; 
in the infinite-frequency limit, $\mae_k\to\Delta_k$, which is the spectrum of the isotropic XX chain, as it should be for the period-averaged Hamiltonian, cf.~Sec.~\ref{app:sec:gft}.
The quasienergy, $\epsilon^\F_{\pm,k}$, by contrast, does not carry a sign since, due to the folding ambiguity, it can be shifted to be positive.  

The many-body Floquet ground state can be computed by placing every pseudo-spin into its lower average-energy state. 
Denoting by $s_k=\mathrm{sgn}(\mae_k)$ the average-energy sign and introducing the angle $\vartheta_k\in[0,\pi]$ through $(\sin\vartheta_k,\cos\vartheta_k)=s_k(A_k,\Delta_k-\omega)/\epsilon_k$, we obtain a BCS-type product state,
\begin{equation}
    \label{eq:gft:circXY_GS}
    \ket{n^\F_\text{GS}[t]} = \prod_{k>0}\Bigl[\cos\tfrac{\vartheta_k}{2}\,\ket{0_k0_{-k}} - e^{-i\omega t}\sin\tfrac{\vartheta_k}{2}\,\ket{1_k1_{-k}}\Bigr]\,,
\end{equation}
whose average energy and average-energy gap in the thermodynamic limit are
\begin{equation}
    \label{eq:gft:circXY_gap}
    \frac{\mae_{\rm GS}}{L} = -\frac{1}{4}\int_{-\pi}^{\pi}\frac{\dd k}{2\pi}\,\abs{\mae_k}\,,
    \qquad
    \Delta\mae = \frac{1}{2}\min_k\abs{\mae_k}\,.
\end{equation}
Here, $\Delta\mae$ is the energy of a single Bogoliubov quasiparticle, i.e., the gap between the ground state and the first excited state of $\mAE$ irrespective of the fermion parity. 
In the ordered phase, where the ground state is twofold degenerate in the thermodynamic limit (one state in each parity sector), this is the gap above this doublet to the even-parity excitation consisting of the zero mode and one bulk quasiparticle. 
Equation~\eqref{eq:gft:circXY_gap} provides the dashed lines in Fig.~\ref{fig:XY}(b); the corresponding finite-size gaps $\Delta\mae$ give the reference values in Fig.~\ref{fig:XY}(a).

The time dependence of Eq.~\eqref{eq:gft:circXY_GS} is a global rotation about the $z$-axis, $\ket{n^\F_\text{GS}[t]}=e^{-i\frac{\omega t}{2}\sum_n S^z_n}\ket{n^\F_\text{GS}[0]}$ up to a global phase; the same holds for every Floquet eigenstate. 
Since the drive term in Eq.~\eqref{eq:gft:circXY} raises the total $z$-magnetization $\sum_nS^z_n$ by two, we have $H(t)=e^{-i\frac{\omega t}{2}\sum_n S^z_n}H(0)\,e^{+i\frac{\omega t}{2}\sum_n S^z_n}$ with a rotation angle of $\omega t/2$; after one period the spins have rotated by $\pi$, which within a parity sector is a global phase, as required for Floquet states. 

Consequently, any order present at $t=0$ persists at all times, up to the corresponding rotation of the observable, $S^x_n\to\cos(\omega t/2)S^x_n+\sin(\omega t/2)S^y_n$; this is why in the main text we focus on the properties of the Floquet ground state at $t=0$. 

\subsubsection{Correlation function and Floquet ground-state phase diagram}
\label{app:subsec:XY_phases}

Let us now determine the order supported by the Floquet ground state.
The factor of $i$ in the drive term~\eqref{eq:gft:circXY} orients the anisotropy at $t=0$ along the $xy$-plane diagonal. Therefore, the pairing in the drive Eq.~\eqref{eq:gft:circXY_fermion} becomes real after rotating all spins by $\pi/4$ about the $z$-axis.
Hence, long-range order in the average-energy ground state appears at $t=0$ along the diagonal axis $x'=(x-y)/\sqrt2$, and can be measured using the correlation function 
$C^{x'}(r)=\mathrm{tr}\bigl(\Pi_0[0]\,S^{x'}_{n_0}S^{x'}_{n_0+r}\bigr)$, with $S^{x'}_n = (S^{x}_n - S^{y}_n)/\sqrt{2}$ and $y'=(x+y)/\sqrt2$. 

For $J,A>0$ the in-plane coupling is antiferromagnetic, and the order is staggered. 
Along the ordering axis $x'$, the Jordan-Wigner string maps the correlation function $C^{x'}(r)$ to the determinant of an $r\times r$ Toeplitz matrix $\mathsf{T}_r(f)$, as for the static XY chain~\cite{schindler2026thesis},
\begin{equation}
    \label{eq:gft:circXY_Cxx}
    \begin{aligned}
    C^{x'}(r) &= \frac{(-1)^r}{4}\det\bigl[\mathsf{T}_r(f)\bigr]\,,
    \\
    \bigl[\mathsf{T}_r(f)\bigr]_{nm} &= \int_{-\pi}^{\pi}\frac{\dd k}{2\pi}\, f(k)\,e^{-i(n-m)k},
    \end{aligned}
\end{equation}
with the symbol $f(k)=s_k\,e^{ik}\,[(\Delta_k-\omega)-iA_k]/\epsilon_k$, built from the pseudo-spin direction $s_k\hat{\boldsymbol n}_k$ of the Floquet ground state; the shift by $-\omega$ relative to the symbol of the instantaneous drive Hamiltonian $H(0)$ comes from the rotating-frame correction (i.e., the Galilean term). 

The asymptotic behaviour of the Toeplitz determinant is governed by the winding of $f(k)$ around the origin in momentum space: 
(i) if $f(k)$ is continuous and winds around the origin of the complex plane as $k$ traverses the Brillouin zone, $\abs{C^{x'}(r)}$ approaches a constant and the order is antiferromagnetic, $C^{x'}(r)\to(-1)^r\,\mathrm{const}$; 
(ii) if $f$ does not wind, $\abs{C^{x'}(r)}$ decays exponentially; 
and, (iii), if $\text{æ}_k$ changes sign for some $k$, i.e., at an average-energy gap closing point, the sign $s_k$ and hence $f(k)$ jump by a factor of $-1$; a symbol with jump discontinuities produces a power-law decay of $|C^{x'}(r)|$.
These are the three types of behaviour shown in Fig.~\ref{fig:XY}(d) of the main text: the F-AFM point (i) retains long-range order, the F-PM point (ii) decays exponentially, and the F-critical point (iii) decays algebraically.

The phase diagram of Fig.~\ref{fig:XY}(f) in the main text is obtained using two features on the single-particle average energy bands~\eqref{eq:gft:circXY_AE}. 
First, the average-energy gap closes when $\mae_k=0$ for some $k$, i.e., when the quadratic form in $z=\cos k$,
\begin{equation}
    \label{eq:gft:circXY_gapclosing}
    q(z) = (J^2-4A^2)\,z^2 + J(2g-\omega)\,z + g(g-\omega) + 4A^2 \,,
\end{equation}
obtained from $\Delta_k(\Delta_k-\omega)+A_k^2$ with $A_k^2=4A^2(1-z^2)$, has a zero in the interval $z\in[-1,1]$; the Floquet ground state is then critical. 
For the value $A/J=1/2$ used throughout this work, the quadratic term in $z^2$ above vanishes, and the condition $z\in[-1,1]$ reduces to $|g(g-\omega)+J^2|\leq J|2g-\omega|$. 
The inequality holds on a region of the $(g,\omega)$-plane rather than on a line, since the gap-closing momentum $k_0=\arccos z_0$ with $z_0=-[g(g-\omega)+J^2]/[J(2g-\omega)]$ moves across the Brillouin zone as the parameters are varied; the equality marks the boundaries of the critical phase, where $k_0$ reaches $0$ or $\pi$, i.e., the straight lines $g=\omega\pm J$ and $g=\pm J$ in Fig.~\ref{fig:XY}(f).

Second, away from gap-closing points, according to Eq.~\eqref{eq:gft:circXY_AE} the phase is set by the winding of the vector $s_k\hat{\boldsymbol n}_k(0)$ in the $xz$-plane as $k$ traverses the Brillouin zone; this occurs iff its $z$-component $\Delta_k-\omega$ changes sign between $k=0$ and $k=\pi$, defining the two points where $A_k$ vanishes. 
Hence, we find that the Floquet ground state is ordered (F-AFM) for $|g-\omega|<J$ and paramagnetic (F-PM) for $|g-\omega|>J$. 

This analysis allows us to identify three regimes. 
(1) In the adiabatic regime, $\omega\ll J$, the winding condition reduces to $g<J$ and we recover the phase diagram of the static XY chain: an ordered phase for $g<J$, a paramagnet for $g>J$, and a critical point in between. 
(2) In the high-frequency regime, $\omega\gg J,g$, the winding condition cannot be met, and only the paramagnetic and critical phases of the isotropic XX chain survive, in accordance with $\mae_k\to\Delta_k$. 
(3) In between, at $\omega\sim\Delta_k$, photon absorption and emission close down the average-energy gap over an extended region, producing the critical phase that separates the F-AFM phase at low frequencies from the F-PM phase; this is the nonequilibrium resonance regime in which the three parameter points of Fig.~\ref{fig:XY} are located. 
The existence of a single resonance regime, rather than a cascade of resonances at $\omega\approx\Delta_k/n$, is a consequence of the integrability of the model: the circular drive couples only to the first harmonic, and the free-fermion structure forbids other multi-photon processes.

\subsubsection{Average-energy versus quasienergy ground state}
\label{app:subsec:XY_quasienergy}

The correlation functions in Sec.~\ref{app:subsec:XY_phases} were measured in the average-energy ground state, and depend on the single-particle average energies $\mae_k$, which carry a sign and thereby determine the filling. 
It is interesting to compare them with the result one obtains by filling the quasienergy bands instead. 

Since the quasienergies $\epsilon^\F_\pm/2 = \pm(\epsilon_k-\omega)/2$ are defined only modulo $\omega$, this requires us to choose a Floquet zone; following common practice, we take the first Floquet zone $(-\omega/2,\omega/2]$ and declare the single-particle states with the lower quasienergy in this zone to be occupied. 
We emphasize that this choice is arbitrary: e.g., the zone $(0,\omega]$ reverses the assignment wherever the two bands sit on each side of $\epsilon^\F=0$; importantly, no choice of zones is distinguished by the dynamics, cf.~Sec.~\ref{app:sec:gft}. 

The quasienergy `gap' of this construction closes when the two bands $\epsilon_k-\omega\in\omega\mathbb{Z}$ coincide modulo $\omega$,
\begin{equation}
    \label{eq:gft:circXY_quasigap}
    (\Delta_k-\omega)^2+A_k^2=(n\omega)^2\,,\qquad n\in\mathbb{Z}_{>0}\,,
\end{equation}
i.e., when $\epsilon_k$ equals an integer multiple of the drive frequency for some momentum $k$; between gap closings, the winding criterion applies as before, but now to the many-body state filled according to quasienergy.  

Figure~\ref{fig:XY_app}(a,b) juxtaposes the two phase diagrams at $A/J=1/2$. 
As expected, they agree in the high-frequency regime, where quasienergies and average energies coincide and the geometric phases vanish, cf.~Sec.~\ref{app:sec:comparison}.
Interestingly, however, they differ in the two other regimes.

In the adiabatic regime ($\omega\ll J$, $g$ arbitrary), the average-energy ground state is gapped Fig.~\ref{fig:XY_app}(c, blue lines) and ordered, as expected by continuity with the static XY chain. By contrast, the quasienergy bands cross multiple times because of increasing folding where $\epsilon_k$ sweeps through multiples of $\omega$ as $k$ varies; the quasienergy `ground state' is then calculated to be critical since the quasienergy bands are flat and degenerate at the edge of the Floquet zone, Fig.~\ref{fig:XY_app}(c, black lines), although no Floquet eigenstate changes character there: the two states whose quasienergies coincide modulo $\omega$ are distinct eigenstates of $H_\F$ with different average energies; which of them is called the ground state depends on the choice of Floquet zone. 

In the resonance regime, conversely, the average-energy gap closes by photon absorption and the Floquet ground state is critical with algebraic correlations, Fig.~\ref{fig:XY}(d), whereas the quasienergy bands remain separated and the quasienergy `ground state' is ordered, Fig.~\ref{fig:XY_app}(d, black lines). 

Hence, there are parameter regimes in which the ground state of $\mAE$ is F-AFM while the quasienergy `ground state' is critical, and vice versa, as stated in Sec.~\ref{sec:FGS_order}. 
The discrepancy is entirely accounted for by the geometric phase, Eq.~\eqref{app:eq:eps_ae_gamma}, which shifts the quasienergy of each Floquet state relative to its average energy by $\gamma_k/T$. 
At the F-AFM point, Fig.~\ref{fig:XY_app}(e), the average-energy band is gapped but $\gamma_k$ sweeps a full $2\pi$ across the Brillouin zone (the values $\pm\pi$ in the plot are the folding of $\gamma_k$ into $(-\pi,\pi]$), so that $\gamma_k/T$ sweeps a full photon energy $\omega$ and the quasienergy of the lower average-energy band is pushed onto the zone boundary; the folded quasienergy spectrum is gapless because the geometric phase, not the average energy, winds as a result of photon absorption~\cite{schindler2025geometric}. 
At the F-critical point, Fig.~\ref{fig:XY_app}(f), $\gamma_k$ varies by less than $2\pi$ between the gap-closing momenta, at which it "jumps" by $\pi$ because the lower average-energy state switches branch; the quasienergy of this state stays at a distance $\omega/4$ from the zone boundary, and the folded spectrum remains gapped although the average-energy gap closes.

\subsubsection{Open boundary conditions}
\label{app:subsec:XY_obc}

The DMRG simulations of Sec.~\ref{sec:XY} use open boundary conditions, for which the exact reference data in Fig.~\ref{fig:XY} are obtained as follows. 
The Jordan-Wigner transformation~\eqref{eq:JordanWigner} maps Eq.~\eqref{eq:XY_chain} with open ends onto a quadratic fermion Hamiltonian in real space, $H(t)\,\hat{=}\,\tfrac12\boldsymbol\Psi^\dagger\boldsymbol{\mathcal H}(t)\boldsymbol\Psi$ with $\boldsymbol\Psi^\dagger=(f_1^\dagger,\dots,f_L^\dagger,f_1,\dots,f_L)$ and a $2L\times2L$ Bogoliubov--de Gennes matrix $\boldsymbol{\mathcal H}(t)$ with the periodicity of the drive. 
Since quadratic Hamiltonians generate Gaussian dynamics, the many-body Floquet problem reduces to the single-particle one: we integrate the Schr\"odinger equation generated by $\boldsymbol{\mathcal H}(t)$ over one period, diagonalize the resulting $2L\times2L$ Floquet unitary to obtain the single-particle Floquet modes $\ket{\phi_\alpha[t]}$, which come in particle-hole conjugate pairs, and compute their average energies $\mae_\alpha=T^{-1}\int_0^T\dd t\,\bra{\phi_\alpha[t]}\boldsymbol{\mathcal H}(t)\ket{\phi_\alpha[t]}$ by numerical integration; this is the single-particle version of Eq.~\eqref{app:eq:average-energies}. 

The many-body Floquet ground state is the Bogoliubov vacuum of the modes with positive average energy, in direct analogy with Eq.~\eqref{eq:gft:circXY_GS}; its average energy is $\mae_{\rm GS}=-\tfrac12\sum_{\mae_\alpha>0}\mae_\alpha$, and the gap of Fig.~\ref{fig:XY}(b) is the smallest nonzero excitation energy, $\Delta\mae=\min_{\mae_\alpha>0}\mae_\alpha$, which reduces to Eq.~\eqref{eq:gft:circXY_gap} in the TD limit $L\to\infty$. With open boundaries, the ordered phase hosts a quasiparticle mode localized at the ends with $\mae_\alpha\to0$ exponentially in $L$; it is responsible for the twofold degeneracy of the ground state, and is excluded from the minimum. 

All observables of the Gaussian state follow from its covariance matrix. 
The entanglement entropy of Fig.~\ref{fig:XY}(c) is obtained from the eigenvalues of the covariance matrix restricted to half the chain, and the correlation function of Fig.~\ref{fig:XY}(d), which involves the Jordan-Wigner string between the two spins, is evaluated by Wick's theorem as the Pfaffian of the corresponding Majorana covariance matrix~\cite{schindler2026thesis}. 
We stress that this procedure produces the \emph{exact} average-energy ground state of the finite chain, without any variational approximation; the agreement of the DMRG data with it in Fig.~\ref{fig:XY}(a)--(e), to within $|\delta\mae_{\rm GS}|/J\lesssim10^{-5}$ at $\chi_X=128$, serves as a benchmark for the algorithm of Appendix~\ref{app:sec:mps_algo}.

\subsection{Continuously driven Ising chain}
\label{app:sec:Ising}

\begin{figure}[t!]
  \centering
  \includegraphics[width=\columnwidth]{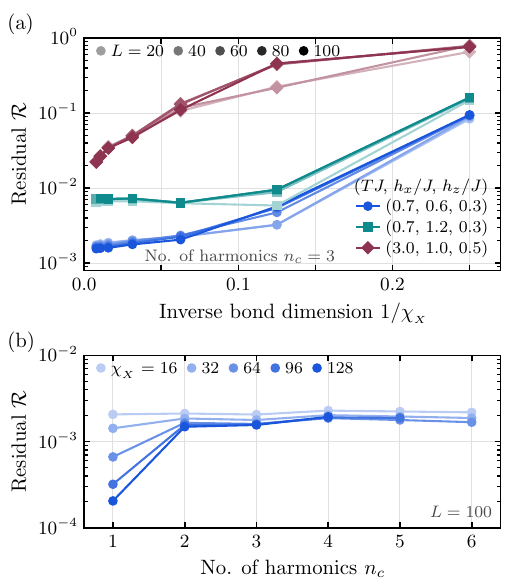}
  \caption{Convergence of the DMRG algorithm to solve the linear equation \eqref{app:eq:normal_eq}, measured by the relative
    residual $\mathcal{R}=\lVert\mathcal{L}^2\lvert X\rangle\rangle-\lvert
    \partial H\rangle\rangle\rVert/\lVert\lvert \partial H\rangle\rangle\rVert$.
    (a) $\mathcal{R}$ against the inverse FDMRG bond dimension $1/\chi_X$ at $n_c=3$. Colour is the parameter set $(TJ,h_x/J,h_z/J)$ and shade is the system size, $L=20$--$100$ from light to dark within each colour. The residual $\mathcal{R}$ curves behave similarly for different system sizes $L$. 
    (b) Residual $\mathcal{R}$ against the number of harmonics $n_c$ at $L=100$ for the F-AFM $(0.7,0.6,0.3)$ set, one curve per bond dimension $\chi_X$. The curves separate significantly only at $n_c=1$ and reach a similar plateau for $n_c\geq2$.
    }
  \label{app:fig:residual}
\end{figure}

\begin{figure*}[t!]
  \centering
  \includegraphics[width=\textwidth]{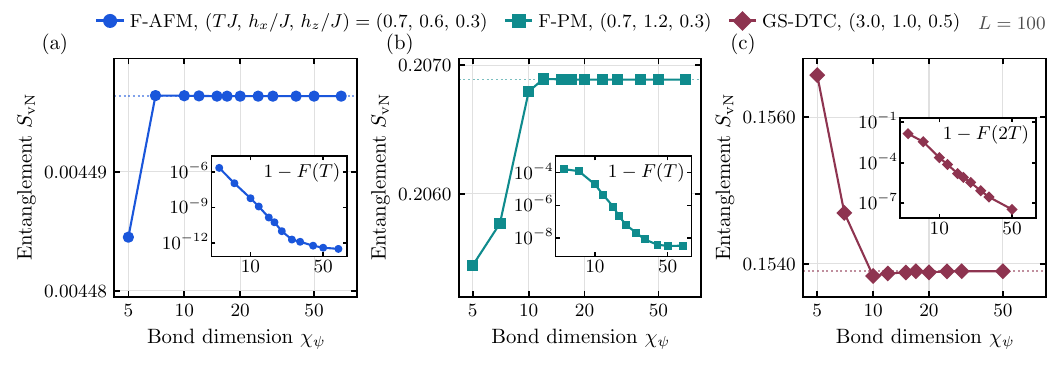}
  \caption{Bond-dimension convergence of the Krylov-refined Floquet eigenstate at $L=100$ and Krylov tolerance $\epsilon_{\mathrm{tol}}=10^{-20}$ with $20$ restart iterations, for the three parameter sets $(TJ,h_x/J,h_z/J)$: (a) F-AFM $(0.7,0.6,0.3)$, (b) F-PM $(0.7,1.2,0.3)$, (c) GS-DTC $(3.0,1.0,0.5)$. Main panels: the half-chain entanglement entropy $S_{\mathrm{vN}}$ against the Floquet ground state bond dimension $\chi_\psi$. The dotted line marks the value at the largest $\chi_\psi$ of each sweep.
  Insets: infidelity of the same states versus $\chi_\psi$ on a log-log scale. The GS-DTC refinement targets the two-period return probability, and hence $1-F(2T)$ is plotted in (c).}
  \label{app:fig:chi_sets}
\end{figure*}

\begin{figure}[t!]
  \centering
  \includegraphics[width=\columnwidth]{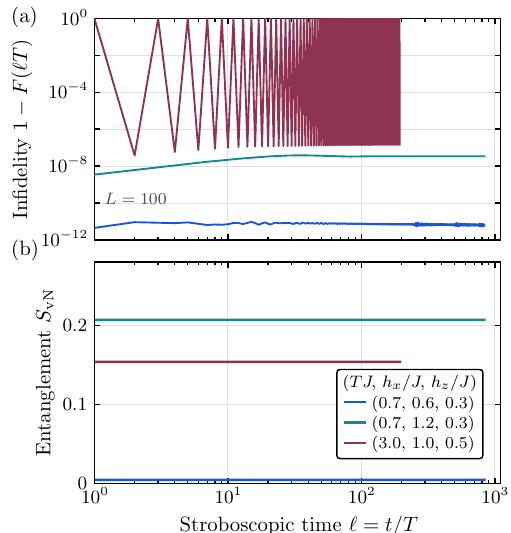}
  \caption{Long-time stroboscopic evolution of the Krylov-refined Floquet eigenstates at $L=100$ for the three parameter sets $(TJ,h_x/J,h_z/J)$ considered in Figs.~\ref{app:fig:residual} and \ref{app:fig:chi_sets}. (a) Infidelity $1-F(\ell T)$ against the stroboscopic time $\ell=t/T$. The infidelity at $(0.7,0.6,0.3)$ and $(0.7,1.2,0.3)$ stays below $\sim10^{-11}$ and $\sim10^{-8}$; the $(3.0,1.0,0.5)$ state returns to itself only at even $\ell$ due to period doubling with an infidelity below $\sim10^{-6}$ and is orthogonal to the initial state at odd $\ell$. 
  (b) The half-chain entanglement entropy $S_{\mathrm{vN}}$ of the same time-evolved states, showing no sign of growth at longer times, as expected for Floquet eigenstates.}
  \label{app:fig:long_time}
\end{figure}

\begin{table*}[t!]
  \centering
  \label{app:tab:gap_extrapolation}
  \begin{ruledtabular}
  \begin{tabular}{llcccl}
    Set & $(TJ,h_x/J,h_z/J)$ & Gap & Fit form & Sizes $L$ (number)
      & Coefficients and $L\to\infty$ limit \\
    \colrule
    F-AFM  & $(0.7,0.6,0.3)$ & $\Delta\mae_{10}$
      & $a_0+A\,e^{-L/\xi}$
      & $16$--$40$ \ ($5$)
      & $A=1.110$, \ $\xi=1.473\pm0.130$ \\
    & & & & & $a_0=\Delta\mae_{10}(\infty)/J=(0.2\pm1.6)\times10^{-9}$ \\[2pt]
    F-AFM  & $(0.7,0.6,0.3)$ & $\Delta\mae_{21}$
      & $a_0+a_1L^{-1}+a_2L^{-2}$
      & $16$--$100$ \ ($8$)
      & $a_2=+8.635$, \ $a_1=-0.09633$ \\
    & & & & & $a_0=\Delta\mae_{21}(\infty)/J=0.27664\pm0.00027$ \\[4pt]
    F-PM   & $(0.7,1.2,0.3)$ & $\Delta\mae_{10}$
      & $a_0+a_1L^{-1}+a_2L^{-2}$
      & $16$--$100$ \ ($8$)
      & $a_2=+6.730$, \ $a_1=+0.3122$ \\
    & & & & & $a_0=\Delta\mae_{10}(\infty)/J=0.21246\pm0.00045$ \\[2pt]
    F-PM   & $(0.7,1.2,0.3)$ & $\Delta\mae_{21}$
      & $a_0+a_1L^{-1}+a_2L^{-2}+a_3L^{-3}$
      & $16$--$100$ \ ($8$)
      & $a_3=-174.3$, \ $a_2=+30.61$, \ $a_1=+0.5723$ \\
    & & & & & $a_0=\Delta\mae_{21}(\infty)/J=-0.0045\pm0.0007$ \\[4pt]
    GS-DTC & $(3.0,1.0,0.5)$ & $\Delta\mae_{10}$
      & $a_0+A\,e^{-L/\xi}$
      & $16$--$40$ \ ($5$)
      & $A=0.1814$, \ $\xi=7.884\pm0.011$ \\
    & & & & & $a_0=\Delta\mae_{10}(\infty)/J=(3.0\pm4.8)\times10^{-6}$ \\[2pt]
    GS-DTC & $(3.0,1.0,0.5)$ & $\Delta\mae_{21}$
      & $a_0+a_1L^{-1}+a_2L^{-2}$
      & $16$--$140$ \ ($12$)
      & $a_2=+8.909$, \ $a_1=+1.139$ \\
    & & & & & $a_0=\Delta\mae_{21}(\infty)/J=0.0638\pm0.0017$ \\
  \end{tabular}
  \end{ruledtabular}
  \caption{Finite-size extrapolation of the two spectral gaps
    $\Delta\mae_{10}=\mae_1-\mae_0$ and
    $\Delta\mae_{21}=\mae_2-\mae_1$ shown in
    Fig.~\ref{fig:Ising_FAFM}(a) (F-AFM, F-PM) and Fig.~\ref{fig:Ising_GS-DTC}(a)
    (GS-DTC). Open gaps (and the closing gap $\Delta{\mae}_{21}$ in the F-PM) are extrapolated with a polynomial in $1/L$,
    $\Delta\mae(L)/J=\sum_{k=0}^{p}a_k L^{-k}$, whose constant term $a_0$
    is the $L\to\infty$ value marked by a star in the figures. The splitting of
    the symmetry-broken doublet ($\Delta{\mae}_{10}$ in the F-AFM and the GS-DTC point) closes exponentially
    and is fitted against $\Delta\mae(L)/J=a_0+A\,e^{-L/\xi}$; only system sizes above the numerical resolution floor are used. All uncertainties are least-squares errors.}
\end{table*}

In this section, we discuss three truncation controls to achieve the Floquet ground state with good accuracy. The bond dimension $\chi_X$ of the MPS form $|X\rangle\!\rangle$ of the matrix-product operator $\hat X$, the number of harmonics $n_c$ retained to represent the
linear equation Eq.~\eqref{app:eq:normal_eq}, and the bond dimension $\chi_\psi$ of the Floquet eigenstate itself considered in the Krylov-Arnoldi optimization. In the following, we study these truncations for the three parameter sets considered in the main text: the Floquet antiferromagnet
(F-AFM, $(TJ,h_x/J,h_z/J)=(0.7,0.6,0.3)$), the Floquet paramagnet (F-PM, $(0.7,1.2,0.3)$), and the ground-state discrete time crystal point (GS-DTC, $(3.0,1.0,0.5)$).

\subsubsection{Linear solve} 

Figure~\ref{app:fig:residual}(a) shows the relative residual $\mathcal{R}$ of the linear equation against the inverse bond dimension $1/\chi_X$. For every parameter set, $\mathcal{R}$ falls monotonically. In the F-AFM case at $L=100$ it drops from $9.5\times10^{-2}$ at $\chi_X=4$ to $1.6\times10^{-3}$ at $\chi_X=128$, and the F-PM and GS-DTC cases reach $6.4\times10^{-3}$ and $2.2\times10^{-2}$,
respectively. For the three parameter sets, $|X\rangle\!\rangle$ needs different $1/\chi_X$ to converge to the same $\mathcal{R}$. Among these three sets, the low-frequency GS-DTC point needs the largest $\chi_X$ at fixed accuracy.
Importantly, the shading in Fig.~\ref{app:fig:residual}(a) shows that $\mathcal{R}$ decreases almost similarly with $\chi_X$ for different system sizes $L$. So, the cost of a given accuracy, in terms of $\chi_X$, does not grow with $L$, which makes the larger system size calculations of the main text affordable.

Figure~\ref{app:fig:residual}(b) presents the convergence with the number of harmonics $n_c$ for $L=100$ at the F-AFM point. For $n_c\geq2$, all curves for different $\chi_X$, $\mathcal{R}$ plateau to similar values.

\subsubsection{Eigenstate bond dimension in Krylov-Arnoldi optimization} 

Figure~\ref{app:fig:chi_sets} shows the convergence of the Floquet ground states after Krylov-Arnoldi optimization for the same parameters considered above. For each data point, we perform $20$ restart iterations with a fixed $\chi_\psi$. In the main panels, the half-chain entanglement entropy $S_{\mathrm{vN}}$ of the Krylov-refined Floquet ground states is shown against $\chi_\psi$ for $L=100$, with the infidelity of the same states in the corresponding insets. The entropies saturate with increasing $\chi_X$ and at values that are small:
$S_{\mathrm{vN}}=0.004496$ (F-AFM), $0.206887$ (F-PM) and $0.153898$ (GS-DTC). The insets show that the infidelity keeps improving with increasing $\chi_\psi$, as expected. At the largest $\chi_\psi$ considered in each case, the infidelity reaches $3.3\times10^{-13}$ for F-AFM,  $3.5\times10^{-9}$ for F-PM and $3.5\times10^{-8}$ fro GS-DTC.

\subsubsection{Long-time stability} 

The quality of convergence can be further tested by evolving the state for a long time and checking how close it returns to itself under the stroboscopic dynamics after many time periods. Figure~\ref{app:fig:long_time}(a) shows the infidelity of the Krylov-refined states for several hundred drive periods. Here we use the TEBD method with time step $\delta t\,J=0.01$ and fourth-order Trotter decomposition of the exact Hamiltonian $H(t)$. The F-AFM infidelity remains below $10^{-11}$ with no visible drift; the F-PM infidelity remains below $10^{-7}$ and the GS-DTC infidelity stays below $10^{-6}$. In the GS-DTC case, the infidelity reaches unity at odd $\ell$, which reflects the period-doubling behavior. Figure~\ref{app:fig:long_time}(b) shows the half-chain entanglement over the same runs; it stays constant over stroboscopic time, so the matrix-product representation never degrades, and no re-truncation error is accumulated during the time evolution.

\subsubsection{Finite-size extrapolation of the average-energy gaps} 

Table~\ref{app:tab:gap_extrapolation}
presents the quantitative details of the extrapolations drawn in Fig.~\ref{fig:Ising_FAFM}(a) and
Fig.~\ref{fig:Ising_GS-DTC}(a). Two functional forms are used: (i) an excitation gap changes polynomially in $1/L$ and remains finite in the TD limit, whereas (ii) the splitting of a symmetry-broken doublet closes exponentially in the system size $L$. The extrapolated open gaps are: $\Delta\mae_{21}(\infty)/J=0.27664\pm 0.00027$ at the F-AFM point, $\Delta\mae_{10}(\infty)/J=0.21246\pm 0.00045$ at the F-PM point, and at the GS-DTC point $\Delta\mae_{21}(\infty)/J= 0.0638\pm0.0017$. The closing gaps behave as expected for a degenerate doublet: the exponential fits give decay lengths $\xi=1.473 \pm 0.130$ at the F-AFM point and $\xi=7.884 \pm 0.011$ at the GS-DTC point. For the exponentially decaying cases, we consider fewer point (smaller $L$'s) to fit the function, since for larger system sizes the exponentially suppressed splitting reaches the numerical accuracy floor. In summary, in the F-AFM and GS-DTC phases, the lowest two Floquet states form a doublet that is degenerate in the thermodynamic limit and separated from the rest of the excitation spectrum by a finite average-energy gap, while in the F-PM phase the lowest average-energy state is unique.

\begin{figure}[t!]
  \centering
  \includegraphics[width=\columnwidth]{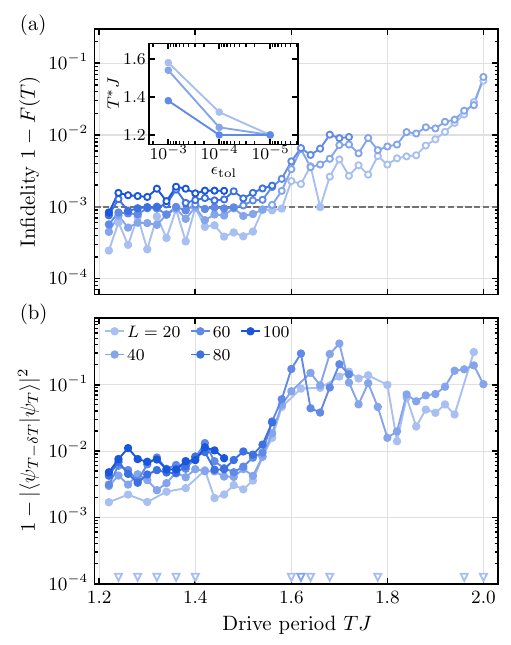}
  \caption{Quasistatic continuation of the Floquet ground state in the drive period, at $h_x/J=0.6$, $h_z/J=0.3$ (the F-AFM point), $J=1$, $n_c=3$. The state at $TJ=1.2$ is prepared using DMRG followed by Krylov refinement; the drive period is then increased in steps of $\delta T\,J=0.02$ with no further DMRG applications. 
  At each step, the converged state of the previous period is used as the Arnoldi seed, and is refined by restarted Krylov Arnoldi at fixed maximum bond dimension $\chi_\psi=70$ and fixed maximum Krylov dimension $m=30$, with at most $10$ restarts; Krylov iterations are terminated as soon as the tolerance falls below $\epsilon_\mathrm{tol}$. 
  The quantities plotted therefore describe the quasistatically continued state, rather than an eigenstate determined independently at each period. 
  Colour shade is the chain length, $L=20$--$100$ from light to dark.
  (a) Krylov one-cycle stroboscopic infidelity $1-F(T)$ against the drive period $T$, with $\epsilon_\mathrm{tol}=10^{-3}$ (dashed line).  Filled symbols mark steps that reached the tolerance within the restart budget and at the fixed Krylov dimension, whereas open symbols mark steps that exhaust it. 
  Inset: period $T^{*}$, up to which every step converged, against $\epsilon_\mathrm{tol}$, for $L=20$, $40$, $60$.
  (b) Step-to-step distance $1-|\langle\psi_{T-\delta T}|\psi_T\rangle|^2$ measured as infidelity between consecutive periods. Steps for which the Arnoldi algorithm returned the initial seed unchanged cannot be shown on a logarithmic axis and are drawn as open triangles at the bottom. Below $TJ\simeq1.5$ the step-to-step distance stays below $10^{-2}$; near $TJ\simeq1.55$ it jumps by more than an order of magnitude to $\sim10^{-1}$.  Past the jump, the continuation likely ceases to follow a single eigenstate.
  }
  \label{app:fig:adiabatic}
\end{figure}

\subsubsection{Quasistatic continuation in the drive period}
\label{app:adiabatic}

We tried to locate the heating crossover mentioned in the main text by continuing the Floquet ground state in the drive period $T$ and asking where the continuation breaks down. The initial state at $TJ=1.2$ is computed using DMRG on the variationally approximated $\tilde\mAE$ (with $n_c=3$) followed by Krylov refinement, at $h_x/J=0.6$, $h_z/J=0.3$ (a F-AFM point). The period is then increased in steps of $\delta T\,J=0.02$ with no further DMRG applications: at each step the converged state of the previous period is used as an Arnoldi seed and refined by the Krylov Arnoldi algorithm at fixed maximum bond dimension $\chi_\psi=70$ and fixed maximum Krylov dimension $30$, with at most ten restarts and a fixed tolerance $\epsilon_{\mathrm{tol}}$. Figure~\ref{app:fig:adiabatic} shows the infidelity versus drive period for $\epsilon_{\mathrm{tol}}=10^{-3}$ and system sizes $L=20$--$100$. The filled symbols mark steps that fulfill the tolerance within the restart budget; open symbols mark steps that exhaust it before reaching convergence.
We observe:

(i) Stable continuation up to $TJ\simeq1.5$. Below this period, the one-period stroboscopic infidelity $1-|\langle\psi_{T-\delta T}|\psi_T\rangle|^2$ has a floor close to $10^{-3}$ [Fig.~\ref{app:fig:adiabatic}(a)] and the step-to-step distance $1-|\langle\psi_{T-\delta T}|\psi_T\rangle|^2$ lies within a plateau below $10^{-2}$ [Fig.~\ref{app:fig:adiabatic}(b)]. A plateau in the latter is the signature of the continuation of a single eigenstate deforming smoothly with $T$.

(ii) The infidelity then grows by about two orders of magnitude with no sharp feature. The step-to-step distance, by contrast, jumps: after $TJ\simeq1.5$ it rises from the plateau to $\sim 10^{-1}$. This suggests that beyond this point the continuation no longer follows a single eigenstate.

(iii) The infidelity increases monotonically with $L$ at every period, across $L=20$--$100$. Within the lower plateau of the infidelity, the larger the $L$, the earlier it caps the restart budget (goes from solid to open symbols). This is the finite-size precursor of the ergodic regime: the continuation survives to later periods the smaller the system size.

(iv) The crossover period cannot be sharply determined from the data, see Fig.~\ref{app:fig:adiabatic}(a) [inset] showing the period $T^{*}$ up to which every step converged. With decreasing tolerance, $T^{*}$ also decreases.

\subsubsection{Computing the one-cycle Floquet holonomy}

Here we give the details of the computation of the non-Abelian holonomy in the many-body Floquet ground state doublet discussed in the main text.

The one-cycle holonomy within the degenerate ground-state doublet is defined as 
\begin{equation}
    W(T,0)=M(T,0)\exp\bigl(iT\,\mathrm{diag}(\mae_1,\mae_2)\bigr)\, ,
\end{equation}
where $M_{ab}(T,0)=\langle n^\F_a(0)|n^\F_b(T)\rangle$, $a,b\in\{1,2\}$, is the overlap matrix of the two Floquet ground states $|n^\F_a(T)\rangle$ that are degenerate in average energy, and the exponential removes their dynamical phases. At the GS-DTC point in the $L=100$ site chain -- $TJ=3.0$, $h_x=J$, and $h_z=0.5J$ -- we find numerically
\begin{equation*}
W(T,0) = \begin{pmatrix}
    \begin{array}{@{}r@{}} 0.000001910 \\ {}+0.000000832\,i \end{array} &
    \begin{array}{@{}r@{}} 0.925777571 \\ {}+0.378068489\,i \end{array} \\[2.2ex]
    \begin{array}{@{}r@{}} 0.929926569 \\ {}+0.367745116\,i \end{array} &
    \begin{array}{@{}r@{}} 0.000003371 \\ {}+0.000001447\,i \end{array}
    \end{pmatrix}.
\end{equation*}

The Berry-phase splitting $\Delta\gamma$ is the difference between the arguments of the two eigenvalues of $W(T,0)$. For the matrix above we obtain $\Delta\gamma/\pi = 1 - 5\times10^{-8}$, i.e.\ the $\pi$-splitting characteristic of the DTC phase.

Repeating the procedure at a few different drive periods $T$ within the corresponding phase of matter produces the data sets displayed in Fig.~\ref{fig:Ising_GS-DTC}(f) for the GS-DTC and Fig.~\ref{fig:Ising_FAFM}(d, inset) for the F-AFM.  


\bibliography{references.bib}

\end{document}

%% file: preamble/macros.tex
\newcommand{\identity}{\mathds{1}}

\newcommand{\hilbert}[1]{\mathbb{#1}}

\renewcommand{\i}{\text{i}}

\newcommand{\im}{\text{im}\,}

\newcommand{\floquetbra}[1]{\left< \left< #1 \right| \right.}
\newcommand{\floquetket}[1]{ \left| \left. #1 \right>\right> }
\newcommand{\floquetbraket}[2]{\left< \braket{#1}{#2} \right> }

\newcommand{\timeket}[1]{\left| #1 \right)}

\newcommand{\sambe}[1]{\hat{#1}} 

\newcommand{\floquethilbert}{\hilbert{F}}

\newcommand{\Hctrl}{H_\text{ctrl}}
\newcommand{\ctrl}{\text{ctrl}}

\newcommand{\HCD}{H_\text{CD}}
\newcommand{\agp}{A}

\newcommand{\mAE}{\text{\AE}}
\newcommand{\mae}{\text{\ae}}

\newcommand{\K}{\text{K}}
\newcommand{\F}{\text{F}}

\newcommand{\HF}{H_\F}
\newcommand{\UF}{U_\F}

\newcommand{\HK}{H_\K}

\newcommand{\AF}{\agp_\F}
\newcommand{\AK}{\agp_\K}

